\documentclass[acmsmall]{acmart}
\AtBeginDocument{%
  \providecommand\BibTeX{{%
    \normalfont B\kern-0.5em{\scshape i\kern-0.25em b}\kern-0.8em\TeX}}}

\acmJournal{TOIS}%TSAS}
\usepackage{makecell}
\usepackage{graphicx,multirow}
\usepackage{natbib}
\usepackage{subcaption}
\usepackage{algorithm}
\usepackage{algorithmic}
\usepackage{enumitem}
\usepackage{xcolor}
\usepackage{lineno}
\newenvironment{breakablealgorithm}
  {% \begin{breakablealgorithm}
   \begin{center}
     \refstepcounter{algorithm}% New algorithm
     \hrule height.8pt depth0pt \kern2pt% \@fs@pre for \@fs@ruled
     \renewcommand{\caption}[2][\relax]{% Make a new \caption
       {\raggedright\textbf{\ALG@name~\thealgorithm} ##2\par}%
       \ifx\relax##1\relax % #1 is \relax
         \addcontentsline{loa}{algorithm}{\protect\numberline{\thealgorithm}##2}%
       \else % #1 is not \relax
         \addcontentsline{loa}{algorithm}{\protect\numberline{\thealgorithm}##1}%
       \fi
       \kern2pt\hrule\kern2pt
     }
  }{% \end{breakablealgorithm}
     \kern2pt\hrule\relax% \@fs@post for \@fs@ruled
   \end{center}
  }
\makeatother

\begin{document}

\title{A Greedy Approach for Increased Vehicle Utilization in Ridesharing Platforms}

\author{Aqsa Ashraf Makhdomi}
\email{makhdoomiaqsa@gmail.com}
\orcid{0000-0001-8736-1316}
%\email{webmaster@marysville-ohio.com}
\author{Iqra Altaf Gillani}
\email{iqraaltaf@nitsri.ac.in}
\affiliation{%
  \institution{NIT Srinagar}
  \streetaddress{Hazratbal}
  \city{Srinagar}
  \state{Jammu \& Kashmir}
  \country{India}
  \postcode{190020}
}

%\affiliation{%
 % \institution{NIT Srinagar}%National Institute of Technology}
  %\streetaddress{Hazratbal}
  %\city{Srinagar}
  %\country{India}}
%\email{iqraaltaf@nitsri.ac.in}

\renewcommand{\shortauthors}{Makhdomi, Gillani}

\begin{abstract}
In recent years, ridesharing platforms have become a prominent mode of transportation for the residents of urban areas. One of the fundamental challenges faced by these platforms is providing efficient route recommendations to drivers. Existing studies in this direction have primarily focused on recommending routes based on the expected passenger demand.
Despite the existing works, statistics have suggested that these services cause increased greenhouse emissions as they do not utilize the vehicle capacity efficiently. To address this, we propose to recommend routes that will fetch multiple passengers simultaneously which will result in increased vehicle utilization and decrease the effect of these systems on  environment.  We establish that route recommendation is NP-hard and develop a \emph{k}-hop-based sliding window to reduce the search space from the entire road network to a window. We further show that maximizing expected passenger requests within a window is submodular which provides greedy algorithms as a solution to optimize the objective function. In addition to route recommendation, we address the challenge of determining the minimum number of vehicles required to fulfill all passenger requests in a given area.
Extensive simulations on the datasets of New York City and Washington DC demonstrate superior performance by our proposed model.
\end{abstract}




\maketitle

\section{Introduction}

Due to the prominent development of  internet and GPS-enabled services, the dynamics of ride-hailing platforms have changed completely. People have now become used to travelling over these services for their day-to-day activities. However, this increased surge  has resulted in their shortage over peak hours. In order to avoid the absence of vehicles during peak hours and provide efficient utilization of vehicles, ridesharing has been proposed as a solution wherein the different users with similar routes 
share a single vehicle. This potentially   brings up   many   benefits   for an  urban  city and results in alleviating  traffic  congestion, providing eco-friendly rides, and reducing  the waiting time of passengers \cite{Alonso:NAS_2017}. 
Owing to 
the benefits made by these platforms, 
they have received significant attention from researchers around the globe. Various methods have been proposed for effective matching \cite{Li:ElsevierTransport_2020},  route planning \cite{Tong:ACMTrans_2022}, and route recommendation  \cite{Yuen:ACMWWW_2019} in ridesharing platforms.

%Although ridesharing services are designed to decrease greenhouse emissions by pairing multiple passengers with similar time schedules in a single vehicle, it has been found they are not effectively utilized and they do not utilize the available vehicle capacity in an efficient manner. 
Despite their popularity, it has been found that the ridesharing services do not utilize the available vehicle capacity in an efficient manner.
%and their usage results in higher CO$_2$ emissions, vehicle miles travelled, and vehicle hours delayed in comparison to private vehicles \cite{Article:ridesharingdisadv}. The higher emissions are due to the cruising of drivers without having a passenger in the vehicle, which results in more distance travelled in comparison to private vehicles. It has been found that transport network companies spend $42\%$ of their service time cruising around or waiting for a passenger, which results in an estimated $50\%$ more emission of CO$_2$ in comparison to private vehicles \cite{Article:UberPollution_2020,Diao:Nature_2021}.
%These statistics demand an efficient recommendation strategy for ridesharing platforms which will direct routes to the driver that have an increased probability of finding passengers. This will decrease the distance travelled by the vehicle in search of passengers and will result in lower emissions of greenhouse gases. In this work, we propose a route-recommendation system for ridesharing platforms, that will predict future requests through the use of Graph Neural Networks \cite{Ashraf_archive:2022} and recommend the routes that have a higher probability of finding the  passengers earlier on their way. This will result in a decrease in extra miles travelled by the vehicle and thereby contribute to eco-friendly rides.
%The other limitation of current ridesharing platforms is that they do not utilize the available vehicle capacity in an efficient manner. 
Statistics have suggested that  ridesharing platforms share only $15\%$ of their rides \cite{Article:UberPollution_2020}  whereas $73\%$ of rides can be shared with a slight increase in waiting time of passengers \cite{Cai:ElsevierEnergy_2019} which can result in 
the efficient utilization of vehicles.  
Therefore, it becomes important to design effective recommendation systems that predict future requests and pair up multiple passengers with similar schedules in a single vehicle which will reduce the fossil fuel consumption that contributes both to local air pollution and climate change.

%One fundamental problem in ridesharing platforms is plan- ning the routes shared among the requests (e.g., passengers) for the vehicles (e.g., drivers). Different from other trip planning queries in spatial databases [17, 31, 11, 34, 26], the shared-route here is a sequence of origins (e.g., pickup locations) and destinations (e.g., delivery locations), which also satisfies the constraints (e.g., deadline constraint) set by the platform. Moreover, these shared-routes are usually planned based on certain optimization objectives.

 % while considering the detour constraints of the users in the vehicle.

Existing works in this direction have recommended the routes with higher expected passenger demand \cite{Yuen:ACMWWW_2019}. These works have predicted the number of passenger requests that arrive at different locations and directed the routes with the highest number of expected requests to the drivers. 
However, passenger mobility patterns can be better analyzed if the origin (the place from where the passenger request arrives), as well as the destination (the place to which the passenger wants to travel) of requests, can be predicted in advance. 
When the origin and destination locations of passenger requests are predicted, ridesharing platforms gain the ability to identify passengers with similar or overlapping travel routes. As a result, it can recommend routes to drivers where multiple passengers have similar routes and a single vehicle can accommodate  them, which reduces the overall number of vehicles required to serve the same number of passengers. 
To the best of our knowledge, none of the existing works so far have recommended the routes to ridesharing platforms keeping in consideration the origin and destination of requests.

%The route recommendation system for origin-destination prediction of requests involves searching the origin of all requests and see if the passengers with similar routes can be paired in a single vehicle. The trip planning queries in spatial databases \cite{Sharifzadeh:vldb_2008,Mahin:ACMTrans_2019} cannot be applied to our proposed model as it involves  searching the origin and destination of all the requests, while considering some requests are in the same vehicle. We demonstrate that the problem of route recommendation using origin-destination of requests is NP-Hard and no polynomial time algorithm exists that can solve it on a deterministic machine. To solve it, we develop a \emph{k}-hop-based sliding window approach which reduces the search space from the entire road network to a window and thereby decreases the time-complexity of the model. Moreover, the underlying objective function of the proposed function, which utilizes the origin and destination of requests, is submodular and greedy algorithms are known to provide well-known approximation guarantees for these functions and we have applied them to maximize our objective function within the corresponding window. 
We demonstrate that the problem of route recommendation using the origin and destination of passenger requests is NP-Hard, and no polynomial time algorithm exists that can solve it on a deterministic machine. In order to overcome it, we develop a \emph{k}-hop-based sliding window approach that will reduce the search space from the 
entire road network to a window and thereby improve the time complexity of the model.
Moreover, the underlying objective function of the proposed function, which utilizes the origin and destination of requests, is submodular and greedy algorithms are known to provide well-known approximation guarantees for these functions and we have applied them to maximize our objective function within the corresponding window.

After applying the greedy strategy, we determine the optimal fleet size required by the ridesharing companies.  The optimal fleet size represents the minimum number of vehicles that ridesharing companies must possess to effectively service all passenger requests within a specific city. 
The determination of the optimal fleet size establishes a lower limit on the vehicle count necessary to fulfil the requests within the designated area and gives ridesharing companies an overview of resources required over different time instants. It is determined by converting the graph that stores the origin and destination of requests, into a vehicle count graph, and applying minimum path cover on it.

In this paper, we  develop an eco-friendly route recommendation system for ridesharing platforms that decreases the extra miles travelled by vehicles by predicting the origin and destination of passenger requests
%areas that would fetch more requests 
and thereby pairs these requests effectively in a single vehicle. %Moreover, while recommending routes we consider passenger satisfaction as an essential factor and ensure their detour ratio which is the ratio of the distance travelled between the source and destination of passengers to the distance of the shortest path between them is bounded by a threshold value. 
The proposed approach utilizes a  $k$-hop window, to decrease the complexity of route recommendation from the road network to a small area and slides this window forward in the direction of expected passenger requests. 
Through this approach, we calculate the fleet size required by our proposed model to cover the entire area which determines the resource utilization of ridesharing platforms over different time periods of the day.

Our key contributions can be summarized as follows:
%\vspace{-5mm}
\begin{itemize}
    \item We 
    propose an eco-friendly route recommendation system for the drivers of the ridesharing platforms which reduces pollution and promotes sustainability. %hazardous effects of these platforms 
    %and creates a sustainable environment.
\item We show that the proposed problem of route recommendation is NP-hard. We overcome the computational complexity by developing a \emph{k}-hop-based sliding window algorithm that reduces the exponential search space from all possible paths in the entire road network to a window.
\item We demonstrate that the underlying objective function of our proposed model is submodular and thereby  greedy algorithms can be used to optimize them, as they are known to provide well-known approximation bounds for submodular functions.
\item A vehicle count graph is constructed to determine the minimum fleet size required to cover the requests that arrive in the ridesharing platforms. This is done by reducing the origin-destination request arrival graph and applying the minimum path cover to it. % for determining the fleet size.  
%\textcolor{blue}{see this point}
\item We evaluate the performance of our proposed model extensively on real-world datasets from New York and Washington DC across different metrics. The experimental results demonstrate that our proposed route recommendation system improves vehicle utilization, and decreases the count of vehicles on the road.

\end{itemize}

\section{Related Work}
The eco-friendly nature of route recommendation systems can be reviewed from two aspects: $1$)  Effective vehicle utilization, and $2)$ Optimal fleet size. %, and $3)$ Determining the optimal fleet size.
 
 %The first case confers to  both ridesharing and ride-hailing platforms and its proposed solutions  overcome the problem of  drivers roaming around without having a passenger in their vehicle. The second case works from the moment the passenger is in the vehicle and it needs to be paired up with other passengers in order to utilize the available vehicle capacity efficiently.  The works done in this direction are designed for ridesharing systems, as the ride-hailing platforms work in solo mode and have a single passenger that needs to be delivered to the destination, and thus there is no concept of effective vehicle utilization.%\textcolor{blue}{IAG: Last line is not clear. Rewrite it. There is confusion of what you are suppporting.}

 \subsection{Route recommendation}

%Ride-hailing platforms have received significant attention from the research community and several works have been done to optimize their functionality. 

Ride-hailing platforms have received considerable attention in the research community, with numerous efforts focused on optimizing their functionality. Existing studies in this direction have designed matching algorithms \cite{Xu:ACM_SSDM_2020,Ta:IEEETrans_2018,Sun:KDD_2022,Shi:KDD_21} which match the passenger requests with drivers, route planning frameworks \cite{Wang:VLDB_2022,Cheng:ACMSIGMOD_2017,Tong:vldb_2018,Tong:ACMTrans_2022} which insert a new request into the existing route of drivers without changing the origin and destination of passengers already in the vehicle, and route recommendation systems \cite{Garg:ACMKDD_2018,Ji:ElsevierKnowledge_2020,Yengejeh:IEEETrans_2021,Yuen:ACMWWW_2019,SCHREIECK:ElsevierTransport_2016,Thangaraj:IEEEConf_2017,Ta:IEEETrans_2018,Liu:IEEETransMC_2023} that recommend routes to  drivers to ensure they get passengers quickly. In our proposed model, we provide the design of a route recommendation system.

The current route recommendation systems aim to achieve diverse objectives, such as 
reducing the cruising distance of drivers without having a rider in the vehicle \cite{Garg:ACMKDD_2018,Yuen:ACMWWW_2019}, decreasing the waiting time of passengers \cite{Ji:ElsevierKnowledge_2020,Yengejeh:IEEETrans_2021}, and increasing the profit of drivers \cite{Verma:aai_2017,Qu:ACMKDD_2014,Garg:ACMKDD_2018,Qu:IEEETrans_2020,Jiang:Springer_2018,Ding:IEEConf_2013,Guo:ACM_Trans_2021,Yuen:ACMWWW_2019}.   These works have predicted the passenger demand using various deep learning frameworks \cite{Wang:ACMKDD_2019,Wang:ACMTrans_2022} and recommended the routes with the highest passenger demand. Some of these works \cite{Yuen:ACMWWW_2019,SCHREIECK:ElsevierTransport_2016,Thangaraj:IEEEConf_2017,Ta:IEEETrans_2018} have enabled ridesharing and paired multiple riders in a single vehicle. 

 Despite the prior research, statistical analysis reveals that ridesharing platforms currently share a mere $15\%$ of their rides \cite{Article:UberPollution_2020}, while a significant $73\%$ of rides could be shared with a slight increase in the waiting time of passengers \cite{Cai:ElsevierEnergy_2019}. This suggests that there is considerable room for enhancing the utilization of vehicles through the effective design of recommendation systems. Our proposed model overcomes this problem and utilizes the vehicle capacity efficiently by predicting the origin and destination of requests and thereupon pairs the passengers in a vehicle based on their shared routes.

\subsection{Optimal fleet size.}
The determination of the optimal fleet size plays a pivotal role in comprehending resource utilization within ride-hailing platforms. It offers valuable insights to ride-hailing companies regarding the number of vehicles needed to effectively service passenger requests in a specific geographic area. Extensive research has been conducted to determine the optimal fleet size, predominantly employing simulation-based methodologies \cite{Gurumurthy:Elsevier_2018,Vosooghi:Elsevier_2019,Cap:RSS_2018,Fagnant:SpringerTranportation_2018}. 
However, these simulation-based approaches have certain limitations. They heavily depend on initial driver positioning and necessitate multiple rounds of simulations to arrive at the optimal fleet size. %Recognizing this challenge, recent research by Vazifeh \textit{et al.} \cite{Vazifeh:Nature_2018} introduced a network flow-based model to determine the optimal fleet size for ride-hailing platforms.
%Different from others
Recognizing this challenge, Vazifeh \textit{et al.} \cite{Vazifeh:Nature_2018} introduced a network flow-based model to determine the optimal fleet size for ride-hailing platforms.
Their proposed model formulated the optimal request calculation as a minimum path cover on a graph %By solving this graph-based problem, they
and determined the minimum number of vehicles required to effectively service passengers operating in ride-hailing (solo) mode. 
However, their model did not consider vehicle sharing and focused solely on the one passenger-one vehicle scenario.
To address the limitations of previous models, Qu \textit{et al.} \cite{Qu:IEEETrans_2022} designed an optimal vehicle-sharing model specifically for ridesharing platforms (when multiple passengers share a vehicle). However, their model is designed for driver-passenger matching and does not explicitly consider route recommendation. %calculated fleet size when the vehicle was shared amongaimed to match drivers and passengers within a bipartite graph and maximized the overall utility. While their model focused on optimizing ride-matching, it did not explicitly consider route recommendation.
In contrast, our proposed model determines the optimal fleet size for ridesharing platforms and it considers the sharing of vehicles among passengers when routes are recommended to drivers. We aim to calculate the number of vehicles required to efficiently service passengers by reducing the origin-destination request arrival graph to a vehicle count graph and thereafter applying minimum path cover (MPC) to it.

\section{Problem formulation}
In this section, we will introduce definitions and formulate our problem. \looseness=-1

\textbf{DEFINITION 1} (GRID). 
%\textcolor{blue}{IAG: Definition should be brief and formal, not general text.} 
\textit{The road network is divided into a grid which is a collection of $n$ non-overlapping grid cells and is represented as $g=\{g_1,g_2,...,g_n\}$.}
%As the road network is vast, it is modelled by  dividing the entire area into a grid which is a collection of $n$ non-overlapping grid cells and is represented as $g=\{g_1,g_2,...,g_n\}$.}

\begin{figure}[h]
    \centering
\includegraphics[width=0.19\textwidth]{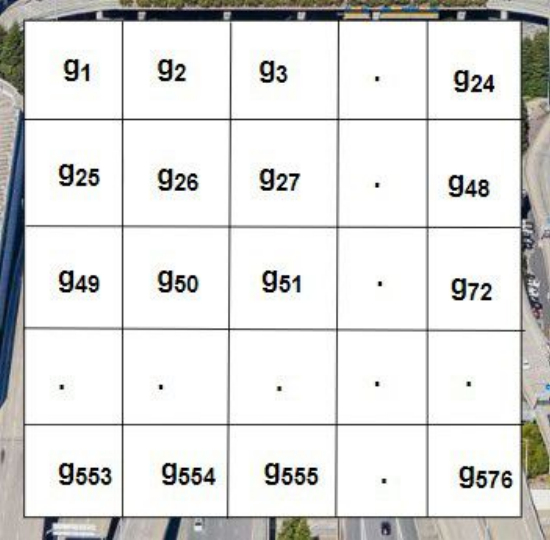}
\vspace*{1mm}
    \caption{Road network represented in the form of a grid}
    \label{fig:mesh1}
\end{figure}

%Each grid cell is defined by its latitude, and longitude pair, and the length between  different grid cells is modelled through the haversine distance between the latitude  and longitude pair of corresponding grid cells.  %This distance is denoted by $L(s)$. 
%Further, 
Within a grid, two grid cells are said to be connected if they are adjacent to each other. 
Figure \ref{fig:mesh1} displays the part of the road network divided into a grid. % with grid cells. 
The road network is a combination of $576$  grid cells and the grid cell $g_\mathrm{1}$ is adjacent to grid cells $g_2,g_\mathrm{25}$, and $g_\mathrm{26}$. 

\textbf{DEFINITION $2$} (ROUTE).\textit{
A route corresponds to a path on the graph and it is a sequence of connected grid cells $P=<g_1,g_2,...g_l>$, where $ l \le n $.} 

\textbf{DEFINITION $3$} ( RIDE-ORDER ).
\textit{A  ride-order $o$ is represented as $<o_s,o_d,o^t>$ where $o_s,o_d$ % \, \epsilon V$  
denote the origin (source) and destination of  ride-order  respectively and $o^t$ denotes the time at which the ride-order is made.}

In this text, we have used ride-order, passenger request and request interchangeably.
%The order set $\mathbb{O}$ contains all the orders which can be paired in a single vehicle.  The orders that can be paired in a single vehicle are determined through their detour ratio, which is the ratio of the distance travelled by the passenger between two nodes to the ratio of the shortest path between the nodes. %which keeps a bound on the extra distance  travelled by each passenger. It is described in detail afterwards. %is the ratio of the distance travelled by the passenger between two nodes to the ratio of the shortest path between the nodes. It is described in detail afterwards. 
%\textcolor{blue}{IAG: Give brief idea about what it is before we introduce formal definition.}
\begin{figure}[t!]
\centering
%\begin{minipage}{.5\textwidth}
  \vspace*{-4mm}\centering  
  \includegraphics[width=0.7\linewidth]{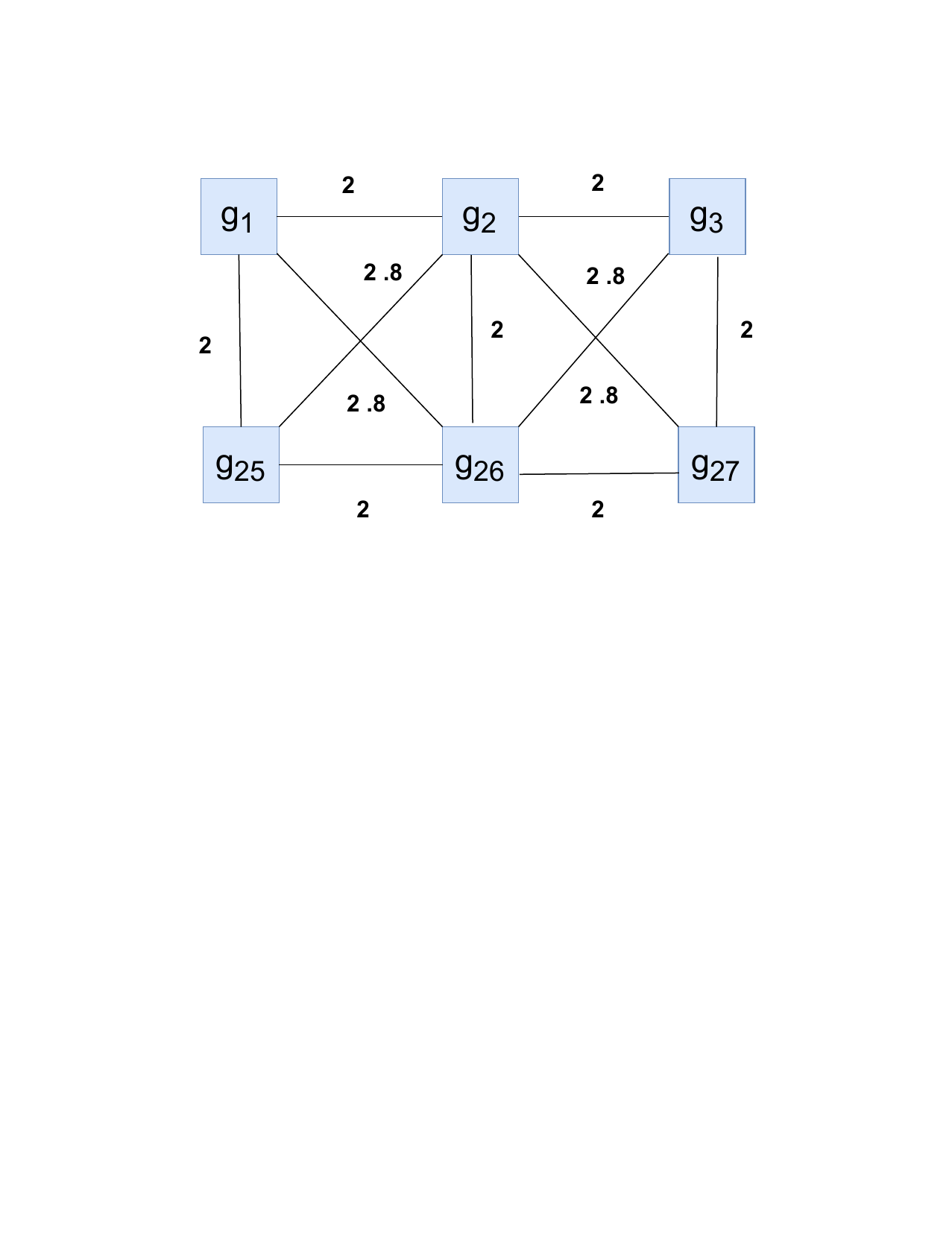}
  \vspace*{-68mm}
    \caption{A particular instance of the road graph $G^R$. In this graph, the vertices represent the grid cells and the edge weight represents the distance between grid cells. The distance is measured between the center points of grid cells. }
    \label{fig:roadinstance}
    \end{figure}
%\end{minipage}%
\begin{figure}
%\vspace*{-1mm}
\begin{minipage}{.4\textwidth}
  \begin{subfigure}{\linewidth} % Add width for subfigure
  \centering
  \vspace*{-4mm}  
  \includegraphics[width=1.5\linewidth]{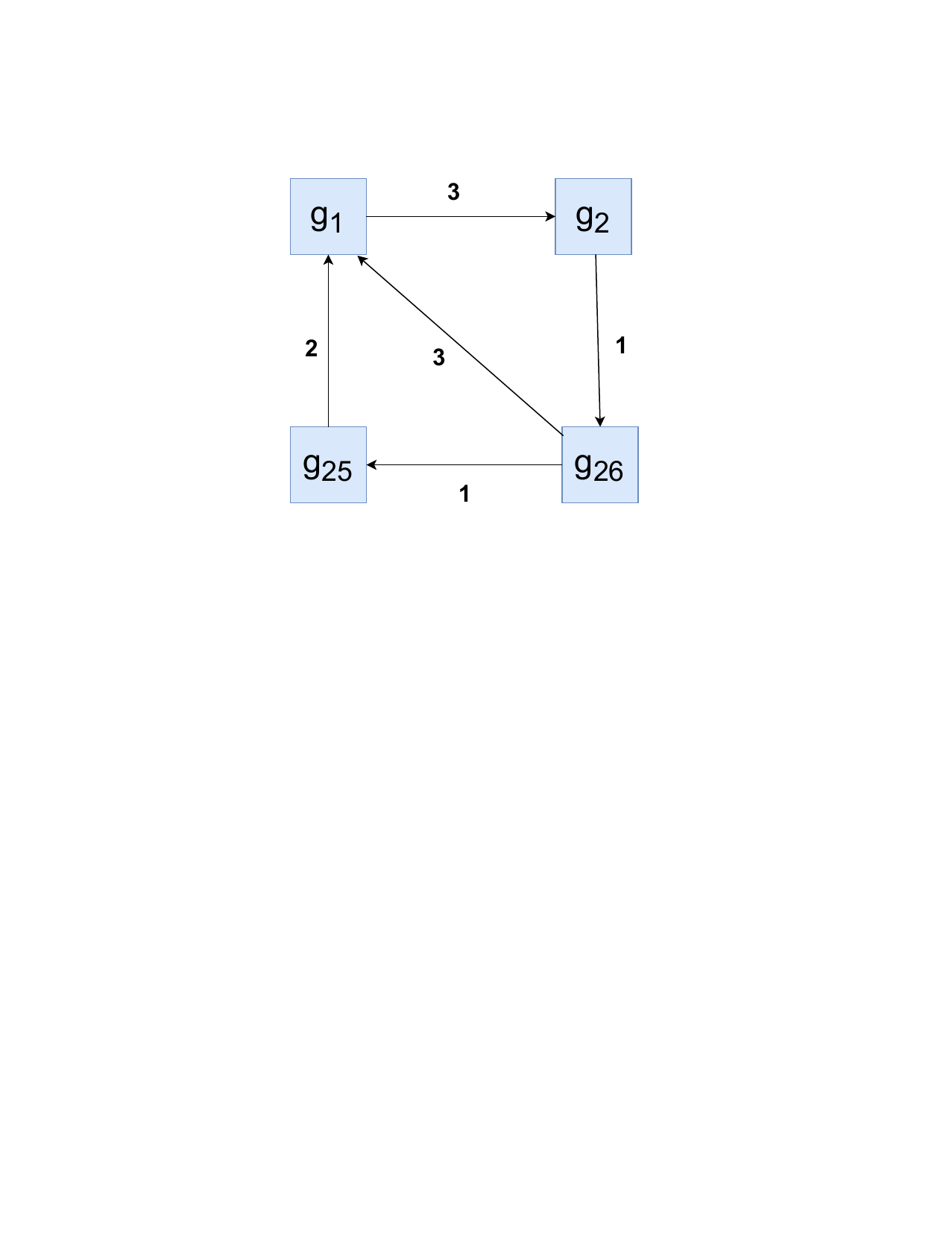}
  \vspace*{-65mm}
    \caption{A particular instance of the request graph $G^Q$. }
    \label{fig:reqinstance}
  \end{subfigure}
\end{minipage}
\begin{minipage}{.56\textwidth}
  \begin{subfigure}{\linewidth} % Add width for subfigure
  \centering
  \vspace*{-9mm}  
  \includegraphics[width=1.1\linewidth]{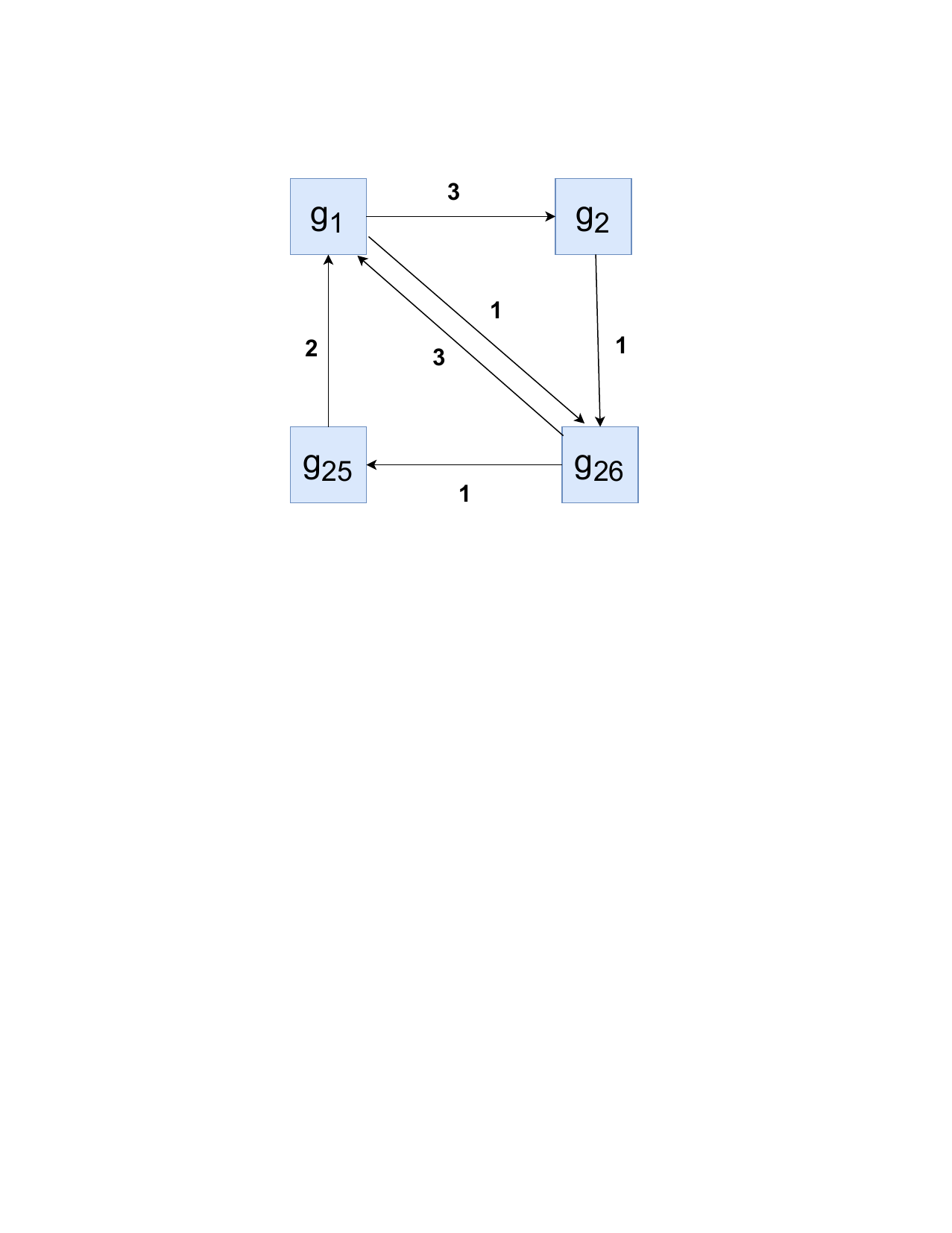}
  \vspace*{-67mm}
    \caption{A particular instance of the target graph $G^T$.}% In this graph, the vertices represent grid cells, edges represent the direction of requests, and the edge weight represents the number of requests that can arrive between the grid cells. For simplicity, we have ignored the edges with $0$ weight.}
    \label{fig:targetinstance}
  \end{subfigure}

\end{minipage}
%\vspace*{-2.7mm}
  \caption{In these graphs, the vertices represent grid cells, edges represent the direction of requests, and the edge weight represents the (a) predicted and (b)  actual number of requests between the grid cells. For simplicity, we have ignored the edges with $0$ weight.}\end{figure}

%    \caption{Target request graph}
%    \label{fig:targetinstance}

\subsection{Graphical modelling}
Our proposed model is represented by a family of subgraphs $G=\bigcup\limits_{k \in \{R,Q,T\}} G^k= \bigcup \limits_{k \in \{R,Q,T\}} (V,E^k,w^k)$ where the nodes $V$ in all the subgraphs  represent the grid cells,  and the edges $E^k$ of subgraph $k$ determine connections between various nodes. We have used the terms nodes or grid cells interchangeably. 
One of the subgraphs $G^k=G^R$, called the road subgraph  is used to represent the structure of the road. The road network is assumed to be divided into a grid with $n$ non-overlapping grid cells as shown in Figure \ref{fig:mesh1}.  The vertices $V$ of the subgraph $G^R$ represent the grid cells, and the edges $E^R$ determine connections between adjacent grid cells. The weight of an edge denoted by $w^R$ represents the distance between the central points of adjacent grid cells. %distance between adjacent grid cells is monitored through the center. and it models the distance between different grid cells. 
Figure \ref{fig:roadinstance} shows the part of the road network in terms of the road graph. In this graph, vertices $\{g_1,g_2,g_3,g_{25},g_{26},g_{27}\}$ represent grid cells and the edges exist between  grid cells if they are adjacent. For instance, vertices $g_1$ and $g_{27}$ are not adjacent, so there is no edge between them. The grid cells $g_1$ and $g_2$ are adjacent, so an edge exists between them with the weight of edge set equal to the distance between the central points of the grid cells. % which  is $2 \, km$ for the grid . 

The other subgraph is the request graph $G^k=G^Q$ which represents the expected passenger requests that arrive in the ridesharing platforms and these requests are predicted through the GNN based architecture described in \cite{Ashraf_archive:2022}.
Each expected request has an origin (the place from which the request arrives) and destination (the place to which the request is headed). The origin and destination of these predicted requests appear on the grid cells and they are represented through the vertices $V$ of the request graph. The edges $E^Q$ of request graph represent the direction of predicted requests i.e, if there is an edge between grid cells $g_i$ and $g_j$, it implies that the passenger has to travel from $g_i$ (origin) to $g_{j}$ (destination). Each edge is associated with an edge weight $w^Q$ that represents the expected number of passenger requests between the grid cells. % i.e., the vertices of this subgraph are the grid cells. 
%Since each grid cell can be the origin to some passenger and the destination to some other, this graph is a complete graph where its vertices represent the grid cells and edges $E^Q$ determine the flow of requests i.e, if there is an edge between grid cells $g_1$ and $g_{26}$, it implies that the passenger has to travel from $g_1$ (origin) to $g_{26}$ (destination). Each edge is associated with an edge weight $w^Q$ that determines the number of passenger requests who want to travel between those grid cells.
%Thus, we can say request graph represents the expected number of requests that arrive between any two grid cells of the road network, which are predicted through GNN based model described in \cite{Ashraf_archive:2022}. %\textcolor{red}{can rmove this line} %Moreover, it is a complete graph as requests can arrive between any two points in the road network. %The expected number of requests between any two grid cells is represented through the edge weight $w^Q$ are 
%This graph contains the predicted the number of requests  that arrive between any two grid cells through the  GNN-based model described in \cite{Ashraf_archive:2022}, and uses the predicted number of requests as the edge weight $w^Q$. It is used as input by the route recommendation algorithm to direct the routes with the highest number of predicted passengers. 
Apart from the request graph, there is a target graph $G^k=G^T$ that contains the actual number of requests between any two grid cells. Its vertices $V$ represent the grid cells, edges $E^T$ represent the flow of requests, and the edge weight $w^T$ represents the actual number of requests between the grid cells. The target graph is not available to the ride-hailing companies beforehand, as these companies do not know the number of requests that will arrive between any two grid cells. This graph is used for monitoring the performance of the proposed model and determining the optimal fleet size as will be discussed in Section \ref{subsec:fleetsize}.
%On the other hand, these platforms predict passenger requests using various deep learning frameworks and recommend routes according to the predicted demand.
The request graph contains the expected number of passenger requests which are predicted through the GNN-based model described in \cite{Ashraf_archive:2022}, and it is used for recommending the routes with the highest expected passenger requests to the drivers.
%and the target graph is used for determining the optimal fleet size which will be discussed in Section \ref{subsec:fleetsize}. %This graph is used for analyzing the performance of the proposed model. % and it is also used for calculating the optimal fleet size which will be discussed in Section \ref{subsec:fleetsize}.% and these requests are predicted through Graph Neural Network (GNN) based model described in \cite{Ashraf_archive:2022}.

Figures \ref{fig:reqinstance} and \ref{fig:targetinstance} show the request graph and the corresponding target graph.  These graphs are complete graphs as the requests can appear between any pair of vertices. For clarity, we have ignored the edges with  weight $0$ in both of these subgraphs.  %The vertices of these subgraphs represent the grid cells as the requests have their origins and destinations at the grid cells, and the edges determine the flow of requests. % but the set of edges and their weights are different in  these subgraphs.
%In the request graph displayed by Figure \ref{fig:reqinstance},  the edge weight between the grid cells $g_1$ and $g_2$ is $3$, which displays that $3$ requests appear that have their origin at $g_1$ and destination at $g_2$. 
%Figure \ref{} shows an instance of the target graph, which stores the actual number of requests that arrive in ride-hailing platforms.
As can be seen through these figures, the edge weights of target graph are similar to that of request graph, except the edge $(g_1,g_{26})$ which has a value of $1$ in the target graph and $0$ in the request graph. %(as there is no edge between $g_1$ and $g_{26}$)
 It displays that there was $1$ request between the grid cells $g_1$ and $g_{26}$ but the Graph Neural Network-based architecture described in \cite{Ashraf_archive:2022} misclassified it and predicted that no request will arrive between grid cells $g_1$ and $g_{26}$.

 \textbf{DEFINITION 4} (DETOUR RATIO AND ORDER SET).
 \textit{The detour ratio is the ratio of the distance travelled  between the grid cells $o_s$ and $o_d$ ($|P_{G^R}(o_s,o_d)|$),  to the distance of the shortest path between these grid cells.
 %endpoints. 
 It is mathematically represented as: }
\begin{equation}
\alpha(P_{G^R},o_s,o_d)=\frac{|P_{G^R}(o_s,o_d)|}{|SP_{G^R}(o_s,o_d)|}
\label{eq:detsimplified}
\end{equation}
\textit{where $|P_{G^R}(o_s,o_d)|$ denotes the distance travelled when traversing through path $P$ on subgraph $G^R$ with end vertices $o_s$ and $o_d$ and $|SP_{G^R}(o_s,o_d)|$  denotes the distance of the shortest path between the grid cells $o_s$ and $o_d$.}  % min_{P_{G^R} \in \mathcal{P}_\mathrm{s,d}} |P_{G^R}(o_s,o_d)|$\mathcal{P}_\mathrm{s,d}$ represents the set of all possible paths between $o_s$ and $o_d$.}

When the routes are recommended to a driver it can't pick up all the passenger requests on the route. This is because the requests have  different destinations, and reaching each destination would likely require a deviation from the original route, resulting in increased travel distance for the passengers. In order to ensure passengers are satisfied and the distance travelled by them is bounded within a specified value, we add a constraint that specifies the detour ratio of all the passenger orders taken, should not exceed a threshold value $t$.  %. This is because thedestinations of different ride orders on that route would be different and reaching the destination of each order will probably take a fraction of different route which will result in an increased distance travelled by the passengers 

%as it will result in longer journeys for some passengers. In order to ensure passengers are satisfied and the distance travelled by them is bounded within a specified value, we add a constraint that specifies the detour ratio of all the passenger orders taken, should not exceed a threshold value $d$.
%\vspace{-4mm}
\begin{equation}
  \forall_{o_s}  \alpha(P_{G^R},o_s,o_d)\leq t
  \label{eq:constraint}
\end{equation}
%\vspace{-4mm}

This constraint ensures that only those ride-orders are taken whose paths are similar and the distance travelled by each ride order beyond its shortest path is bounded. %The orders that can be paired in a single vehicle without violating the detour constraints are denoted by the order set $\mathbb{O}$. 
The orders that can be paired in a single vehicle without violating the detour constraints are denoted by the order set
 $\mathbb{O}$. % of the proposed model contains all the ride orders that can be paired in a single vehicle.  These orders can be paired on the condition that the detour ratio of any order does not violate the detour constraint  specified by  Eq. \eqref{eq:constraint}. 
 In order to understand the detour ratio and order set consider the following example.

\begin{figure}[t!]
  \centering
\vspace*{-4mm}  \includegraphics[width=0.85\linewidth]{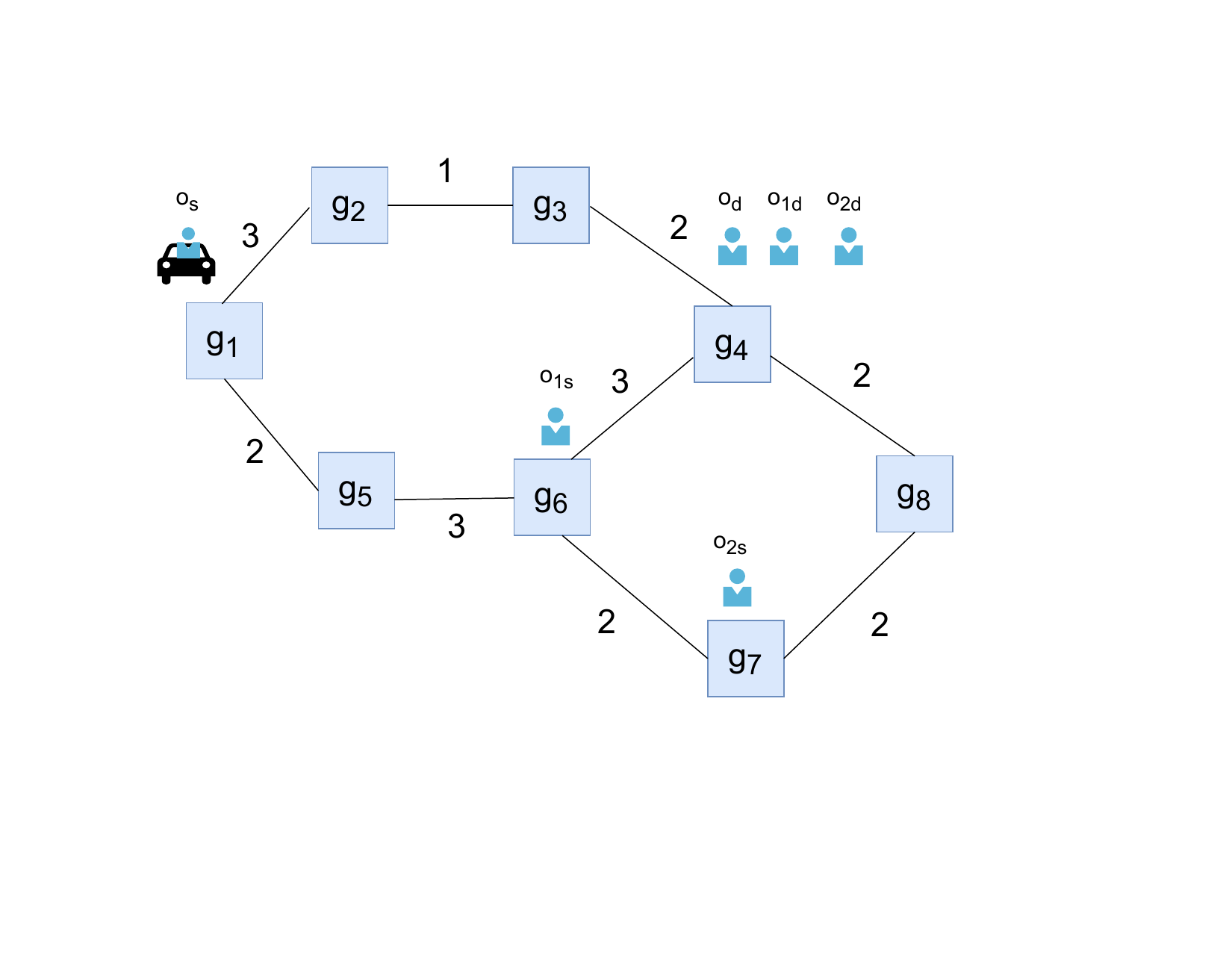}
  \vspace*{-22mm}
  \caption{A particular instance of the road network with the driver and passenger orders. The nodes represent grid cells and the edge weight represents the distance between the center points of grid cells. The vehicle is at grid cell $g_1$ and servicing the passenger who has to travel from $g_1$ to $g_4$. Meanwhile, the requests for two orders arrive at $g_6$ and $g_7$, and they have to reach $g_4$.}
  \label{fig:road}
%  \Description{Road}
\end{figure}

\textbf{EXAMPLE 1
:}
Consider the road network shown in Figure \ref{fig:road}. The nodes represent grid cells and 
the edge weights represent the distance between the center points of different grid cells.
A vehicle is at node $g_1$ and is currently servicing a passenger $o$ whose source ($o_s$) and destination ($o_d$) are $g_1$ and $g_4$ respectively. The current order set $\mathbb{O}$ of the vehicle is
$\mathbb{O}=\{o\}$ where $o=<g_1,g_4,o^t>$. Meanwhile, the request for $2$ orders arrives while the vehicle is in transit and has not yet reached their starting point.  $o_1$ arrives at $g_6$ ($o_{1s}$) and wants to go to $g_4$ $(o_{1d}$), $o_2$ arrives at $g_7$ ($o_{2s}$) and wants to travel to $g_4$ ($o_{2d}$). These orders are represented as $o_1$$=$$<g_6,g_4,o^t>$ and $o_2=<g_7,g_4,o^t>$ respectively.
Assume the detour ratio is $1.5$. %We will first determine if order $o_1$ can be taken by the platform.
In order to determine whether the order $o_1$ can be taken by the platform, 
we need to ensure the detour ratio of passenger $o$ who has to travel from $g_1$ to $g_4$ is not violated by taking the order $o_1$.  If order $o_1$ is taken, the path followed by the vehicle will be $P=\{g_1,g_5,g_6,g_4\}$.
We need to determine if the detour ratio of order $o$ gets violated through this modified path which incorporates $o_1$. The detour ratio of order $o$ is the ratio of the distance travelled in the current path $P$ (after incorporating order $o_1$) to the distance of the shortest path between the grid cells $g_1$ and $g_4$.
The distance travelled with the modified  path $P$ after incorporating $o_1$ is $8$ (sum of edge weights in the path $P$), and the distance of the shortest path between $g_1$ and $g_4$ is  $6$ (through the path $\{g_1,g_2,g_3,g_4\}$). So the detour ratio of passenger $o$ is $\alpha=\frac{8}{6}=1.33$ which is less than the threshold value of $1.5$. Thus the order $o_1$ is accepted by the platform, as it does not violate the detour ratio of passenger $o$ which is already in the vehicle. The order set gets modified as $\mathbb{O}=\{o,o_1\}$.
%Then order $o_1$ can be taken by the platform and the path followed by the vehicle when dropping $o$ and $o_1$ is $P=\{g_1,g_5,g_6,g_4\}$ whose sub-paths are $P_1=\{g_6,g_4\}$ (which serves $o_1$) and the path $P$ itself (which serves $o$). The sub-path $P_1$ is the shortest path between $g_6$ and $g_4$ so it satisfies the detour ratio. For sub-path $P$, the distance travelled is $8$, and the distance of the shortest path between $g_1$ and $g_4$ is  $6$ (through the path $\{g_1,g_2,g_3,g_4\}$). So the detour ratio is $\alpha=\frac{8}{6}=1.33$ which is less than the threshold value of $1.5$. Thus the order $o_1$ is accepted by the platform.
Similarly, to determine whether the order $o_2$ can be taken by the driver we check the detour constraints of both the ride-orders $o$ and $o_1$ in the vehicle.
The path that would be followed by the driver if $o_2$ is taken is $\{g_1,g_5,g_6,g_7,g_8,g_4\}$. Let's first determine the detour ratio of $o_1$ and see if it gets violated by taking $o_2$. The path followed by ride-order $o_1$ if $o_2$ is taken will be $\{g_6,g_7,g_8,g_4\}$. The distance corresponding to this path is $6$ and the distance of the shortest path between $g_6$ and $g_4$  is $3$ (through path $\{g_6,g_4\})$. The detour ratio of ride-order $o_1$ is $\frac{6}{3}=2$ which is greater than the threshold of $1.5$. Thus the detour constraint of the order $o_1$ which is already in the vehicle is violated if we take the new order $o_2$. The order $o_2$ is thereby not accepted by the platform and the order set remains $\mathbb{O}=\{o,o_1\}$. \looseness=-1

\subsection{Computing the origin and destination of passenger requests}
%The passenger requests we use the historical data with the past ride-orders origin and destination pairs. 
The requests in ridesharing platforms follow a specific pattern and depend upon requests originating from subsequent areas (spatial dependencies) or on previous request patterns (temporal dependencies). To model these spatio-temporal dependencies and predict the future origin and destination of requests, we use a Graph Neural Network (GNN)-based model described in \cite{Ashraf_archive:2022} which analyzes the passenger mobility patterns and predicts the requests that can arrive in the future. Though there have been various approaches that have predicted the origin and destination of passenger requests \cite{Wang:ACMKDD_2019,Wang:ACMTrans_2022}, however, the GNN-based model captures the non-recurring trends in data apart from the other recurring trends that have been predicted by previous studies. This results in efficient prediction and allows our proposed model to recommend the route with the highest number of passengers. %These predictedrequests are 

\subsection{Problem Statement}
The main objective of our proposed model 
is to develop an eco-friendly route recommendation system that reduces the hazardous emissions of ridesharing platforms by predicting the origin and destination of requests and utilizing them to make efficient use of vehicles. In order to utilize the vehicles all the way, the proposed approach recommends the route that has the higher number of expected passenger demand (origin of requests) on its way. 
However, while recommending routes with more passengers, the vehicle can deviate too much from the shortest path of passengers and cause inconvenience to them. In order to ensure that vehicle does not deviate and passengers on board are satisfied by the platform service, the detour ratio which is the ratio of the length of the path taken by the passengers in the vehicle 
to the length of the shortest path between their origin and destination should be bounded by a threshold.  
Thus the objective of our proposed approach is to select a route that has  the highest number of expected passengers with a constraint that the detour ratio of all the passengers is satisfied. Moreover, there is a constraint on the number of passengers a driver can take  since the capacity of the vehicle is limited. % there is another constraint on the number of ride orders that can be taken by the driver.
The problem can be described mathematically as: 
\begin{equation}
    P_{G^Q}^{*}= arg\, max_{P_{G^Q}}\{ \,\; \mathbb{E}[|P_{G^Q}|] \}%| \mathcal{O},o^t]
    \label{eq:obj}
\end{equation}
subject to
\begin{equation}
  \forall_{o_s}  \alpha(P_{G^R},o_s,o_d)\leq t
  \label{eq:constraint1}
\end{equation}
\begin{equation}
   |P_{G^T}| \leq c
   \label{eq:capcontraint}
  \end{equation}
where
\begin{equation}
  |P_{G^Q}|=\sum_{g_i\, \in \, P}\sum_{g_j\, \in \, \mathcal{F}} w^{Q}_{{ij}}
\label{eq:objsimplified}
\end{equation}
\begin{equation}
  |P_{G^T}|=\sum_{g_i\, \in \, P}\sum_{g_j\, \in \, \mathcal{F}} w^{T}_{{ij}}
\label{eq:objtargtesimplified}
\end{equation}

Eq. \eqref{eq:obj} denotes the objective function of our proposed model which is to select a path $P$  from the request graph $G^Q$ that has the highest number of expected passenger requests. The expected number of requests in a path is calculated  through the summation of edge weights $w^{Q}_{ij}$ from all the grid cells $g_i$ in the path $P$ i.e, $(g_i\,\in \, P)$ to the set of their forward nodes ( $g_j\, \in \,\mathcal{F}$) which lie on the path to the destination,  as is defined through Eq. \eqref{eq:objsimplified}. For instance, the expected number of requests in path $\{g_1, g_\mathrm{2}, g_\mathrm{26}\}$ in Figure \ref{fig:reqinstance} is $w^Q_\mathrm{(1)(2)}+w^Q_\mathrm{(1)(26)} +w^Q_\mathrm{(2)(26)}=3+0+1=4$.
%The edge weight $w_{\mathrm{ij}}$ represents the expected number of passengers that can arrive between the grid cells $g_i$ and $g_j$ of request graph $G^Q$. In order to calculate the expected number of passengers that can arrive in path $P$, we take each grid cell $g_i$ and calculate the expected number of requests that originate from it and are destined towards the set of its forward nodes $\mathcal{F}$ that lie in its path towards the destination as is defined through Eq. \eqref{eq:objsimplified}. For instance, the expected number of requests in path $\{g_1, g_\mathrm{2}, g_\mathrm{26}\}$ in Figure \ref{fig:reqinstance} is $w^Q_\mathrm{(1)(2)}+w^Q_\mathrm{(1)(26)} +w^Q_\mathrm{(2)(26)}=3+0+1=4$
While selecting the path with the maximum expected requests there is a constraint specified by Eq. \eqref{eq:constraint1} which states that for each ride order taken, its detour ratio  should be bounded by a threshold $(t)$ in order to ensure they  are satisfied. Moreover, there is a constraint on the vehicle side denoted by Eq. \eqref{eq:capcontraint} which states that the number of passengers taken $|P_{G^T}|$  should not be more than the capacity $c$ of the vehicle. The total number of passengers in a vehicle is calculated through the summation of edge weights of the target graph %with the forward set of nodes as was done in case of request graph. '
$w^{T}_{ij}$ from all the grid cells $g_i$ in the path $P$ i.e, $(g_i\,\in \, P)$ to the set of their forward nodes ( $g_j\, \in \,\mathcal{F}$) which lie on the path to the destination,  as is defined through Eq. \eqref{eq:objtargtesimplified}. \looseness=-1% For instance, the total passengers in the path $\{g_1, g_\mathrm{2}, g_\mathrm{26}\}$ in Figure \ref{fig:targetinstance} is $w^T_\mathrm{(1)(2)}+w^T_\mathrm{(1)(26)} +w^T_\mathrm{(2)(26)}=3+1+1=5$
%This completes our problem formulation. 

\section{Baselines}
%\textcolor{cyan}{Separate section: Baselines}

In this section, we will discuss the existing works, identify their limitations and highlight how our proposed model overcomes them.
\subsection{Existing Works}
There are three baselines for our proposed model. One of the baselines is the trivial baseline that recommends the route with the shortest distance between the source and destination. This approach does not involve any element of prediction and traverses a sequence of vertices where the distance between the source and destination vertices is the shortest. The other baselines are SHARE \cite{Yuen:ACMWWW_2019}, and insertion-based route planning framework \cite{Tong:ACMTrans_2022} which predict the passenger demand on the road and recommend the routes that may be slightly longer than the shortest path but have high expected demand. \looseness=-1  
\subsection{Limitations}
The shortest path algorithm does not consider the expected number of requests. Whenever a request arrives, this approach recommends the shortest path between the source and destination locations of the request. It results in inefficient utilization of vehicles since there is a small probability of having another request on the shortest path between source and destination. %This results in inefficient utilization of the vehicle as the vehicle can travel a small distance beyond its shortest path if that path would fetch him another request. This path beyond the shortest path can be recommended by predicting the requests that may arrive in future.   %the driver can travel some distance beyond its shortest path if that path would provide him with more passengers.

\begin{figure}[t!]
  \centering
\vspace*{-4mm}  \includegraphics[width=0.85\linewidth]{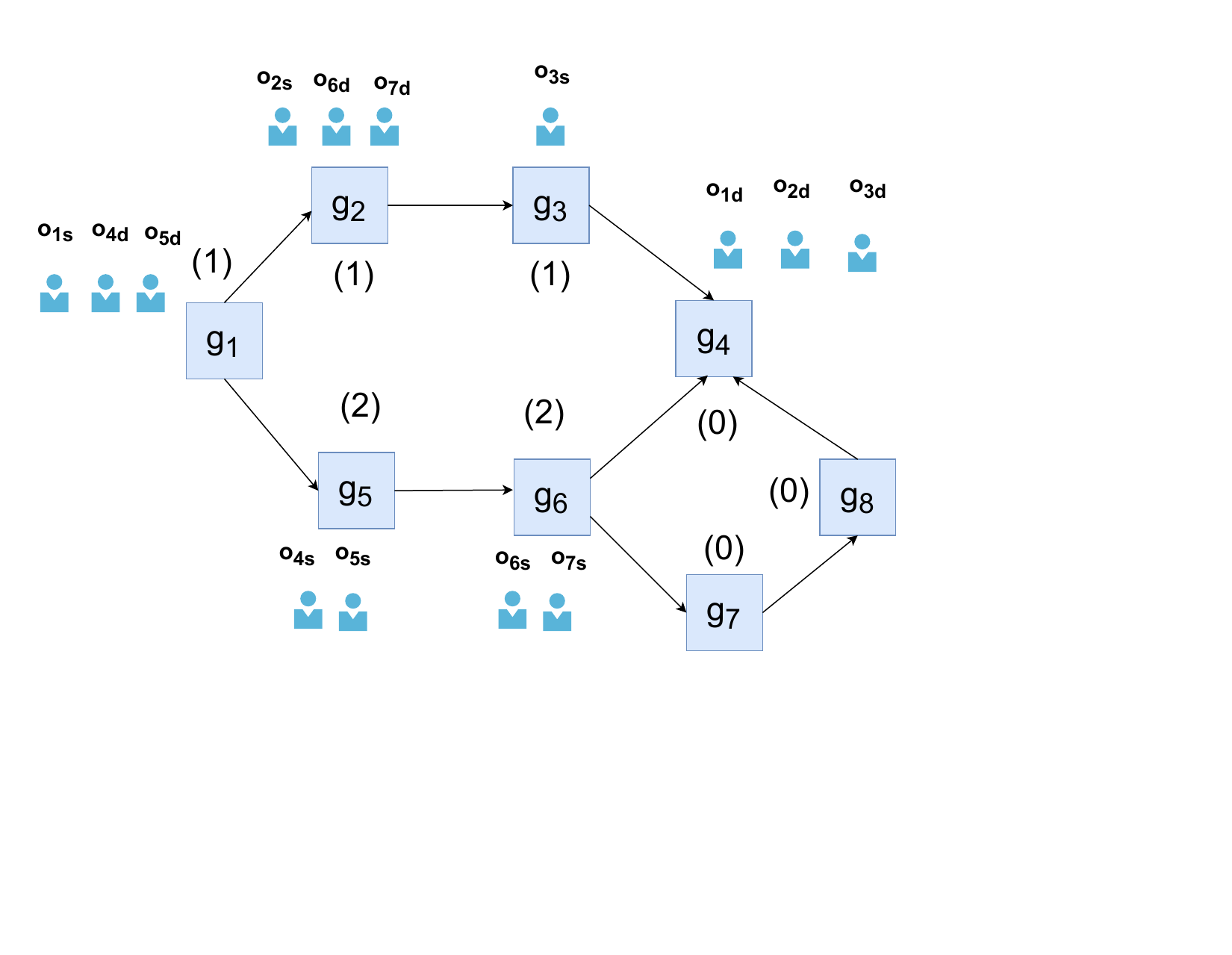}
  \vspace*{-22mm}
  \caption{Performance of demand and origin-destination prediction. In this graph, nodes represent grid cells and the node weight represents the expected number of passengers on the grid cell. The origin and destination of passenger $i$ are denoted as $o_{is}$ and $o_{id}$ respectively. }
  \label{fig:baseline}
  \vspace*{-6mm}
%  \Description{Road}
\end{figure}

SHARE and the route planning-based frameworks,  predict future requests and recommend routes that may be slightly longer than the shortest path, but they have a higher probability of finding more passengers.  However, these methods only predict  the origin of requests, which is also called demand prediction,  and not where they are headed towards. This could result in the 
recommendation of routes that do not utilize the vehicle capacity efficiently. Consider the part of the road network captured through Figure \ref{fig:baseline}, where the nodes represent grid cells and the node weight determines the expected number of passengers that may arrive at that node. In this road network, we have to recommend a route between grid cells $g_1$ and $g_4$, and there is already a passenger in the vehicle who has to travel between these grid cells. 
The expected demand between the grid cells $g_1$ and $g_4$ is maximum along the path $P=\{g_1,g_5,g_6,g_4\}$ and is equal to the  summation of node weights in the path  $P$ 
i.e., $1+2+2+0=5$. This path would be recommended by the optimal demand prediction algorithm. However as can be seen through the origin and destination of requests (represented by $o_\mathrm{si}$ and $o_\mathrm{di}$ for $i^\mathrm{th}$ request), the corresponding destination of  requests originating from nodes $g_5$  are $g_1$, and the destination of requests from $g_6$ are $g_2$  and none of the requests is directed towards $g_4$. 
Thus the path $P$ would contain only one passenger who has to travel from $g_1$ to $g_4$, and will not utilize the vehicle capacity efficiently. 
On the other hand, the origin and destination of requests %which is shown through $o_{si}$ and $o_{di}$ for request $i$ 
displays that the requests from $g_2$ and $g_3$ are directed towards $g_4$. Thus the optimal path between nodes $g_1$ and $g_4$ is $\{g_1,g_2,g_3,g_4\}$ %\textcolor{red}{chnage a-b distance to 1 and d-g to 1} 
which contains $3$ requests and they have their destination at $g_4$, and this path can be recommended by utilizing the origin and destination of requests and not only their origins as was done in the existing baselines. \looseness=-1%Thus, we can say the demand prediction graph shows the areas with higher demand, but it cannot result in effective pairing of requests in a single vehicle as it does not show the destination of requests and does not tell about the common sharing of routes. 

From the above explanation, we can follow that passenger mobility patterns are better exploited by using the origin as well as the destination of requests and we provide a route recommendation system for ridesharing platforms that anticipates the passenger's origin and destination and overcomes the limitation of previous models.
We will describe it next.

\section{Our Proposed Model}
In this section, we will  describe the working of our proposed model. 

\subsection{Complexity analysis}
\label{complexity}
Our proposed model recommends routes to drivers that will maximize the expected number of passengers on the way and utilize the vehicle effectively. %The driver wants the recommendation from the place where it is located and we assume there is a passenger in the vehicle %The  location from which the driver wants to get the route recommended is considered the starting location and the location of the last passenger in the vehicle is considered the destination location.
%The algorithm recommends a route between the source and destination that maximizes the expected number of requests. 
In order to recommend a route, the algorithm needs to analyze all the expected requests between the source and  the destination.
This problem is NP Hard as it can be reduced from the Longest Path problem. \looseness=-1

\textit{Longest Path problem:}

Given a weighted graph $G=(V,E,w)$  with non-negative edge weights, the Longest Path problem is to find the longest simple path from the source vertex $(s)$ to the destination vertex $(d)$.

\textit{Max-Request Path problem}

Given a weighted complete graph $G=(V,E,w)$  with non-negative edge weights, the Max-Request Path problem is to find a simple path with the maximum number of expected requests from the source vertex $(s)$ to the destination vertex $(d)$.

Route recommendation for ridesharing platforms is actually  Max-Request Path problem, where the weight of each edge $w$ is the expected number of passengers that can arrive between the source vertex $s$ and destination vertex $d$. The route which has the highest number of expected  passengers is the Max-Req path. The Longest Path problem is known to be NP-hard as it can be reduced from the Hamiltonian path \cite{Cormen:book_2022}. \looseness=-1

\textit{Theorem: Finding Max-Request path is NP-Hard}

\textit{Proof:}
Given an arbitrary instance of the Longest Path Problem, we reduce it to an instance of the Max-Request Path problem through the following procedure. Consider the given graph $G=(V,E,w)$ in the Longest Path problem, we design a graph $G'=(V',E',w')$ such that $\forall v \, \in \,V$ , we have a vertex $v'$ in $V'$ and $\forall e=(u,v) \,\in \,E$ we have an edge $e'=(u',v')\, \in \, E'$ with edge weight $w'=w$. For the edges $(u,v) \, \notin \, E$ in $G$, we have the edge $(u',v')$ in $G'$ with the edge weight $w'=0$. 
In $G'$ the Max-Request problem is to identify the path from a source $s$ to a destination $d$ with the maximum number of expected requests. Since each edge $e'=(u',v')$ with edge weight $w'$ selected in the path in $G'$ corresponds to selecting the edge $e =(u,v)\,\in \, E$ with weight $w$, if the Max-Request problem is solved, the resultant path is the longest path from $s$ to $d$ in a graph $G$ and hence  we would have solved the Longest Path Problem.  

Considering the NP-Hardness optimal algorithms are not possible. Thus we apply heuristics to solve the problem at hand. \looseness=-1
\subsection{Heuristic}
In order to reduce the exponential nature of the problem, we use a $ k$-hop-based sliding window approach, where the route is recommended in the reduced space whose size is determined by the hop count $k$ of the window. In this space, there are few requests and they can be analyzed and paired up effectively. The route that maximizes the expected number of requests and satisfies the detour constraints  is recommended within the window. The window is then slid forward in the subsequent steps and the route is recommended from thereon until the destination point is reached.  %We will explain this procedure in detail next.
\subsubsection{How sliding window-based route recommendation algorithm works}
In this subsection, we will describe the working of $k$-hop based sliding window.
The driver wants a recommendation mechanism from the place where he is currently located, and without loss of generality, we assume there is already a passenger in his vehicle who has to travel between the specified source and destination points. %If there is not a passenger in the vehicle, the destination point can be set as the point that will have the highest expected requests from the current point of driver.  In both the cases, we assume the source and destination are fixed and route needs to be recommended between them.
The driver can follow the shortest path between the source and destination points of the passenger who is in the vehicle, but there is a small probability that the  shortest path will contain any  other request. This will result in inefficient  utilization of the vehicle as there will be only one passenger in the vehicle whereas the vehicle can be filled up to its capacity $c$.  %The passenger wants to travel between his source an
In order to utilize the vehicle capacity,
the recommendation mechanism will provide a route to the driver that might be slightly longer than the shortest path but it will have a higher probability of finding  passengers. % while satisfying the detour constraints of the passenger who is already in the vehicle. 
This will result in the efficient utilization of vehicles and contribute to eco-friendly rides. % who has to travel from 

In order to recommend a route with the highest number of expected passengers between the source and destination points of the passenger who is already in the vehicle while satisfying the detour constraints, we need to search the entire search space  
and find the best-constrained route among all the possible routes that exist in the graph. This results in exponential complexity as was proved in subsection \ref{complexity}.
%check the origin and destination of all the expected passenger requests that can arrive at the grid cells and see if these requests can be paired with the passenger who is already in the vehicle without violating his detour ratio. However, checking the origin-destination of all the requests %through brute force w in the road network results in exponential complexity, as the given problem  is NP-Hard. % as the request graph is a complete graph and recommending a route  re are exponential number of requests that arrive at the nodes of the road network. 
To reduce the complexity of route recommendation, %the window is created around the source node i.e., the point where the request arrives and it is $g_{26}$ here. 
%Since there can be an exponential number of paths between the source and destination vertices of the passenger, 
the proposed approach uses a % an proximation algorithm that 
window to create a small area around the source node (the point at which the request of passenger who is the vehicle arrived)
and recommends the optimal path within this area which reduces the complexity of recommendation from an exponential frame to a smaller area.  %In this window, there are small number of grid cells and the optimal route c%route can be recommended effectively. %This window creates a smaller area where we see the origin and destination of all the requests and merge these requests with the passenger already in the vehicle. This window is then moved forward and the optimal path is selected in it. This process continues till the destination of last passenger in the vehicle is reached.
%Initially, the window is created around the  source of the first request (request that has to travel between source and destination points), and the middle point of the window is placed at it. The proposed model selects the highest request path that satisfies the detour constraints of passenger, within the window.  
The window is then slid forward in the direction of the destination(s) of the passenger(s) already in the vehicle, and the highest expected requests within the window that satisfies the detour constraints of passengers are checked till the window reaches the destination point of the last passenger in the vehicle. 

\begin{figure}[t!]
    \centering
    \vspace*{-4mm}
    \includegraphics[width=0.7\textwidth]{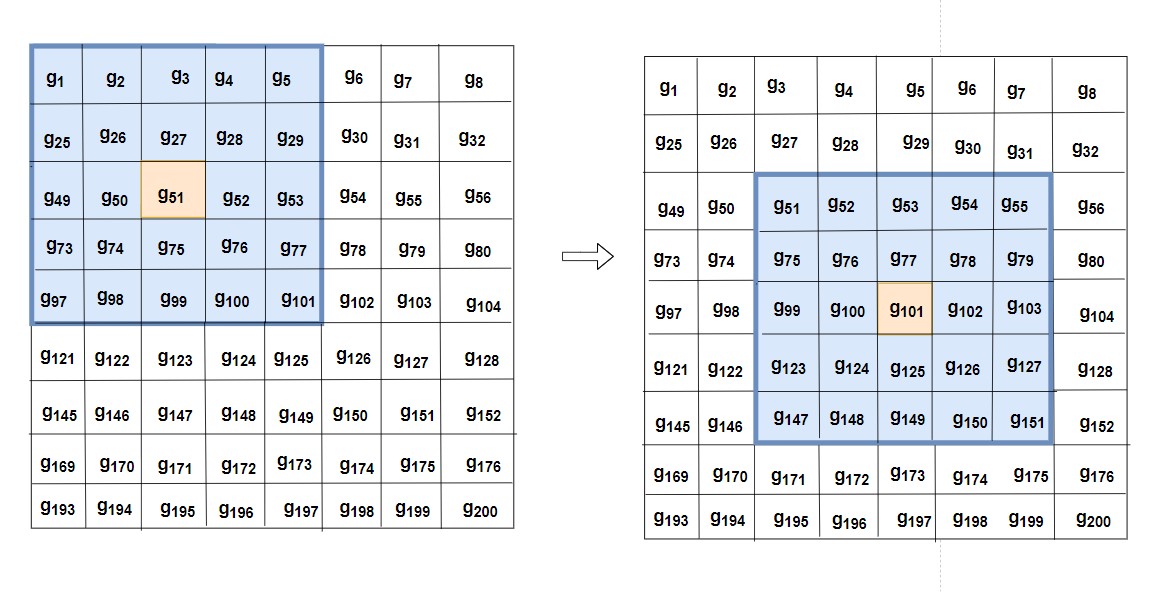}
    \vspace*{-2mm}
    \caption{Road network represented in terms of  a window that is slid in the direction of passengers already in the vehicle. The road network is replicated $2$ times to show how the window moves at each step. Initially, the window (highlighted in blue) is created around $g_{51}$ (highlighted in orange) which is the source node. The window is then slid and the next window is created around grid cell $g_{101}$.}
    \vspace*{-3mm}
    \label{fig:sliding_window}
\end{figure}

To understand this point, consider Figure \ref{fig:sliding_window}. The driver is at grid cell $g_{51}$ and it has a passenger in it whose destination is $g_{176}$. The shortest path between these grid cells is $\{g_{51},g_{76}, g_{101},g_{126},g_{151},$\\$g_{176}\}$. %, which has a length of $12\,km$ (since the grid cells are assumed to be separated by $2\,km$).
The proposed model selects a path that might be slightly longer than the shortest path but will have a higher number of expected passengers. %In order to select a path with the highest number of expected passengers, we need to check the origin and destination of all the passenger requests that arrive at the grid cells and see if these requests can be paired with the passenger in the vehicle without violating its detour ratio. However, checking the origin-destination of all the requests %through brute force w in the road network results in exponential complexity, as there are exponential number of paths in the road network between any source and destination. 
In order to select a path with the highest number of expected passengers, the window is created around the source node which is the point where the first request arrives and it is $g_{51}$ here (highlighted in orange). The middle point of the window is placed at the source node $g_{51}$. The window is assumed to be $2$- hop, which means the two neighbors from each side are taken, % there are $8$ neighbors, $1$ from each side 
as displayed through Figure \ref{fig:sliding_window} (window is highlighted in blue). By using this window, we constrain the search space around the source node to this window of $2$ hop and see the maximum expected request path  within this window. Let's assume the highest expected request path in this window is $\{g_{51},g_{75},g_{100},g_{101}\}$ (how this path is obtained will be described later on).
Since the maximum expected request path within the window  is achieved,  the window is slid forward with the destination point of this window which is $g_{101}$ as the source point of the next window, as is shown by Figure \ref{fig:sliding_window}. The window after sliding is centred around $g_{101} $ and the highest expected request path within its $2$ hop neighbours is chosen. % which is $\{g_{101},g_{125},g_{150},g_{151}\}$. 
The window continues moving until the destination point of the passenger already in the vehicle, which is $g_{176}$ is reached.
Thus, by applying the window we reduce the search space from the entire road network which has an exponential complexity to a set of neighborhood grids around a node and move this window at each step in the direction of the destination of passengers already in the vehicle. \looseness=-1
%While moving the window at each step, the proposed model needs to take into account the detour constraints of passengers already in the vehicle and ensure their value lies below the threshold specified in order to ensure passenger satisfaction. We will describe it later on. 
 
 Till now, we have described the mechanism for reducing the search space from the road network to a window. Next, we need to determine how the highest expected request path is obtained within the window.  We can use a brute-force approach to determine the optimal path in a window and slide the window towards the destination in the next step. However, the brute force approach leads to an exponential search space in the size of the number of elements $e$ in the window ($2^e$),  which implies that the number of elements needs to be very small. 
 We need to apply an algorithm that finds the route within a window in an efficient manner. To determine the route in the window we analyze the structure of our objective function and see if it can be solved  effectively without the use of  a brute force approach that is impractical for large window size. We analyze our objective function in the next subsection. \looseness=-1

\vspace{-2mm}

\subsection{Analysis of Objective function}
Eq. \eqref{eq:obj} determines the objective function of our proposed model. It states that we want to recommend a path that has the highest number of expected passenger requests.  %\textcolor{blue}{IAG: This definition doesn't look correct.} 
As can be seen clearly from Eq. \eqref{eq:objsimplified}, this function is a linear combination of edge weights in a path and  we know that linear functions are  \textit{submodular} in nature.

%\textcolor{blue}{IAG: This is not straightforward to see that it is linear, especially in absence of a definition there. Also, the second claim may not be true in general. We can have counterexamples for it.} 

\subsubsection{Submodular}
A function is said to be submodular  if the addition of an element leads  to decreasing  difference in the incremental value of the function. 
Let $Z$ be a finite set. A function $f: 2^Z$$\rightarrow$$R$ is said to be \emph{submodular} if for all subsets $X \subseteq Y \subseteq Z$  and $\forall a\,\epsilon \,Z\,\backslash \, Y $:
\begin{equation}
    f(X \cup \{a\})-f(X) \geq f(Y \cup \{a\})-f(Y)
    \label{eq:subdef}
\end{equation}
In simple words, submodular functions  state that  the marginal benefit of adding an element to the smaller set    is at least as high as the marginal benefit of adding it to a bigger set. %Equivalently, a function is submodular if for every $X,Y \subseteq Z$ %\begin{equation}
 %   f(X)+f(Y) \geq f(X \cup Y) +f(X \cap Y)
  %  \label{eq:submod}
%\end{equation} 
Apart from being submodular, our objective function is also monotone. \looseness=-1

\subsubsection{Monotone}

A function is monotone if $f(A)\leq f(B)$ implies $A \subseteq B$. We know that our objective function is linear which adds the requests repeatedly after each iteration. Since the request value is non-negative, when an element is added to a set it cannot decrease the value of the set. By this argument, we can say that our function is monotone.  \looseness=-1

\subsubsection{Optimization of monotone submodular functions}

The greedy algorithm has proved to be a natural method for maximizing a monotone submodular function subject to certain constraints. 
In various  
settings, the approximation ratios provided by greedy algorithms
are best-known \cite{Conforti:ElsevAppliedmaths_1984}. Further, the simplicity of greedy algorithms makes them useful in route recommendation systems and their performance has been found to be at par with other  algorithms \cite{Liang:SIGIR_2018}. \looseness=-1

As our objective function  determines the route followed by the vehicle based on the expected requests that arise within that route, and it is submodular and monotone, greedy provides an obvious choice for our proposed system. Within a window, in order to recommend a path among exponential paths, we switch from the brute force approach to the greedy approach due to the submodular and monotone nature of our objective function.  While directing the route, two key factors need to be taken into account: the expected passenger requests in the path and the detour constraints of the passengers in the vehicle. Now, as the subgraph instance is reduced  to the size of a window, we need to find the path in this window with the above objective and constraints through the greedy strategy. 
Firstly, we will describe the simple greedy strategy wherein the node with the maximum expected requests is selected among the neighborhood nodes and we will show that it does not utilize the window appropriately. After that, we will propose two variants of the greedy approach. 

\subsection{Simple Greedy}

%As the name suggests, this approach looks for local optimum at each step, and selects the  node with the highest requests among connected nodes.

The simple greedy approach finds a route with the highest expected requests that satisfies the detour constraints of passengers by looking only at the directly connected nodes.
As the name suggests this approach  looks for local optimum at each step and selects the adjacent node with the highest expected requests from the current node.
The process continues until the driver wants the recommendation mechanism.  If we consider the road network displayed by Figure \ref{fig:sliding_window}, and assume the driver is at grid cell $g_{51}$, the simple greedy approach will select the highest expected requests among the $8$ neighborhood nodes of $g_{51}$ which are $\{g_{26},g_{27},g_{28},g_{50},g_{52},g_{74},g_{75},g_{76}\}$. Suppose the request graph $G^Q$ displays that the highest expected requests are from $g_{51}$ to  $g_{76}$. The proposed approach will move the window to $g_{76}$ and select the highest expected requests from $g_{76}$ among its $8$ adjacent nodes. This process will continue until the point driver wants the recommendation mechanism.

If we carefully analyze the procedure we can see that this approach does not violate the detour constraints of any passenger. This is because at each step the source and destination of passengers are one grid cell away, as we look for maximum expected requests from a grid cell among its neighbors and move directly towards that grid cell.
%away which is always their shortest path. 
However, the main drawback of this approach is that it only considers the directly connected nodes for selecting the path and does not utilize the window size properly. Even if the window covers the whole road network this approach will still look at the directly connected nodes to determine whether the route has the higher flow of requests or not.
%\textcolor{cyan}{}

\subsection{Greedy Variants}

With the simple greedy approach, we 
look at the neighborhood nodes in order to get the maximum request path. This reduces the search space to $1$ hop and does not utilize  the origin and destination of the majority of requests in the window. In order to utilize the origin-destination of requests appropriately, we use two variants of the greedy algorithm which consider the entire window in their initial step, with a view to obtain the maximum request path,
and subsequently,  consider the neighborhood nodes in order to keep the road segments connected.

%The greedy algorithm finds the route with the highest expected requests that satisfies the detour constraints of the passengers in the window. The window is then slid forward till the destination point of all the passengers in the vehicle is reached. 
%As the name suggests the greedy algorithm looks for the local optimum at each step and finds the route with the highest expected requests within the detour constraints of all the passengers onboard. However, this approach would reduce the search space to $1$ hop and look for the highest expected request path among the neighborhood nodes of the current grid cell. In order to improve the search space, the proposed approach uses two variants of greedy strategy
%The proposed approach does not apply a simple greedy strategy to select the node greedily at each step among the neighborhood nodes. Rather, it utilizes the full size of window and selects the highest expected request path in a window.
%With the simple greedy approach, we  look at the neighborhood nodes in order to get the maximum request path. This reduces the search space to $1$ hop and does not utilize  the origin and destination of the majority of requests in the window. 
%In order to do it, %it utilize the origin-destination of requests appropriately, 
%we use two variants of the greedy algorithm which consider the  entire window in their initial step, with a view to obtain the maximum expected request path,and subsequently,  consider the neighborhood nodes in order to keep the road segments connected. 

In both of these approaches, we consider the driver's starting position as the point on a graph from which the route is to be recommended. After that, we select the node or location within the window that will have the maximum number of expected requests from the starting location. This step utilizes the full window 
and selects the node with the maximum expected requests from the source node as the %
destination or endpoint of the window i.e., the route recommended in the window has the source as the drivers starting location and the destination as the point from which the source has the maximum expected requests.   
Since the route needs to be connected, we need to check if the destination selected in the window is directly connected to the source point of the window.
If that is the case, then the path in the window is complete and the window is moved forward. If the destination is not connected to the source, then we need to find a set of connected nodes between the source and destination of the window, and the connected nodes should have high expected requests and the detour of passengers onboard should not be violated. There are two approaches to selecting the set of connected nodes between source and destination, %with maximum requests and detour constraints, 
namely Backward Greedy and Forward Greedy. These approaches select the  nodes between the source and destination either through backtracking from the destination point of the window until the source is reached or by following the path from source to destination respectively. They are described in detail in the next subsections.%We will describe these approaches in the next subsections.

\subsubsection{Backward Greedy (BG)}
After selecting the node that has the maximum expected requests from the source as the destination point of the window, we need to select the set of connected nodes between these points and this set of nodes should have higher expected requests within the detour constraints of the passengers onboard.
The first approach to selecting these nodes is  Backward Greedy, and as the name suggests this approach selects the set of nodes greedily in backward direction i.e., after the source and destination points of the window are fixed, this approach checks the nodes that are directly connected to the destination and greedily returns the one that has the highest expected requests to the destination. If the selected node that has the highest expected requests to the destination is connected directly to the source, then the path is complete and the window is slid. However, if it is not connected, then the procedure is repeated and the nodes with maximum expected requests to the already connected nodes with the destination are returned. In this way, we continue till the source node is reached.  Thus in this approach, we backtrack from the destination until we reach the source node with the connected set of nodes. 

Consider the sliding window shown in Figure \ref{fig:sliding_window}. In this figure, we will show how the route is recommended by using the backward greedy approach within a window when the driver is at the grid cell $g_{51}$. %which shows a particular window and we will show how the route is recommended within the window by using a backward greedy approach. 
 %Suppose the driver is at the node $g_\mathrm{51}$ %\textcolor{blue}{IAG: Where? Give reference to some figure.} and it needs route recommendation from that point. We have considered a $2$-hop-based window for this example. The window is created with the driver's starting position which is $g_{51}$ as the middle point of the window, and the $2$-hop neighbors around it are selected for route recommendation. %, as can be seen through Figure \ref{fig:window}.
Initially, we consider all the expected requests that can arrive from the starting point of the driver  and return the one with the maximum value in the window.
For instance, in the above window, if the request graph $G^Q$ displays that the maximum expected requests from $g_{51}$  are towards $g_\mathrm{101}$, we will consider $g_\mathrm{101}$ as the destination node of the current window. As the route needs to be connected, we check the maximum expected requests from the directly connected nodes of  $g_\mathrm{101}$ in the window i.e., $\{ g_\mathrm{76},g_\mathrm{77},g_\mathrm{100}\}$ to $g_\mathrm{101}$. In this case, if the expected requests to $g_\mathrm{101}$ are highest from  $g_\mathrm{76}$, then the path is returned and the window is slid as  $g_\mathrm{76}$ is directly connected to the source node $g_\mathrm{51}$. However, if the highest expected requests come out  from  $g_\mathrm{77}$ or $g_\mathrm{100}$ then we need to continue backtracking until we reach  the source node, i.e., we need to find the highest expected requests to these nodes ($g_{77}$ or $g_{100}$) among their directly connected nodes, and continue this procedure till the nodes directly connected to the source are returned. %\textcolor{red}{can remove these lines}After finding a path in the window, we slide the window and create a new window  where the middle point of the new window will be $g_\mathrm{101}$ and its $2$-hop neighbors will be used to build the window around it.  This procedure is repeated till the destination point of all the passengers is reached.

\begin{figure}[t!]
    \centering    \includegraphics[width=0.24\textwidth]{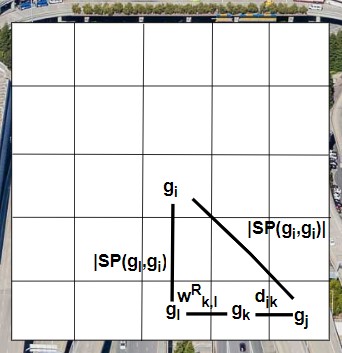}
    \vspace*{-0.1mm}
    \caption{Road network in the form of a window. The driver is assumed to be at $g_k$ and its adjacent grid cell $g_l$ is checked for inclusion in the path between $g_i$ and $g_j$.  }
    \label{fig:detour_exp_simplified}
\end{figure}

It can be seen from the above procedure that the origin and destination of all expected requests except the source and destination points of the window are just a $1$ hop away. This can be followed directly, as the first step utilizes the full window size and looks for the maximum expected requests from the source within the window, %of size $k$, 
and the subsequent steps keep the path connected and look for expected requests among the directly connected nodes. 
%If we carefully analyze this procedure
From the above analysis, we can follow that the detour ratio of the requests that are directly connected will not be violated as %our proposed model keeps the path between    
their source 
%\newpage
\begin{breakablealgorithm}
\vspace{1mm}
  \hspace{-2mm}\textbf{Algorithm 1:} Backward greedy approach for finding the highest expected request path within the detour constraints of passengers onboard  through $k$-hop sliding window approach
  \vspace{1mm}
  \hrule  
  \begin{algorithmic}[1]
  \vspace{2mm}
\STATE \textbf{Input:} Drivers starting location $o_s$, detour ratio $\alpha$ , number of hops $k$, request graph $G^Q$, road graph $G^R$
\STATE   \textbf{Output:} Highest expected requests path $p^{*}$ within the detour constraints of all the passengers in the vehicle
   \STATE $p^{*}=[]$
   \STATE $b=[]$  %// array for backtracking from destination node of window to source node
   \STATE  $p^{*} \leftarrow o_s$%Add $o_s$ to path 
   \WHILE {destination is not reached}
   %$i =1$ to $\frac{n}{k}$}  
   \STATE Create a $k$-hop-window with $o_s$ as the middle point of the window
  \STATE Calculate requests from the source node $o_s$ to all other nodes in the window  %maximum number of requests from source
 \STATE Select the node with maximum requests ($o_d$) from the source ($o_s$) as the endpoint of the current window 
 \STATE $b \leftarrow o_d$%Add $oi$ as a last element to the path
 \IF{ $o_d$ is directly connected to source}
 \STATE $p^{*} \leftarrow reverse(b)$
 \STATE $o_s \leftarrow o_d$
 \STATE Slide the window
  \ELSE
    \WHILE {$o_d$ is not equal to $o_s$}
   
   \STATE Find the maximum directly connected element $o_j$ from endpoint $o_d$ that satisfies detour ratio
   \STATE $b\leftarrow o_j$
    \STATE $o_d \leftarrow o_j$
  \ENDWHILE   
  \ENDIF
  \ENDWHILE
   \end{algorithmic}
\vspace{4mm}
\end{breakablealgorithm}
 and destination are just $1$ hop away and our proposed model will always move to subsequent grid cells directly without any detour.   However, the detour ratio of the passenger on board who has to travel from the source point of the window and reach the destination point in the same window can be violated if we explore nodes from  destination to source  without taking into consideration the extra distance travelled at each step.
In order to ensure that the detour ratio of the passenger who has to travel between the source $g_i$ and destination $g_j$ points of the window is satisfied the ratio of extra distance travelled  by him when the set of connected nodes is selected between these points, to the distance of the shortest path between these points %that can be travelled at each time step 
should be bounded by a threshold value. % Thus, when  the set of connected nodes are selected between the source and destination points of the window, the proposed model needs to ensure they do not violate the %detout % without violating the 
%detour constraint of the passenger who has to travel from $g_i$ to $g_j$.
If the driver is assumed to be at grid cell $g_k$ , and the distance of $g_k$ from the destination point $g_j$ is $d_{jk}$, and the maximum expected passenger requests among the adjacent grid cells of $g_k$ arrive at $g_l$ (see Figure \ref{fig:detour_exp_simplified}), then the driver can move to grid cell $g_l$ from $g_k$ if the extra distance  travelled by moving through the modified path (which incorporates $g_l$) 
does not violate the detour ratio of a passenger who has to travel from $g_i$ to $g_j$ i.e., %the detour ratio  $\alpha$ of the passenger who has to travel from $g_i$ to $g_j$  should be bounded by a threshold value $t$ if the driver moves through grid cell $g_l$.
%Since the backward greedy approach selects a destination point and backtracks to the source through the destination point, we will move backwards from destination to source and see if the requests can be taken at grid cells without violating the detour constraints of the passenger  who has to travel from $g_i$ to $g_j$. For the adjacent node $g_k$ of $g_j$, the proposed model would take them if the edge weight between $g_k$ and $g_j$ satisfies the following equation.
%\vspace{-2mm}
\begin{equation}
    \alpha(P_{G^R},g_i,g_j) \leq t
\end{equation}
We know the detour ratio of the passenger is the length of the path travelled by the vehicle between its source and destination points
to the length of the shortest path between those points. The length of the shortest path between the grid cells $g_i$ and $g_j$ is denoted by $|SP(g_i,g_j)|$.
The length of the path taken by the vehicle between grid cells $g_i$ and $g_j$ changes when the driver is assumed to be at grid cell $g_k$ %request arrives who has to travel from $g_k$ to $g_l$ and it 
and its adjacent grid cell $g_l$ is checked if it could be taken in the path, and it becomes  $d_{jk}+w^R_{k,l}+ |SP(g_l,g_i)|$, where $d_{jk}$ is the distance travelled by vehicle between the destination point $g_j$ and the point at which driver is currently located $g_k$, $w^R_{k,l}$ is the distance between the current position of the driver which is $g_k$ and its adjacent grid cell $g_l$, and $|SP(g_l,g_i)|$ is the length of the shortest path between the grid cells $g_l$ and $g_i$. % as we assume the driver would follow the shortest path after moving from grid cell $g_l$ i.e, between the grid cells $g_l$ and $g_i$.
The detour constraint specifies that the length of the modified path to the original path should be bounded by threshold $t$ i.e., 
\begin{equation}
     \frac{d_{jk}+w^R_{k,l} +|SP(g_l,g_i)|}{|SP(g_i,g_j)|}\leq t
    \label{eq:distextra}
\end{equation}

This constraint states that the driver can move to grid cell $g_l$  if the distance  between the current position of driver $g_k$ and the grid cell $g_l$ which is denoted by $w^R_{k,l}$, is less than $t\cdot |SP(g_i,g_j)|-d_{jk} -|SP(g_l,g_i)|$. %That is, if the edge weight between grid cells $g_k$ and $g_l$ is less than the threshold distance, it implies the diver can move through this grid cell without violating the detour constraints of the passenger who has to travel from $g_i$ to $g_j$. However, if the edge weight between grid cells $g_k$ and $g_l$ is more than the threshold distance then the driver cannot move to this grid cell as it would violate the detour constraints of the passenger who has to travel from $g_i$ to $g_j$. \textcolor{red}{can remove}
After the route in the window is complete, the window is slid %towards the final destination of the driver 
%\textcolor{blue}{IAG: or backward?} 
%\textcolor{cyan}{Forward. Actually, we mean, the window is moved towards the final destination point, the point till which the driver wants  recommendation mechanism. Since we have used the word destination for the endpoint of the window also, so it will create confusion if we write destination again}
and the same process is repeated with the destination point of the previous window as the source of the next window. This process continues until the last passenger in the vehicle reaches his destination. %driver wants  a recommendation mechanism. 

Algorithm 1 presents the pseudocode of the Backward Greedy algorithm. After creating the $k$-hop window around the source node $o_s$, the node with maximum expected requests from the source node is returned ($o_d$) as the destination point of the window and added to the backtrack array $b$ (lines 5-8). 
If $o_d$ is directly connected to the source then the source is updated and the path is returned by backtracking from the destination point to the source through the array $b$,  and the window is slid (lines 9-12), else we will loop till the nodes  connected to the destination point are directly connected to the source node $o_s$ and keep appending them to the backtrack array $b$ (lines 13-17). This procedure is repeated  till the last passenger in the vehicle reaches his destination. %the point the driver wants the recommendation mechanism.
%\textcolor{blue}{IAG: This needs more clearance. Also, next give no. or name to Algorithm, and then write next we present the formal algorithm. Introduce forward greedy in the section itself. Regarding algorithm: what is b? Mention its use while mentioning it first time in comments}
%\textcolor{red}{read this}
%\vspace{-5mm}
\subsubsection{ Forward Greedy (FG)}

Forward Greedy has an approach similar to the Backward Greedy variant with a slight change in procedure. This change occurs in the selection of nodes between the source and destination. While the Backward Greedy selects the directly connected nodes from 
the destination node and backtracks to the source node, the Forward Greedy does the reverse and follows the path from the source to the destination. In this approach after selecting the source and destination points, instead of looking at the  requests from the nodes connected to the destination,   we look at the requests from the source node and continue the procedure till the destination node of the window is reached.
Initially, we check the adjacent nodes of the source and 
%to its adjacent nodes and 
return the one which has the highest number of expected requests directed from the source node. 
After that, we check if the node connected to the source that has the maximum requests from the source is connected directly with the destination node of the window. If that is the case, we move the window forward. However, if that is not the case we repeat the procedure until we get a  node that is directly connected to the destination.  After the route in the window is complete, we move the window with the destination of the current window as the source of the next window. In this way, we continue until the final destination point is reached.

Like Backward Greedy, this approach does not violate the detour ratio of directly connected requests but can violate the constraints of
requests that have to travel from the source point to the destination point. In order to ensure the distance travelled by the passenger is bounded when the highest expected request path is selected between its source and destination points, it should satisfy the following detour constraint: %threshold specified by Eq. \ref{eq:distextra}.
\begin{equation}
     \frac{d_{ik}+w^R_{k,l} +|SP(g_l,g_j)|}{|SP(g_i,g_j)|}\leq t
    \label{eq:distextraforward}
\end{equation}
%The detour constraint of forward greedy is similar to that of backward greedy and the only difference is the way we select grid cells. While in backward greedy, grid cells are selected from the destination onwards, in forward greedy the grid cells are selected from the source point. Accordingly, the numerator in the equation changes.
where $d_{ik}$ represents the distance between the source grid cell $g_i$ and the grid cell $g_k$ at which the driver is located, $w^R_{k,l} $ represents the distance between grid cell $g_k$ and the grid cell $g_l$ that needs to be checked for detour constraint, and $|SP(g_l,g_j)|$ represents the distance of the shortest path between grid cells $g_l$ and $g_j$. According to this constraint, the driver is permitted to move to grid cell $g_l$ only if the distance between its current position which is grid cell $g_k$ and the target grid cell $g_l$ (represented by $w^R_{k,l}$) is less than $t\cdot |SP(g_i,g_j)|-d_{ik} -|SP(g_l,g_j)|$.

In both the forward and backward greedy approaches, the proposed approach recommends a route between the source and destination points of the window that has the highest expected requests and satisfies the detour constraints of the passenger on board who has to travel from the source to destination points of the window. However, apart from the passenger who has to travel from source to destination points of window, there is another passenger who is already in the vehicle whose source and destination points can lie outside the window. The proposed model needs to ensure that while selecting the highest request path, the detour constraints of the passenger who is already in the vehicle %and the passengers who have to travel between the source and destination points of the window 
lies below the threshold value. It ensures that through the use of equations \eqref{eq:distextra} and \eqref{eq:distextraforward}, where $g_i$ and $g_j$ now correspond to the source and destination points of the passenger who is already in the vehicle. 

\subsubsection{Complexity analysis}
The %time taken to execute our proposed model 
complexity of our proposed model depends upon the following two factors: 1) the time required to execute the operations performed within the window, and 2) the number of times the window slides between the source and destination points of the passenger. 
Within the window, the proposed model selects the source node and finds the node that has the highest expected requests from the source node as the destination node of the window. This operation determines finding the maximum element within the window and its complexity will depend upon the number of elements in the window. Let $e$ denote the number of elements in the window, then the complexity of finding the maximum element in the window is $O(e)$. After selecting the maximum request point as the destination point, the proposed model finds the set of directly connected nodes between the source and destination nodes. In the worst case, the proposed model can visit each and every element within the window which leads to the complexity of $O(e)$. So the complexity of executing the operations within the window is $O(e^2)$.  
We know $e$ denotes the number of elements within a window and its count depends upon the number of hops $k$ i.e.,  $e=8(1+\sum_{i=2}^k(i-1)+1)$. %To illustarte it consider Figure \ref{} where hIf the hop count $k$ is $1$ there are $8$ elements in the window,if hop count is $2$ there will be $24$ elements, and so on. In general, the number of elements $e$ depend upon the hop count $k$ as  $e=8(1+\sum_{i=2}^k(i-1)+1)$.
Thus, the number of elements $e$ is of the order of $k^2$ and the time complexity within a window is $O(k^4)$. The second factor which determines the complexity is the number of times the window is slid. In the worst case, the window is slid over each and every grid cell. If we assume there are $n$ grid cells, the complexity of the sliding window between source and destination points will be $O(n)$. The overall complexity of the proposed approach depends upon the number of times the window is slid and the time required to execute instructions per window and it is $O(nk^4)$. The hop count $k$ is determined through experimental evaluation and it is usually a small constant as the performance of the system improves quadratically initially with the increase in the hop count $k$ and stabilizes after some time (see Section \ref{sec:experiments}). Its value is usually smaller than $6$. So we can say the proposed model runs in linear time. %whose value is less than $5$ as the performance improves rapidly initially with the increase in $k$ and stabilizes after $k$ reaches a constant value. 

\subsection{Fleet size}
\label{subsec:fleetsize}
In this subsection, we will describe the process of obtaining the fleet size. The fleet size is a crucial factor that determines the number of vehicles required to effectively service all passenger requests that arrive within a given time frame. %It provides the ridesharing platforms with an estimate of the resources required to service all of their passengers. 
There are two cases that arise in the determination of fleet size: 1) the fleet size required by our proposed model, which predicts the origin and destination of requests through GNN based architecture described in \cite{Ashraf_archive:2022} and thereafter utilizes the greedy-based sliding window approach to recommend the routes, and 2)  the optimal fleet size, which occurs when the arrival of passenger requests 
is known beforehand.

To determine the fleet size of our proposed model, we employ a random sampling technique where we allocate an imaginary set of vehicles to various grid cells. These vehicles are then assigned routes using the greedy-based sliding window approach described in the previous subsection. After allocating routes to all the imaginary vehicles, we count the number of  vehicles that were actually assigned to the passengers. This count represents the fleet size required by our proposed model to service all passengers within a given area and time frame effectively.

The second case involves determining the optimal fleet size, which represents the \emph{minimum} number of vehicles required by the ridesharing platforms to accommodate all passenger requests when the passenger arrival sequence is known \emph{in advance}.   
Although this scenario rarely occurs in practice, as companies typically do not possess precise knowledge of passenger arrival sequences beforehand, it serves as a benchmark for evaluating the optimal number of vehicles needed by ridesharing platforms. %It also provides insights into the performance of our proposed greedy-based approach compared to the best-known case. 
To derive the optimal fleet size in  the offline case, we utilize the  \emph{target graph} $G^T$ that accurately captures the arrival sequence of passenger requests for the ridesharing platforms. The target graph incorporates all the requests that can arrive within a particular time frame.
%As previously mentioned, our proposed system consists of a family of subgraphs, where the request graph represents the expected number of passenger requests between any pair of grid cells, and the target graph represents the actual number of requests between each pair of grid cells.  While the request graph is used by our proposed model for route recommendation and fleet size estimation in the first case, the target graph is employed specifically for determining the optimal fleet size.  The target graph incorporates all the requests that can arrive within a particular time frame. For instance, the target graph displayed in Figure \ref{fig:targetinstance} contains the requests that arrive within $15$ minutes. We need to determine the number of vehicles needed to service the total requests in this graph %that arrive within $15$ minutes,  which are equal to  the summation of its edge weights i.e., $11$ requests. 
This graph contains some requests which can be merged into the same vehicle without violating their detour constraints. %For instance, the requests between grid cells $g_{26}$ and $g_{1}$, can share the same vehicle with the requests between grid cells $g_{26}$ and $g_{25}$, and the path followed by the vehicle for servicing both the requests simultaneously will be $P=\{g_{26},g_{25},g_{1}\}$. It can be  seen the modified path $P$ for the request(s) that have to travel from $g_{25}$ to $g_1$ remains same as their shortest path, so their detour ratio does not get violated (the modified path for these requests is $\{g_{25},g_1\}$, which is the same as the shortest path between the grid cells $g_{25}$ and $g_1$). However, the path of requests that have to travel from $g_{26}$ to $g_1$ changes after incorporating requests from $g_{25}$ to $g_1$ and becomes $P$. The detour ratio of this passenger is the length of the modified path $P$   which is $4$, to the length of the shortest path between its endpoints $g_{26}$ and $g_1$ which is $2.8$. The detour would be $\frac{4}{2.8}=1.42$ which is less than the threshold value of $1.5$. Thus these orders can be merged into a single vehicle without violating their detour constraints. 
It also contains certain requests which can be serviced by the same vehicle consecutively i.e, after dropping off some passengers, the vehicle arrives at the starting point of the next passenger before his waiting time is over. 
%, where the vehicle after dropping a request, reaches the starting point of the next request before its waiting time is over. %a single vehicle can service the requests 
%In addition to merging the requests based on their detour constraints, certain requests can be serviced by the same vehicle consecutively, where the vehicle after dropping a request, reaches the starting point of the next request before its waiting time is over. For instance, after dropping passengers from $g_1$ to $g_2$, there is a passenger request at $g_2$ who has to travel to $g_{26}$ (see Figure \ref{fig:targetinstance}). If the driver reaches $g_2$ within his waiting time period then this request can be serviced by the same vehicle. Therefore, we need to handle these two cases and consider the number of vehicles required by the ridesharing platforms when some requests can be serviced by the same vehicle. % either when it is running or if i. 
%For determining the optimal fleet size, we need to take these cases into account and
We need to handle these two cases and consider the number of vehicles required by the ridesharing platforms when some requests can be serviced by the same vehicle. % either when it is in transit and merges passengers or .
%When considering these cases, it is essential to take into account the vehicle capacity, as 
While considering these two cases, we need to take into account the vehicle capacity, as there is a %there is an important consideration to take as a single vehic
maximum limit on the number of requests a vehicle can accommodate. %we need to take into account 

%there is an important Moreover, there is a maximum limit on the number of requests a vehicle can accommodate.
In order to derive the fleet size with 
the above two cases 
and 
capacity constraints,
we transform the target graph  which contains the request arrival sequence known in advance, into the vehicle count graph. The target graph $G^T=(V,E^T,w^T)$, being a complete graph, encompasses all possible requests between any pair of grid cells. This graph is transformed into the vehicle count graph $G^C=(V^C,E^C,w^C)$, which specifically denotes the requests that can be serviced by the same vehicle. The transformation process is described next. % involves the following steps:

\begin{figure}[t!]
\centering
%\begin{minipage}{.5\textwidth}
 % \centering  \includegraphics[width=1\linewidth]{target.pdf}
  %\vspace*{-165mm}
   % \caption{Target request graph}
    %\label{fig:targetinstance}
%\end{minipage}%
%\begin{minipage}{.5\textwidth}
 % \centering
\vspace*{-4mm}  \includegraphics[width=0.7\linewidth]{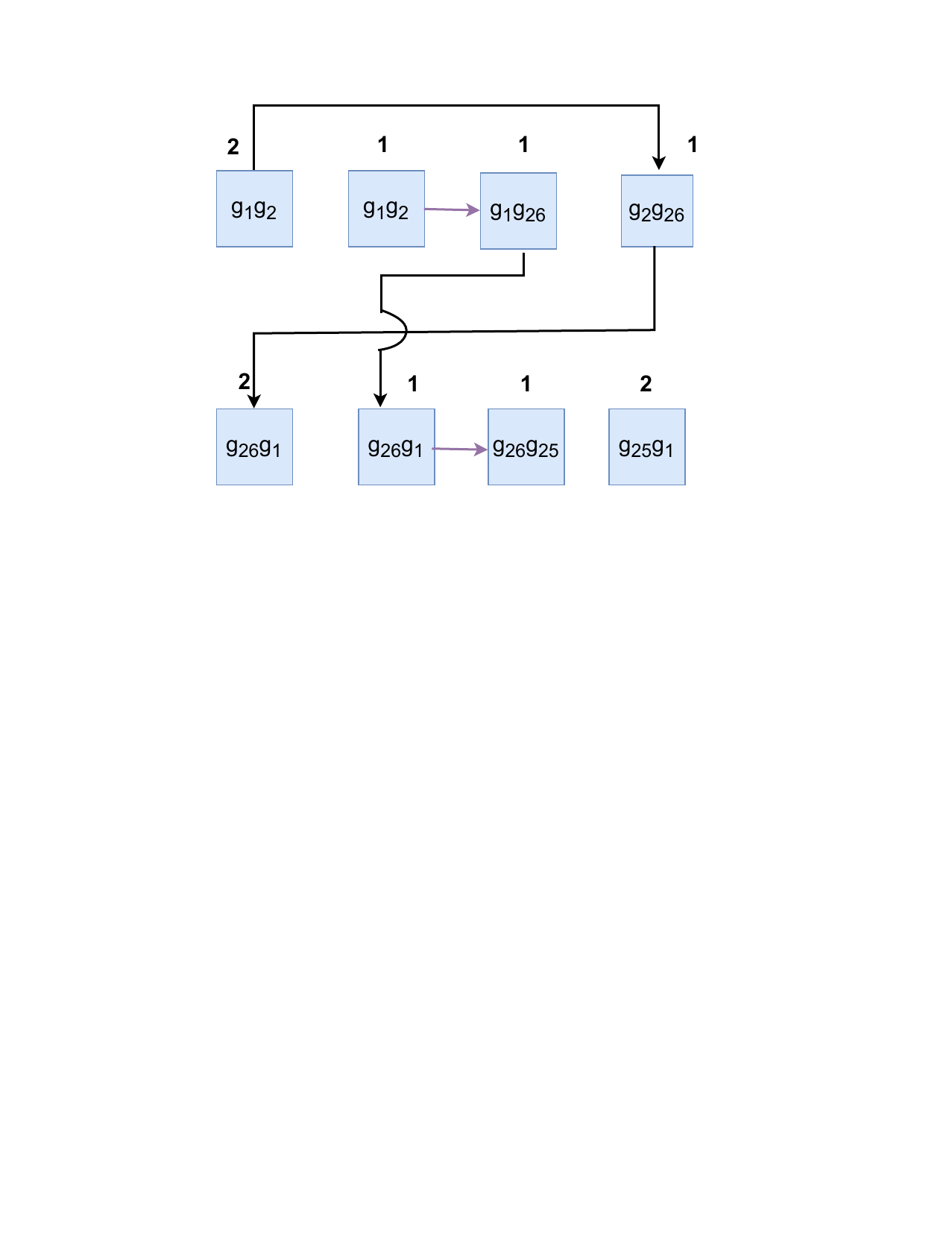}   \vspace*{-71mm}
    \caption{Vehicle count graph}
    \label{fig:vehcount}
  %  \vspace*{-6mm}
%\end{minipage}
\end{figure}

The vertices $V^C$ of the vehicle count graph represent the edges $E^T$ of the target graph in terms of the vehicle capacity i.e, if the edge weight $w^T_{ij}$ between any two grid cells $g_i$ and $g_j$ in the target graph is less than or equal to the vehicle capacity  $c$ , these edges appear directly as vertices in the vehicle count graph with the vertex weight in vehicle count graph set equal to the edge weight in the target graph. However, if the edge weight $w^T_{ij}$  between the grid cells $g_i$ and $g_j$ of target graph is more than the vehicle capacity than the number of vertices in the vehicle count graph will be $\left \lceil{\frac{w^T_{ij}}{c}}
\right \rceil$, and the vertex weight of $\left \lfloor{\frac{w^T_{ij}}{c}}
\right \rfloor$ vertices will be
equal to the vehicle capacity, and one vertex will have a weight of $w^T_{ij}-\left \lfloor{\frac{w^T_{ij}}{c}}
\right \rfloor \cdot  c$.
In order to understand this point consider Figures \ref{fig:targetinstance} and \ref{fig:vehcount} which show the target graph and its vehicle count graph. Let's assume the vehicle capacity is $2$ i.e., the vehicle can at maximum carry $2$ passengers in a single run. The edge $(g_1,g_{26})$ of target graph has a weight of $1$ which is less than the vehicle capacity of $2$, so it directly appears as the vertex $g_1g_{26}$ in vehicle count graph with vertex weight being equal to the corresponding edge weight in the target graph i.e, $1$. The edge $(g_1,g_2)$ of the target graph has a weight of $3$ which is more than the vehicle capacity of $2$, so there are $\left \lceil {\frac{3}{2}}\right\rceil$ vertices in vehicle count graph i.e, $2$ vertices.  The vertex weight of $\left \lfloor{\frac{3}{2}}
\right \rfloor$ i.e, $1$ vertex will be
equal to the vehicle capacity of $2$, and one vertex will have a weight of $3-\left \lfloor{\frac{3}{2}}
\right \rfloor \cdot  2 =1$ in vehicle count graph.

Now, that we have defined the vertices of the vehicle count graph, we will describe the formation of the edges of this graph.
The edges of the vehicle count graph connect the vertices that can be serviced through a single vehicle. A single vehicle can service multiple requests either if it is running with more than one passenger in it, or if it reaches the starting point of next passenger after dropping previous.  %There are two cases that arise in the%in either of the two cases described above i.e, when the vehicle pairs the passengers in a single vehicle when it is running, or if it reaches the location of next passenger request after dropping the previous one. 
 For the first case which confers to the efficient pairing of passengers 
 in a single vehicle when it is operating, we check the vertices of the vehicle count graph with vertex weight less than the vehicle capacity $c$, and connect them through an edge if they can be merged in a single vehicle without violating the detour constraints of any of them. % if the vertices satisfy the detour constraints of passengers they are connected through an edge.   
 %we take into account the detour ratio of passenger(s) onboard and ensure only those vertices of the vehicle count graph are connected through an edge that can be taken by the driver without violating the detour constraints of any of the passenger(s) onboard. 
 For the second case, in which the vehicle can service the passengers consecutively, %i.e., it picks up the next passenger(s) after dropping the previous passenger(s), 
 we connect the vertices through the edge only if the vehicle  arrives at the next passenger's place (grid cell) before its waiting time is over.  Figure \ref{fig:vehcount} displays the vehicle count  graph corresponding to the target graph shown in Figure \ref{fig:targetinstance}. The purple edges represent the vertices that can be serviced through the same vehicle in the sharing mode, % i.e., when the vehicle is running and is able to pair the passengers by satisfying their detour ratio. The 
 and the black-coloured edges connect the vertices which can be serviced by the same vehicle one after the other. %i.e., it picks up the next passenger after dropping the previous one. 
%In our proposed model, we predict the requests that can arrive within $15$ minutes, and we assume that the passenger will wait for a maximum of $10$ minutes. Moreover, the vehicle capacity is assumed to be $2$.   Taking these conditions, the  vertices $g_1g_2, g_2,g_{26}$, and $g_{26}g_{1}$ of the vehicle count graph are connected through the black edges which indicates that the same can vehicle can service them after dropping the previous requests at their respective destinations. 
%In this graph, the connection of vertices $g_1g_2$ and $g_2g_{26}$ indicates that after dropping the passenger from grid cell $g_1$ to $g_2$, the driver is currently at $g_2$ and the request has arrived at  $g_2$ which wants to travel to $g_{26}$.   The distance between grid cells $g_1$ and $g_{2}$ is $2$ $km$ (see road graph displayed by Figure \ref{fig:roadinstance}) and assuming the average speed of $53\, km/hour$ \cite{Article:speed_dc}, the vehicle can reach $g_2$ from $g_1$ within $10$ minutes, i.e, the driver reaches $g_2$ before the waiting time of passenger at that grid cell is over, and can service it.  After connecting the vertices $g_1g_2$ and $g_2g_{26}$ the black edge connects $g_2g_{26}$ and $g_{26}g_1$ in the vehicle count graph. This is because the vehicle can travel from $g_1$ to $g_2$ to  $g_{26}$ with the average speed of $48.28\, km/hour$ speed within $10$ minutes, and take the passenger request at that grid cell.

After formulating the vehicle count graph, we need to determine the minimum number of vehicles required to service all the passenger requests in this graph. The vertices that are isolated i.e., those vertices which are not connected with any other vertex through an edge require a single vehicle.  Each isolated vertex represents a passenger request that couldn't be paired with any other passenger due to detour constraints or the unavailability of a suitable vehicle that could reach within the waiting time constraints of the passenger. %It could also represent that the vertex weight is equalt to vehicle capacity %, or exceeding the vehicle capacity. Hence, each isolated vertex requires a single vehicle.%These are the vertices that could not get paired with any other passenger due to the detour constraints, or the passenger requests at these requests were more than the vehicle capacity, or no other vehicle reached their starting  point without violating their waiting times.
On the other hand, vertices that are connected through an edge will result in a decrease in vehicle count and the connected vertices will require a single vehicle to service them. Determining the minimum fleet size on this graph is reduced to finding the minimum number of the edge sequences such that each vertex precisely belongs to only one sequence, which is equivalent to finding the minimum path cover on the vehicle count graph.  For example, referring to Figure \ref{fig:vehcount}, the vehicle count graph exhibits $3$ edge sequences that cover the graph, ensuring that each vertex is part of exactly one sequence. Therefore, the minimum number of vehicles required is $3$. \looseness=-1%For instance, in the vehicle count graph displayed through Figure \ref{fig:vehcount}, at minimum there are $3$  edge sequences that cover the graph such that each vertex belongs to exactly one sequence are $3$.

\section{Experiments and Results}
\label{sec:experiments}
In this section, we determine through experimental evaluation that our proposed model is eco-friendly, scalable, and can be used by ridesharing platforms for effective route recommendation.

\subsection{Experimental setup}
All the experiments are implemented in Python and performed on a machine with Intel(R) Core(TM) i9-12900  CPU $2.40$ GHz with $32$ GB RAM. \looseness=-1
\subsubsection{Datasets}

\begin{table}[h!]
    \centering
    \vspace*{-1mm}
    \small
        \caption{A Summary of Datasets used}
        \vspace*{-1mm}
    \begin{tabular}{|c|c|c|}
      \hline   Datasets& New York & Washington DC  \\ \hline
        Time Span & 1 month & 1 month \\ \hline
        Grid cell size & $2 *2\, km ^2$ & $2 *2 \,km ^2$ \\ \hline
        Size &$ 20024124$ & $1389234$\\ \hline
Number of grid cells & $576$ &$99$ \\ \hline
Time slot granularity & $15$ minutes & $15$ minutes \\ \hline
    \end{tabular}
    \vspace*{-2mm}
    \label{tab:datasets}
\end{table}

To evaluate the performance of our proposed model, we conducted experiments using two real-world datasets obtained from New York and Washington DC. These datasets contain different passenger distribution patterns and determine the functioning of the system in varied environments. %They have been used by previous studies as benchmarks \cite{}.
They were collected for February  $2016$ and $2017$ respectively. Table \ref{tab:datasets} provides a brief overview of the datasets utilized in our experimental analysis. The datasets were divided into grid cells with a size of $2$ km and a time interval of $15$ minutes. %The selection of the grid cell size is crucial in route recommendation systems, as it impacts the prediction accuracy. If the grid cells are too large, the large areas will be encompassed within a single grid, which oversimplifies the analysis. On the other hand, if the grid cells are too small, the prediction accuracy would decrease considerably \cite{Ashraf_archive:2022}. To strike a balance between these two cases, we followed the approach employed in previous studies on route recommendation. Specifically, we divided the datasets into grid cells of size $2$ km, which has been found to maintain an effective balance  in previous research on route recommendation systems \cite{Wang:VLDB_2020, Tong:VLDB_2017}.
%We evaluate the performance of our proposed model on two real-world datasets generated by New York and Washington DC.  Table \ref{tab:datasets} provides a brief description of the datasets used for the experimental evaluation of our proposed model. The New York City and Washington DC datasets were collected for the month of February $2018$ and $2019$ respectively. The datasets are divided into grid cells of $2\,km$ with a time gap of $15$ minutes.  The size of grid cell is an important parameter in route recommendation. If it is too large  then the whole area will be concentrated within a single grid and if it is too small then the prediction accuracy will decrease \cite{Ashraf_archive:2022}. To maintain a balance between the two cases, as some of the previous route recommendation studies  have we divide the datasets into the grid cells of size $2\, km$ \cite{Wang:VLDB_2020,Tong:VLDB_2017}
%The datasets are divided into the grid cells of length $1.24\, miles$ since it provides   
%\cite{Wang:VLDB_2020,Tong:VLDB_2017}
The rows of the dataset are of the form pick-up time, pick-up latitude and longitude, drop-off latitude and longitude, and passenger count.  %This data about passengers' origin and destination is used for recommending routes to drivers.
This data
about passengers' origin and destination is provided as input to the GNN-based model described in \cite{Ashraf_archive:2022}, which predicts the number of requests that can arrive between any two grid cells within the next $15$ minutes. 
The predicted data 
is fed as input to the request graph $G^Q$ which is used for recommending routes to the drivers. As can be seen from Table \ref{tab:datasets} the size of the dataset is massive and thereupon we have utilized the greedy-based sliding window approach to decrease the search space considerably and speed up the recommendation process.

\subsubsection{Rationale for Grid-based Approach}

The grid-based approach provides a simplified representation of the road network, facilitating the modelling of complex transportation systems. It divides the road into geographical zones, allowing the proposed model to recommend zones with a higher flow of passenger requests efficiently.
This approach has gained popularity in various studies on route recommendation and matching algorithms, as evidenced by works such as \cite{Tong:VLDB_2017,Wang:VLDB_2020,Tong:ACMTrans_2022,Sun:KDD_2022}.

When implementing a grid-based approach, the size of the grid cell becomes an important parameter. If the size is small, there will be an increase in the computational complexity of the model. On the other hand, if the size is  large, a single grid cell covers a substantial number of road segments, which oversimplifies the road network and ignores the important details.
Informed by insights from previous studies \cite{Wang:VLDB_2020, Tong:VLDB_2017}, we have set the size of the grid cell as $2\,km$. This size strikes a balance, aiming to capture relevant details of the road network while maintaining computational feasibility.
%It is impcomplexityortant to note that while the grid-based approach simplifies the representation of road network, this system can be easily applied in practical settings, where the vertices represent road intersections in place of grid cells, and the edges represent road segments.

It is important to note that while the grid-based approach simplifies the representation of road networks, this system can be easily applied in practical settings. In these cases,  the vertices are redefined to represent actual road intersections, %offering a finer granularity in the modelling process. 
%Consequently, 
and the edges connecting these vertices denote the road segments, enabling a transition from a grid-centric model to one that more accurately mirrors the intricacies of real-world road networks. This adaptability enhances the generalizability of the proposed model, allowing it to be  integrated into diverse real-world scenarios with different levels of road network complexity. %where the vertices represent road intersections in place of grid cells, and the edges represent road segments.

%Furthermore, the adaptability of the grid-based approach extends beyond its inherent simplification. In practical settings, the versatility of this system becomes evident as it seamlessly translates to a more detailed representation. In such cases, the vertices of the grid can be redefined to represent actual road intersections, offering a finer granularity in the modeling process. Consequently, the edges connecting these vertices would then represent the road segments, enabling a transition from a grid-centric model to one that more accurately mirrors the intricacies of real-world road networks.

%This flexibility not only underscores the pragmatic nature of the grid-based approach but also highlights its scalability. By accommodating a transition from grid cells to road intersections, our model can be easily tailored to match the level of detail required by various applications, including those mirroring the directed graph representations found in ridesharing platforms. %This adaptability enhances the generalizability of the proposed model, allowing it to be seamlessly integrated into diverse real-world scenarios with different levels of road network complexity.

\subsubsection{Evaluation Framework}
To evaluate the performance of our proposed model, we divide the dataset into training data and test data. The training dataset contains $75\%$ and the rest is used for testing. The model is learned on training data and its performance is evaluated on the test dataset. The data generated by the test set is used as input by the route recommendation algorithm to provide optimal routes to the driver. Thereafter, the performance of the route recommendation system is measured by using the actual data of passengers' origin and destination of the corresponding place.  The working of the proposed route recommendation system depends upon the output of the deep learning model applied for predicting the origin and destination of passenger requests. Although we have used the GNN-based model proposed in \cite{Ashraf_archive:2022} for predicting the passengers' origin and destination, any deep learning model can be applied to do the prediction task. \looseness=-1

\subsubsection{Baselines} 
We compare the performance of our proposed model with the following baselines:

\textbf{Shortest path (SP):} The shortest path algorithm recommends the shortest path  between the source and destination points. %There is no element of prediction here.

\textbf{SHARE \cite{Yuen:ACMWWW_2019}:} We evaluate SHARE which predicts the demand at various nodes and recommends the routes with the highest expected demand without violating the detour constraints of passengers in the vehicle. %SHARE has an element of prediction and predicts the request sequences that can arrive in the future.

\textbf{Unified route planning (URP) \cite{Tong:ACMTrans_2022}:} It is a route planning approach that uses historical data to analyze areas with high demand. Whenever requests appear dynamically it modifies the original route of the driver to incorporate the new request through its insertion operation. %inserts the requests that arrive dynamically through its insertion operation. 

\subsubsection{Metrics}
The performance of our proposed model is measured by the following metrics:

\textbf{Percentage of orders with ridesharing 
:} This metric quantifies the proportion of ride orders that were executed in sharing mode, reflecting the effectiveness of our proposed model in pairing multiple passengers based on their origin and destination information. A higher value of this metric indicates better performance of the route recommendation system in terms of maximizing shared rides. \looseness=-1
%This metric determines the percentage of ride orders that were running in sharing mode. It measures the efficiency of our proposed model in pairing up multiple passengers by using their origin and destination data. Higher the value of this metric, better will be the performance of the route recommendation system. %If the driver travels without a passenger in the vehicle it incurs a financial loss. Thus we measure this loss through the percentage of orders that are unable to find a compatible passenger on their path. Lower the value of this metric,  the better the performance of the model.

\textbf{Passengers per grid 
:}
This metric measures the occupancy level of the vehicle per grid. Its value should be higher for better performance of the system.

\textbf{Vehicle Utilization:} 
This metric measures the effective utilization of the vehicle. It is defined as,
$    VU=\frac{u}{c}$
, where $c$ is the  vehicle's capacity, and $u$ is the number of passengers in the vehicle i.e., this metric determines the utilized capacity of the vehicle. The value of $VU$ can range between $0$ and $1$, where the value $0$ indicates no passenger in the vehicle and the value $1$ indicates that the vehicle was filled completely. 
%We evaluate the performance of our proposed model on this metric and its value should be close to $1$ for good performance.

\textbf{Waiting time:}
%Customers are the priority of any system and the system will perform well if it keeps up with them. Keeping this point in view, we have considered this metric, and the proposed system should not increase it excessively as it will result in the dissatisfaction of customers with the system. 
Customers are the primary stakeholders of any platform, and the performance of the system depends upon its ability to meet their needs effectively. In ride-hailing platforms, when the passengers book a ride, they should get it quickly from the platform and their waiting time should be low. To ensure passenger satisfaction, we evaluate this metric and see how its value gets affected by incorporating greener rides.   %Taking this point in consideration, 
%We evaluate this metric in order to esnure its value should not get affected consdierably by the incorporayion of greener rides. % , we have taken this metric into account. It is essential that the proposed system does not significantly raise this metric, as it would ultimately lead to customer dissatisfaction with the system
\subsubsection{Parameter}
Our proposed model has one parameter $k$ which determines the size of window and needs to be estimated for complexity and accuracy. Its value is determined through experimental evaluation under different conditions. The default value of $k$ is $5$. 

\begin{figure*}[t!]
    \begin{minipage}[b]{1\textwidth}
\vspace*{-20mm}
  %  \centering
 % \vspace*{-16mm}
    \begin{subfigure}[b]{0.4\textwidth}
   % \vspace*{15mm}
    % \vspace*{10mm}
        \includegraphics[width=1.1\textwidth]{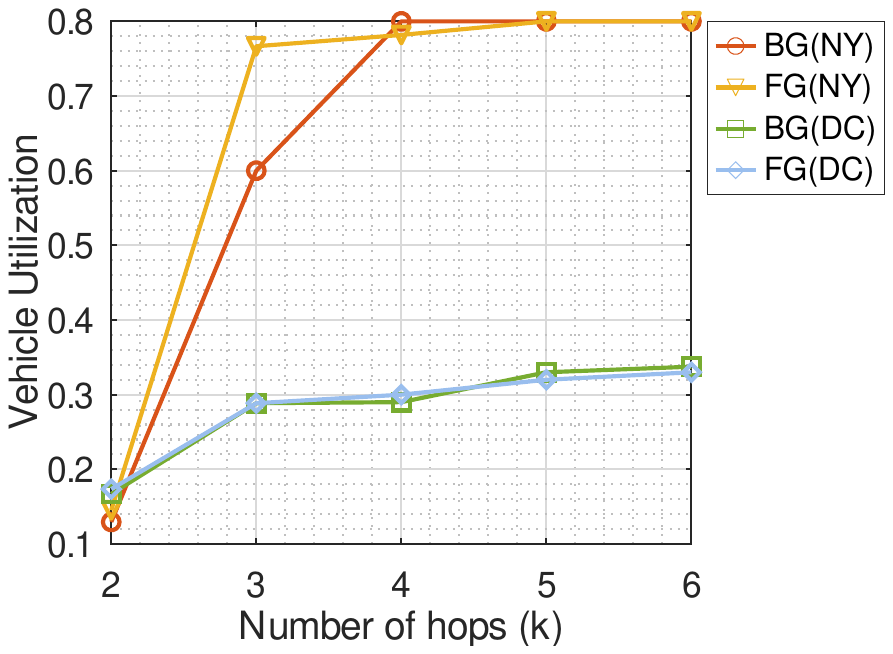}
         %     \setlength{\abovecaptionskip}{-40pt}
      %   \vspace*{-18mm}
          %  \caption{Vehicle Utilization}
        \label{img:vunyk}
     %   \caption{Demand}
    \end{subfigure}
    \hfill
%\vspace{-20mm}
\begin{subfigure}[b]{0.4\textwidth}
%\vspace*{-15mm}
        \includegraphics[width=1.1\textwidth]{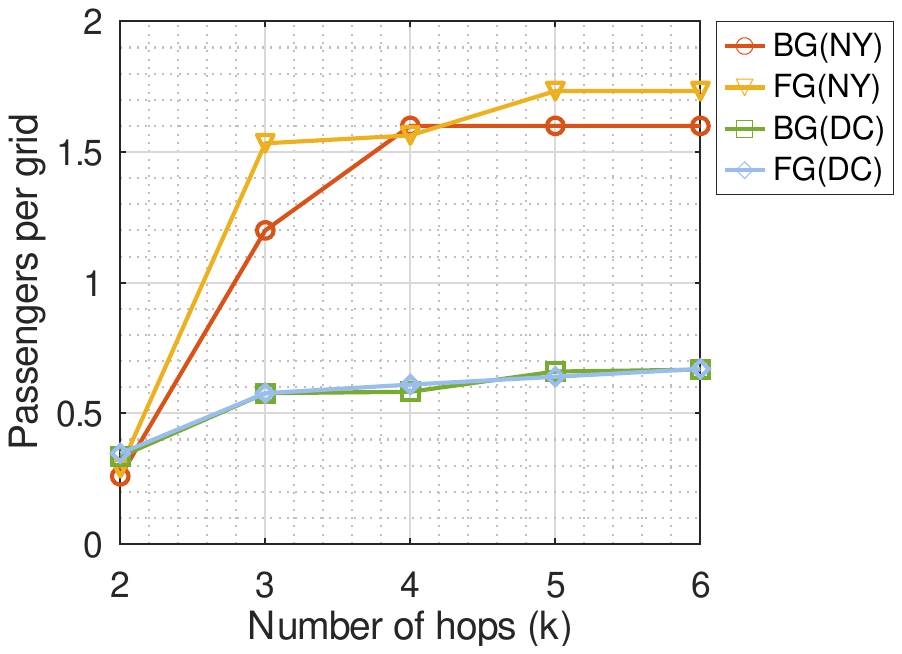}
            % \setlength{\abovecaptionskip}{-40pt}
            %\vspace*{-18mm}
            %\caption{Percentage of orders with ridesharing}
       \label{img:ordersnyk}
    \end{subfigure}
%    \vspace{-30mm}
  \hfill
  \end{minipage} 
      \begin{minipage}[b]{1\textwidth}
\vspace*{-45mm}
 % \vspace*{-80mm}
  \begin{subfigure}[b]{0.4\textwidth}
        \includegraphics[width=1.1\textwidth]{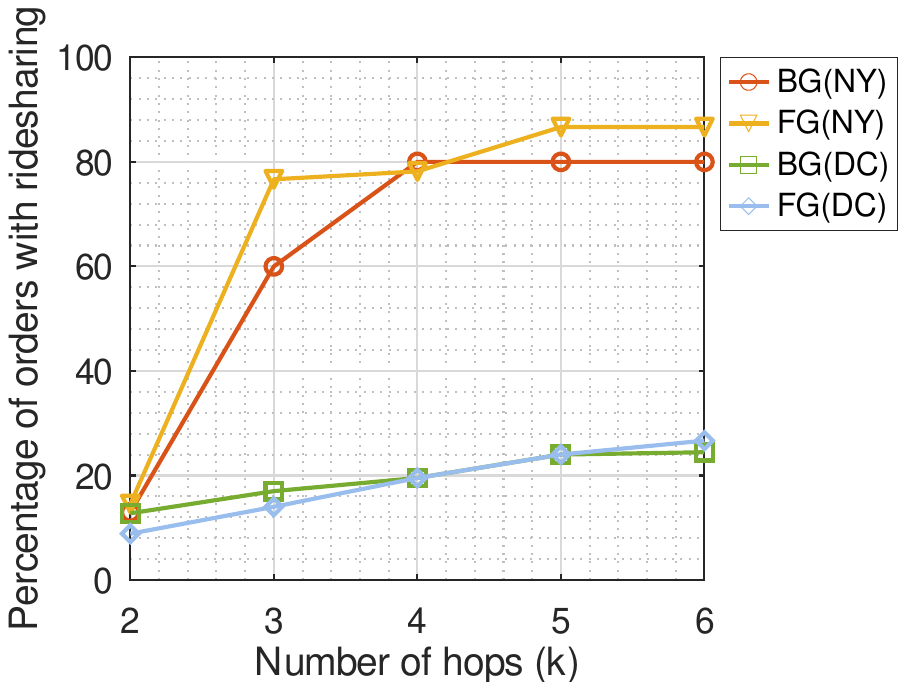}
         %    \setlength{\abovecaptionskip}{-40pt}
  %       \vspace*{-18mm}
     %        \caption{Passengers per grid  }
        \label{img:ppgnyk}
    \end{subfigure}
   % \vspace*{-7.5mm}
      \hfill
  \begin{subfigure}[b]{0.4\textwidth}
        \includegraphics[width=1.1\textwidth]{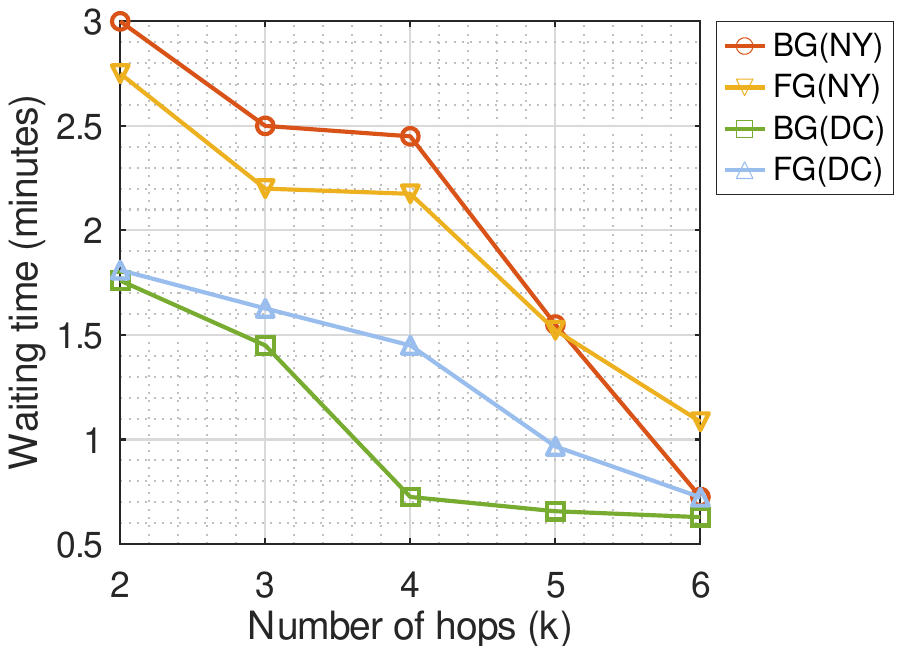}
         %    \setlength{\abovecaptionskip}{-40pt}
        % \vspace*{-18mm}
     %        \caption{Passengers per grid  }
        \label{img:ppgnyk}
    \end{subfigure}
    \vspace*{-25mm}
     \caption{Evaluation of different metrics on New York (NY) and Washington (DC)  dataset by Forward Greedy (FG) and Backward Greedy (BG) approaches based upon the number of hops ($k$)}
    \label{img:k}
    \end{minipage}
  %  \vspace*{-5mm}
%\hfill
 %\hfill
\end{figure*}

\begin{figure*}[t!]
  \vspace*{-27mm}\begin{minipage}[b]{1\textwidth}
  
   % \begin{figure}
    \begin{subfigure}[b]{0.4\textwidth}
    \includegraphics[width=1.1\textwidth]{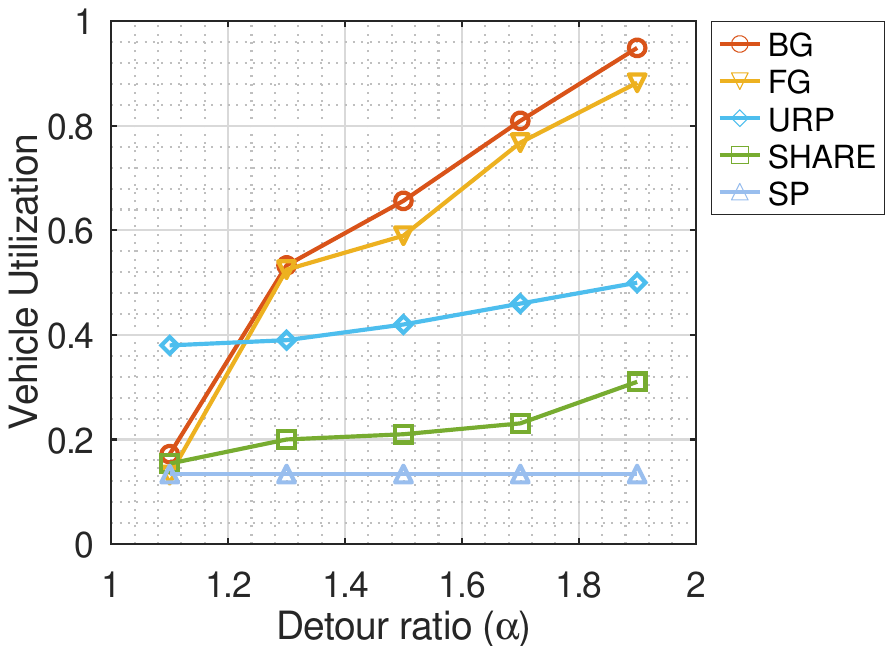}
      \vspace*{-25mm}
      \caption{New York dataset}
      \label{img:vunya}
    \end{subfigure}
    \hfill
    \begin{subfigure}[b]{0.4\textwidth}
      \includegraphics[width=1.1\textwidth]{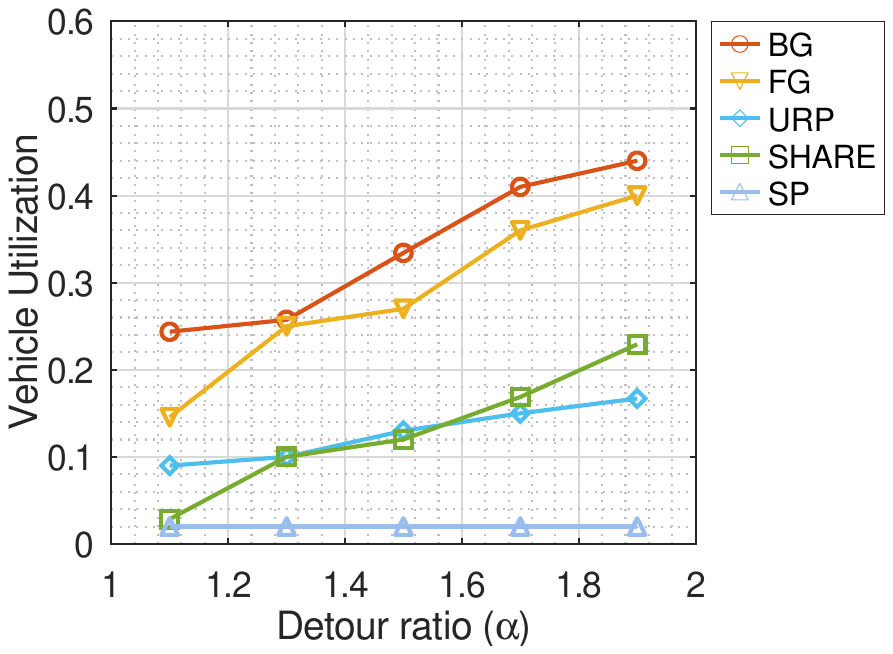}
      \vspace*{-25mm}
      \caption{Washington DC dataset}
      \label{img:vudca}
    \end{subfigure}
    \vspace*{-1mm}
        \caption{Vehicle Utilization with increase in detour ratio ($\alpha$) }
        \label{fig:vu}
    %\end{figure}
    
  \end{minipage}
  \hfill
  \vspace*{-25mm}
  \begin{minipage}[b]{1\textwidth}
    \begin{subfigure}[b]{0.4\textwidth}
      \includegraphics[width=1.1\textwidth]{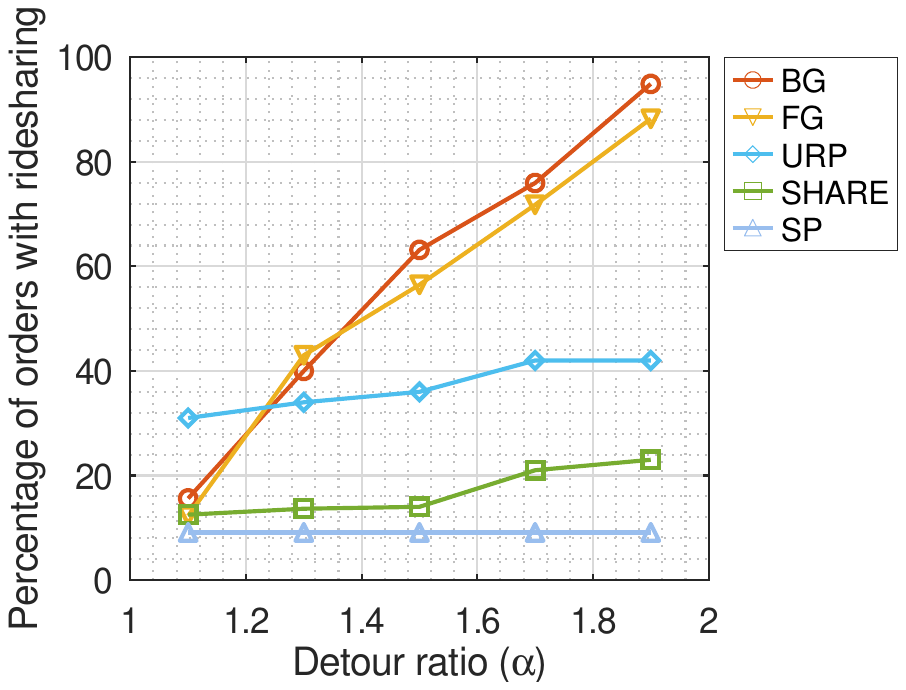}
      \vspace*{-25mm}
      \caption{New York dataset}
      \label{img:ordersnya}
    \end{subfigure}
    \hfill
    \begin{subfigure}[b]{0.4\textwidth}
      \includegraphics[width=1.1\textwidth]{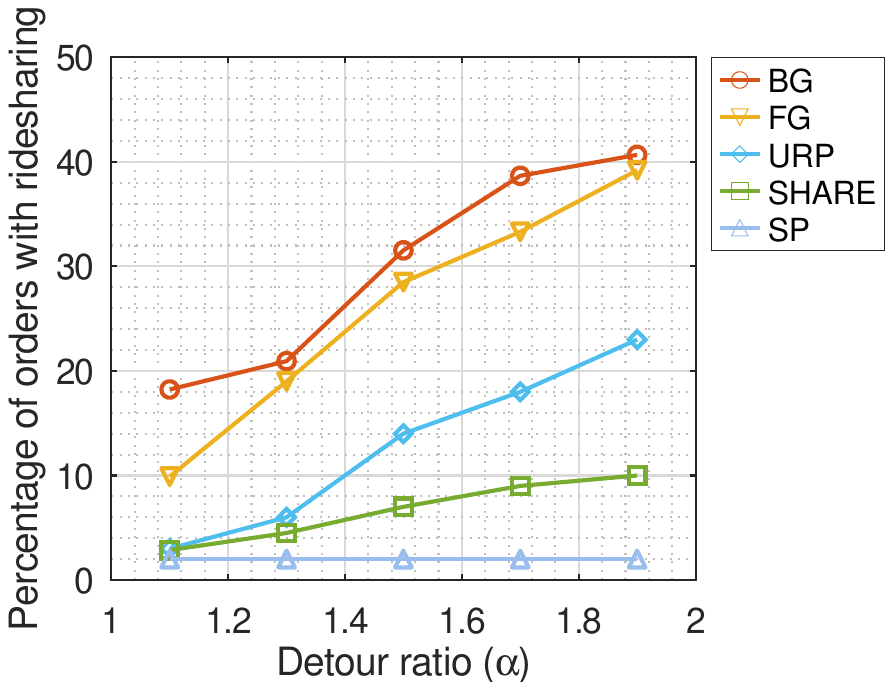}
      \vspace*{-25mm}
      \caption{Washington DC dataset}
      \label{img:ordersdca}
    \end{subfigure}
%        \caption{Percentage of orders with ridesharing}
 %       \label{fig:orders}
 \vspace*{-1mm}
 \caption{Percentage of orders with ridesharing with increase in detour ratio ($\alpha$) }
  \label{img:orders}

  \end{minipage}
  \vspace*{-3.5mm}
 \end{figure*}

\begin{figure*}
  \vspace*{-18mm}
  \begin{minipage}[b]{1\textwidth}
  %\vspace*{-19mm}
    \begin{subfigure}[b]{0.4\textwidth}
  %  \vspace*{-3mm}
      \centering\includegraphics[width=1.1\textwidth]{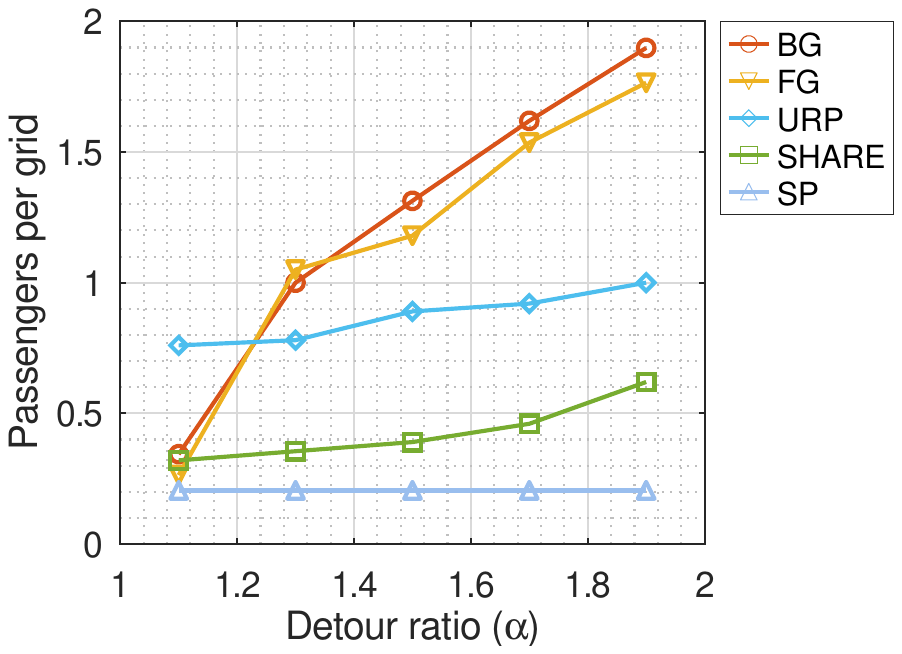}
      \vspace*{-25mm}
      \caption{New York dataset}
      \label{img:ppgnya}
    \end{subfigure}
    \hfill
    \begin{subfigure}[b]{0.4\textwidth}
    \vspace*{-13mm}
      \includegraphics[width=1.1\textwidth]{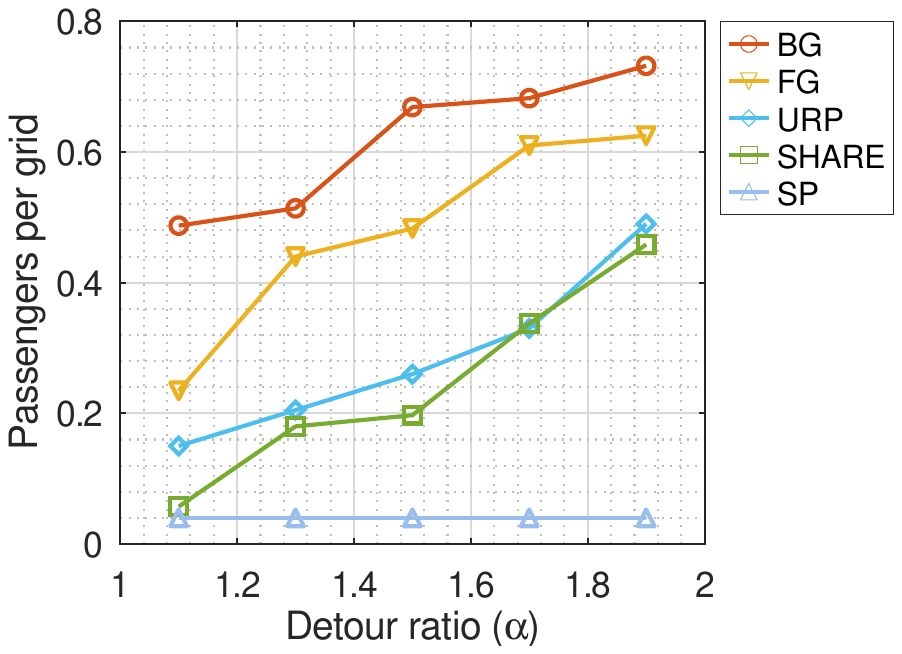}
      \vspace*{-25mm}
      \caption{Washington DC dataset}
      \label{img:ppgdca}
    \end{subfigure}
    \vspace*{-1mm}
  \caption{Passenger per grid with increase in detour ratio ($\alpha$)}
  \label{img:ppg}
  \end{minipage}
  \hfill \vspace*{-25mm}
  \begin{minipage}[b]{1\textwidth}
  %\vspace*{-19mm}
    \begin{subfigure}[b]{0.4\textwidth}
  %  \vspace*{-3mm}
      \includegraphics[width=1.1\textwidth]{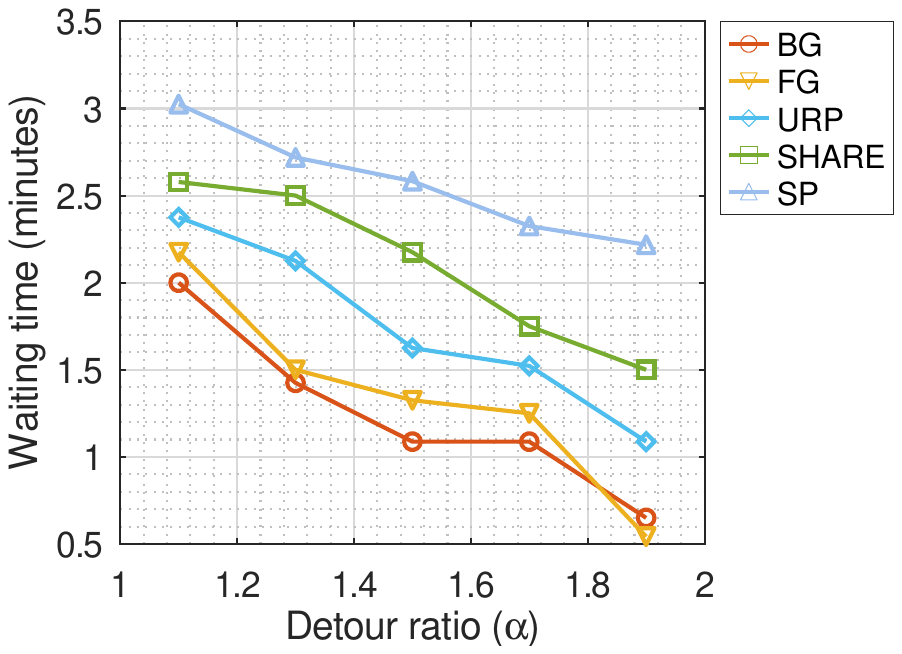}
      \vspace*{-25mm}
      \caption{New York dataset}
      \label{img:wtnya}
    \end{subfigure}
    \hfill
    \begin{subfigure}[b]{0.4\textwidth}
    \vspace*{-13mm}
      \includegraphics[width=1.1\textwidth]{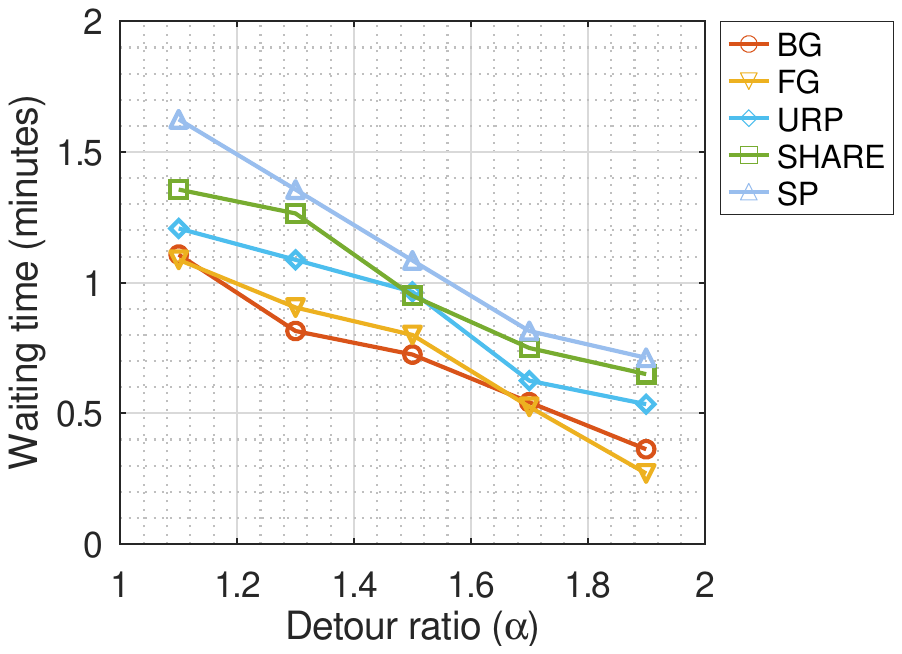}
      \vspace*{-25mm}
      \caption{Washington DC dataset}
      \label{img:wtdca}
    \end{subfigure}
    \vspace*{-1mm}
  \caption{Waiting time with increase in detour ratio ($\alpha$)}
  \label{img:wt}
  \end{minipage}
  \vspace{-3.5mm}
\end{figure*}

%%%%%%%%%%%%%%

%%%%%%%%%%%%%%%%%%%%%

%%%%%%%%%%%%%%%%%%%%%%%%%%%%%%%

\subsection{Results and Discussion}
In this subsection, we will analyze the performance of our proposed model %results
on the range of metrics, and the parameter specified above. 

\subsubsection{Impact of $k$}

The hop count $k$ determines the size of window used for recommending routes to drivers. Figure \ref{img:k} shows the performance of our proposed model on vehicle utilization, percentage of orders with ridesharing, passengers per grid, and waiting time with the increase in $k$. As can be seen through the figures, 
%These figures also display that 
New York City has higher values of vehicle utilization, passengers per grid, and percentage of orders with ridesharing, than Washington DC. This is because of the difference in  demand between the two cities. The higher demand in New York City leads to better utilization of vehicle capacity and is reflected in the metrics displayed in the figure. However, the higher demand leads to more waiting time for passengers in New York City than in Washington DC. When the demand is high, even with effective vehicle utilization all the passengers are not able to access the ride quickly which displays in their higher waiting times.  

These figures also display the improvement in the performance of the proposed model with the increase in the number of hops. %  the performance of our model is found to improve with the increase in the number of hops.
When the number of hops increases, the search space increases which implies that the origin-destination pairs of multiple passengers can be  matched effectively which leads to the improvement in the performance of the model. 
Moreover, it can be seen from the figure performance continues to improve with the increase in the number of hops in Washington DC, whereas the performance remains constant after the number of hops increases beyond $3$ in New York. This is because of the request arrival patterns in these cities. New York City has a higher request count which leads to good performance even with the low hop size, and this performance remains nearly about the same with the increase in hop count after $3$. Whereas Washington DC has fewer requests, and the performance continues to improve with the hop count, as the higher hop count will result in the pairing of more requests.
%These figures also display that New York City has higher values of vehicle utilization, passengers per grid, and percentage of orders with ridesharing, due to the higher number of requests there.
%the performance improves quickly with the number of hops on the Washington DC dataset as compared to New York.
%This is because of the grid granularity difference in the datasets. The count of grid cells is higher in New York dataset  which leads to the diverse spread of requests over them, whereas Washington DC contains only $63$ grid cells. With the few grid cells in Washington DC, even the smaller hops provide good performance, whereas it requires more hops  to converge in New York dataset. 
%and compare the performance with the existing baselines.'

\begin{figure*}
%\vspace*{-35mm}
    \begin{minipage}[b]{1\textwidth}
  %  \centering
  \vspace*{-23mm}
    \begin{subfigure}[b]{0.4\textwidth}
   % \vspace*{15mm}
    % \vspace*{10mm}
        \includegraphics[width=1.2\textwidth]{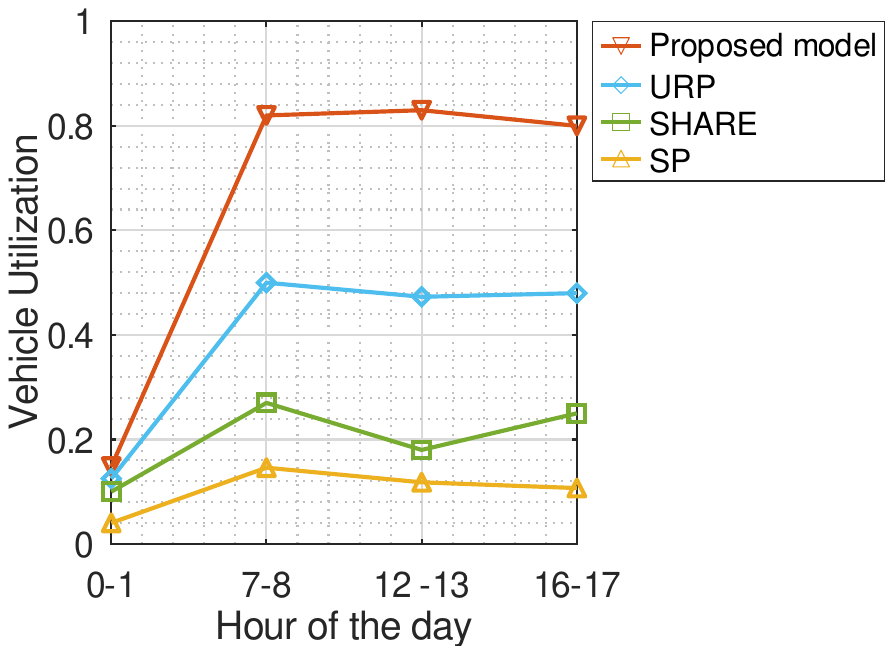}
         \vspace*{-18mm}
          %  \caption{Vehicle Utilization}
        \label{img:vunytime}
     %   \caption{Demand}
    \end{subfigure}
    \hfill
%\vspace{-20mm}
\begin{subfigure}[b]{0.4\textwidth}
%\vspace*{-15mm}
        \includegraphics[width=1.2\textwidth]{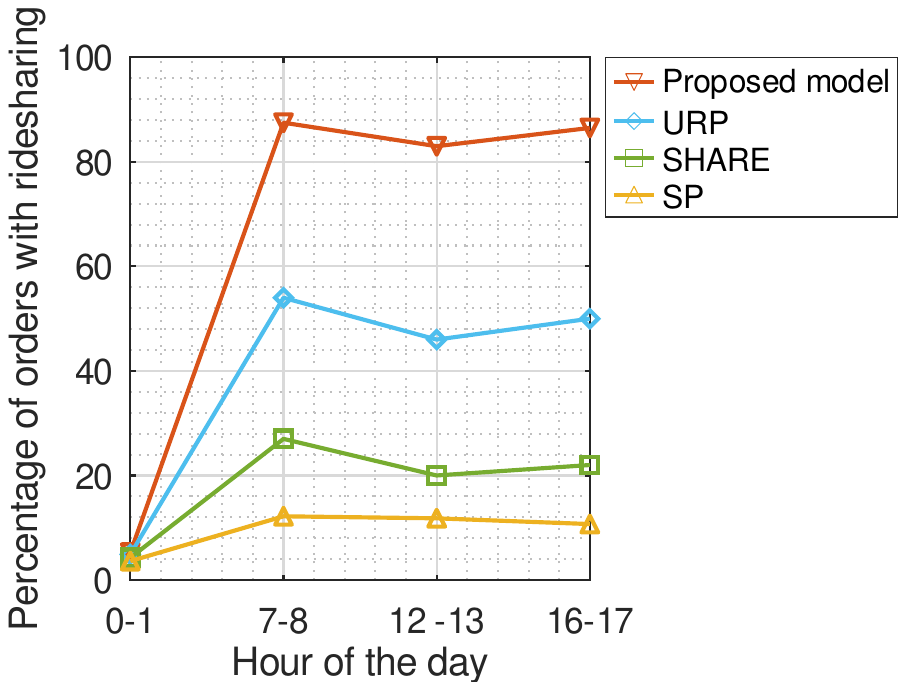}
            \vspace*{-18mm}
            %\caption{Percentage of orders with ridesharing}
       \label{img:ordersnytime}
    \end{subfigure}
%    \vspace{-30mm}
\end{minipage}  \hfill \vspace*{-33mm}
    \begin{minipage}[b]{1\textwidth}  
  \begin{subfigure}[b]{0.4\textwidth}
        \includegraphics[width=1.2\textwidth]{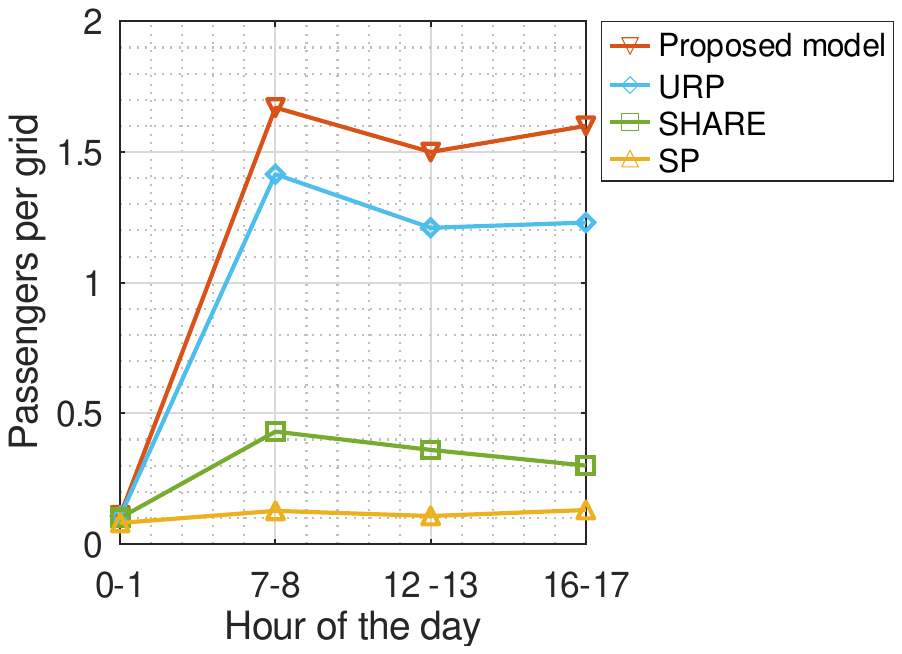}
         \vspace*{-18mm}
     %        \caption{Passengers per grid  }
        \label{img:ppgnytime}
    \end{subfigure}
    \vspace*{-7.5mm}
      \hfill
  \begin{subfigure}[b]{0.4\textwidth}
        \includegraphics[width=1.2\textwidth]{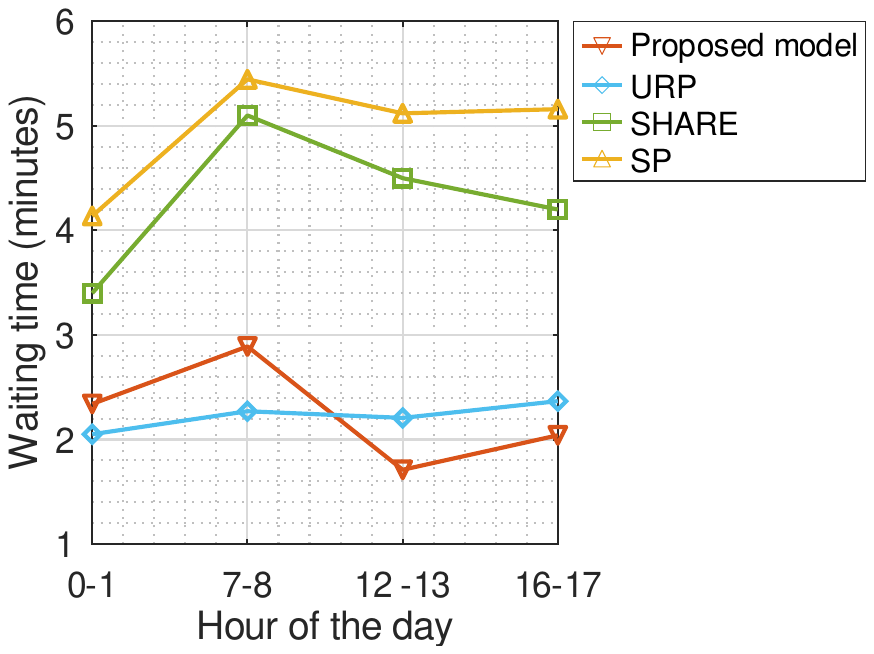}
         \vspace*{-18mm}
     %        \caption{Passengers per grid  }
        \label{img:wtnytime}
    \end{subfigure}
    \vspace*{-4.5mm}
     \caption{Improvement of the proposed model over the existing baselines over different time periods of the day on the New York dataset}
    \label{img:time_ny}
    \end{minipage}
  %  \vspace*{-5mm}
%\hfill
 %\hfill
\end{figure*}

\begin{figure*}
%\vspace*{-35mm}
    \begin{minipage}[b]{1\textwidth}
  %  \centering
  \vspace*{-23mm}
    \begin{subfigure}[b]{0.4\textwidth}
   % \vspace*{15mm}
    % \vspace*{10mm}
        \includegraphics[width=1.2\textwidth]{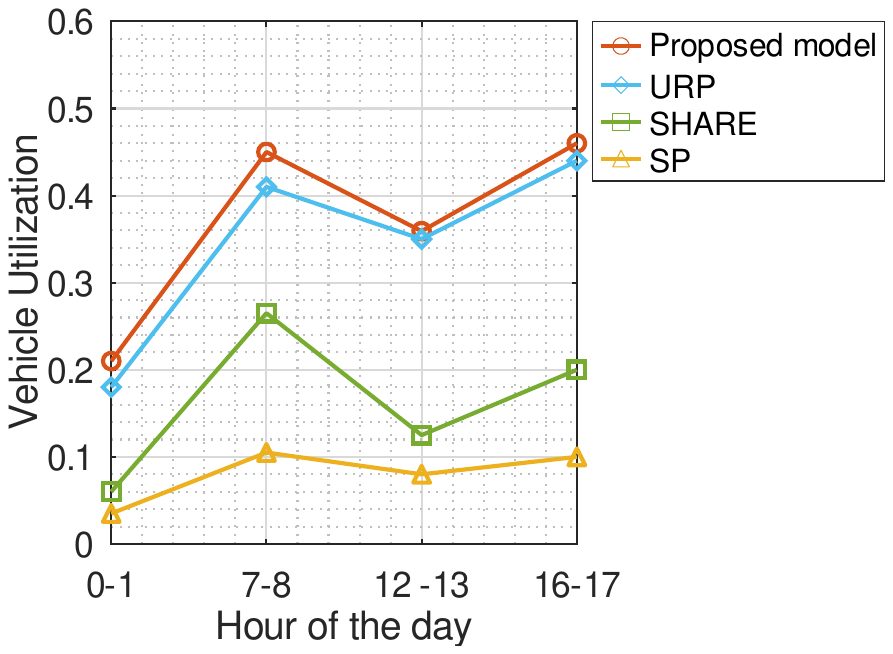}
         \vspace*{-18mm}
          %  \caption{Vehicle Utilization}
        \label{img:vudctime}
     %   \caption{Demand}
    \end{subfigure}
    \hfill
%\vspace{-20mm}
\begin{subfigure}[b]{0.4\textwidth}
%\vspace*{-15mm}
        \includegraphics[width=1.2\textwidth]{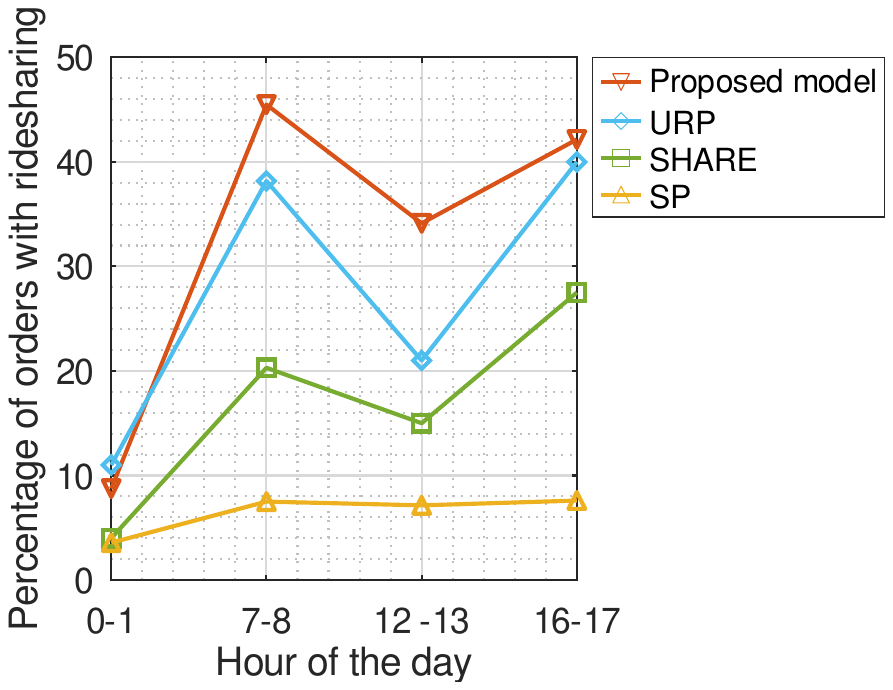}
            \vspace*{-18mm}
            %\caption{Percentage of orders with ridesharing}
       \label{img:ordersdctime}
    \end{subfigure}
%    \vspace{-30mm}
  \end{minipage}  \hfill
\vspace*{-33mm}
  \begin{minipage}[b]{1\textwidth}
   \begin{subfigure}[b]{0.4\textwidth}
        \includegraphics[width=1.2\textwidth]{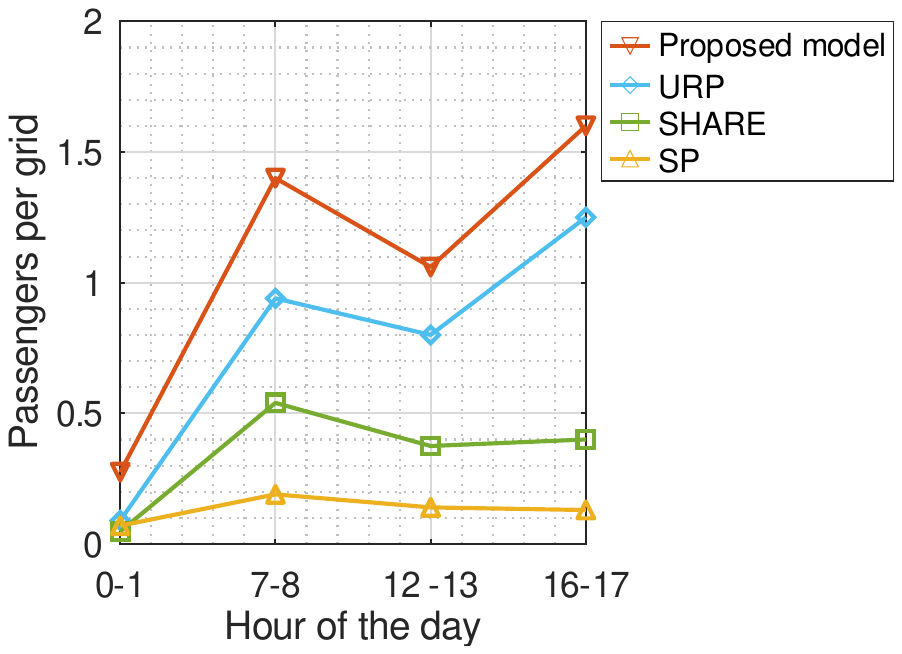}
         \vspace*{-18mm}
     %        \caption{Passengers per grid  }
        \label{img:ppgdctime}
    \end{subfigure}
    \vspace*{-7.5mm}
      \hfill
  \begin{subfigure}[b]{0.4\textwidth}
        \includegraphics[width=1.2\textwidth]{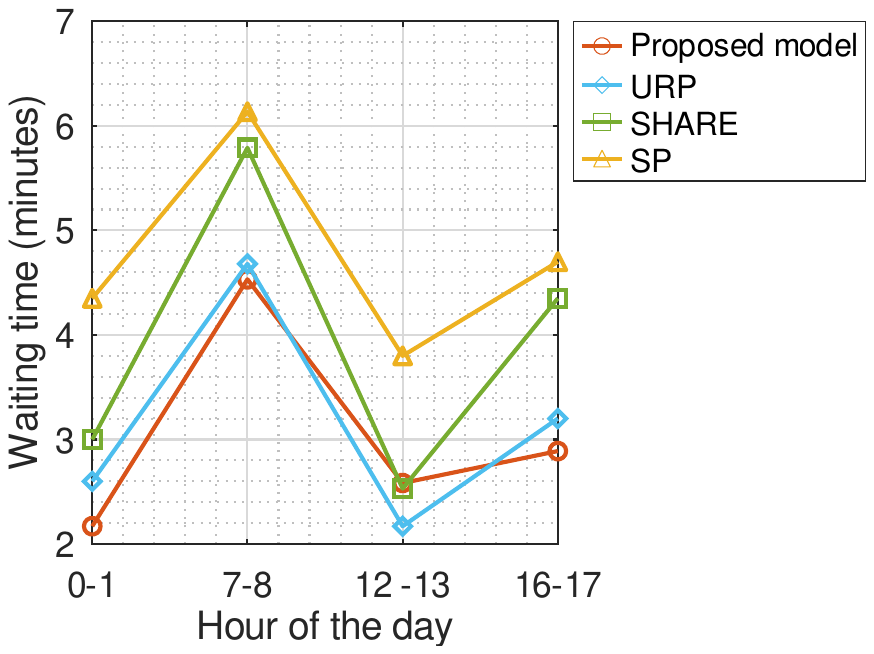}
         \vspace*{-18mm}
     %        \caption{Passengers per grid  }
        \label{img:wtdctime}
    \end{subfigure}
    \vspace*{-4.5mm}
     \caption{Improvement of the proposed model over the existing baselines over different time periods of the day on the Washington DC dataset}
    \label{img:time_dc}
    \end{minipage}
  %  \vspace*{-5mm}
%\hfill
 %\hfill
\end{figure*}

\subsubsection{Impact of detour ratio on the performance of the system and existing baselines}

The detour ratio determines the distance that can be travelled beyond the shortest path of passengers in the vehicle. %In this subsection,we determine the performance of our proposed model with an increase in the detour ratio ($\alpha$).  
Figures \ref{fig:vu}, \ref{img:orders}, \ref{img:ppg}, and \ref{img:wt} show the performance of Forward Greedy (FG) and Backward Greedy (BG) approaches based upon the detour ratio on the New York and Washington DC datasets. The performance improvement of Forward Greedy with an increase in the detour ratio is found to be similar to Backward Greedy on the New York dataset. However, on the Washington dataset, the Backward Greedy is found to perform better. Intuitively, both the Forward Greedy and Backward Greedy should perform similarly. %\textcolor{cyan}{may be changed in the next sentence} 
The superior performance of Backward Greedy on the Washington DC dataset can be attributed to the distribution pattern of requests which helps Backward Greedy  make use of the increase in search space more effectively than that of Forward Greedy. These figures also display that the performance of the proposed system continues to improve with the increase in detour ratio. This follows through intuitive reasoning, as the increase in detour ratio displays that the passengers are willing to travel higher distances beyond their shortest paths which results in searching for routes with higher expected passenger count more effectively and increases the vehicle utilization, passengers per grid, and percentage of orders with ridesharing. It also decreases the waiting time of passengers as effective vehicle utilization results in the reaching of drivers at the passenger areas quickly.
%interpreted due to the distribution pattern of requests which helps Forward Greedy  make use of the increase in search space more effectively than Backward Greedy. 

Apart from comparing the performance of Forward Greedy and Backward Greedy  approaches, Figures \ref{fig:vu}, \ref{img:orders}, \ref{img:ppg}, and \ref{img:wt}
also display the performance of the proposed model and the existing baselines: Unified Route Planning (URP) \cite{Tong:ACMTrans_2022}, SHARE \cite{Yuen:ACMWWW_2019}, and Shortest Path (SP).
The proposed model is found to perform better on all the evaluation metrics and surpasses the existing baselines. 
This is because  we have used the passenger data effectively and experimented with the origin and destination of requests. The proposed model's objective function after incorporating the origin and destination of requests comes out to be submodular on which greedy algorithms are known to provide well-known approximation guarantees, which results in improved performance by the proposed model. \looseness=-1%Our proposed model recommends the routes that have a higher count of expected requests. % origin and destination within the different sub-paths.
%Further, the proposed objective function is submodular which provides the greedy approach as an effective solution for solving the underlying route recommendation problem.

\begin{figure*}
\vspace*{-26mm}
    \begin{minipage}[b]{1\textwidth}
  %  \centering
%  \vspace*{-10mm}
    \begin{subfigure}[b]{0.4\textwidth}
   % \vspace*{15mm}
   \centering
     %\vspace*{-13mm}
        \includegraphics[width=1.2\textwidth]{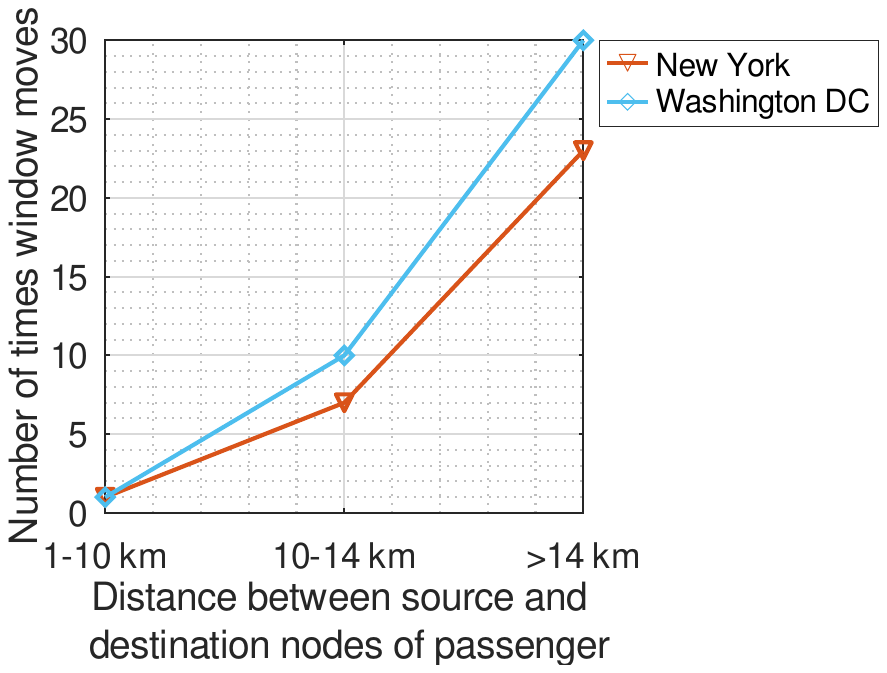}
         \vspace*{-29mm}
            \caption{}
        \label{img:window_distance}
     %   \caption{Demand}
    \end{subfigure}
    \hfill
%\vspace{-20mm}
\begin{subfigure}[b]{0.4\textwidth}
%\vspace*{35mm}
        \includegraphics[width=1.2\textwidth]{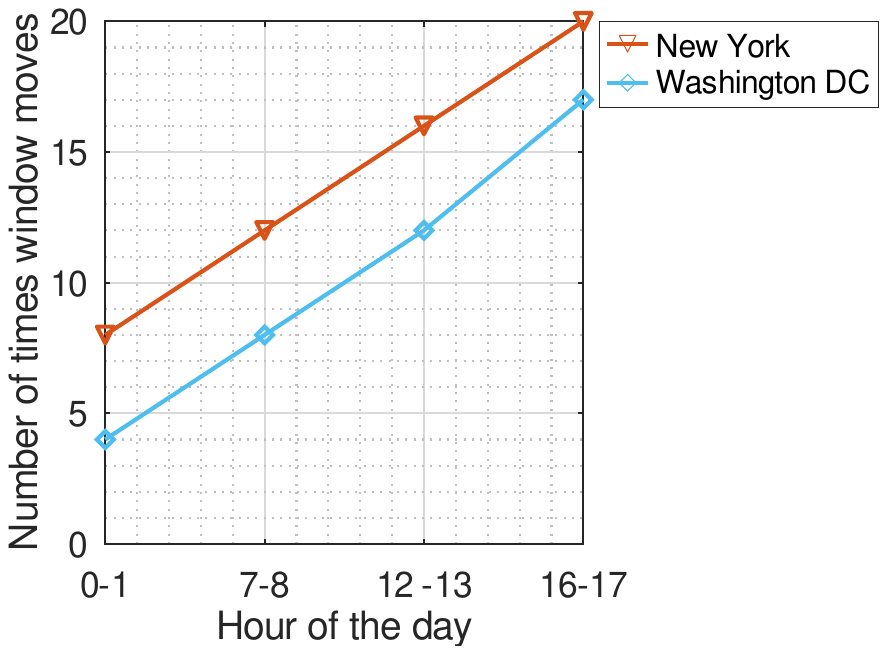}
            \vspace*{-29mm}
            \caption{}
       \label{img:window_time}
    \end{subfigure}
    %\vspace{-5mm}
  %  \caption{Window movements over different times and distances}
%    \vspace{-30mm}
%\end{minipage}
%\hfill  \vspace*{-13mm}
 %  \begin{minipage}[b]{0.22\textwidth}
 
    \end{minipage}
  %  \vspace*{-5mm}
%      \hfill \begin{minipage}[b]{0.22\textwidth}
\vspace{-6mm}
\caption{Window movements at (a) different distances  and (b) over different times }
%\end{minipage}
\label{fig:window_movements}
\vspace*{-3mm}
%\hfill
 %\hfill
\end{figure*}

\subsubsection{Time-frames}

Figures \ref{img:time_ny} and \ref{img:time_dc} display the performance of the proposed model and the existing baselines over different time periods of the day. The performance is evaluated over the night hour ($0-1$) when the demand is low, morning rush hour ($7-8$) when people are moving from their homes to offices, day hour ($12-13$) when the demand is scattered around different places, and the evening rush hour ($16-17$) when people are coming to their homes from respective offices. %The performance improvement of the proposed model is higher in New York City than in Washington DC which can be attributed to the passenger demand in New York City. When the passenger demand is high the origin destination prediction of higher quantity of requests is high which leads to effective functioning of the system. 
During the morning and evening rush hours, the proposed model and existing baselines perform well for all the parameters. However, the proposed models' performance is superior due to the prediction of origin and destination of requests which results in the effective functioning of the system. The performance improvement of the proposed model is particularly high during the morning rush hours than that of the evening rush hour. %In order to analyze this fact, we estimated the demand during those hours. During the morning hour there was a demand of %This is primarily because during the morning hours, the proposed model predicts the flow of requests is from residential areas to office areas, so it recommends routes where multiple passengers from residential areas have a portion of the route in common which results in effective vehicle utilization, higher passengers per grid, and higher orders with ridesharing, whereas during evening rush hours people move from offices to residential areas which are scattered around the city and the routes are relatively less common. Moreover, this functionality can also be due to the route preferences of passengers. 
The analysis behind this functioning can be that during the morning hours, people prefer speed and efficiency and try to reach their offices as quickly as possible, which results in effective vehicle utilization, whereas during the evening hours, people prefer convenience and can follow the preferred routes which results in less utilization than the morning rush hour. The values of different metrics are relatively low during the night hour due to the low demand over those time periods. They are also slightly lower in the day hours which can be attributed to the demand scattered over different places during these hours. Overall, the proposed model exhibits superior performance during different time frames of the day with varying demands which shows its adaptability in different environments.  %which can be slightly longer than the shortest route and . %The reason behind this functioning of the model is during the morning hours people travel from residential areas to their offices. The origins of different users are scattered over different places, even though their destination might be at the same place. Their time of leaving might be different, whereas during the evening hours, everyone leaves at nearly the same time, and even if the destination of requests is different the part of routes among them will be more common which results in improved values of these parameters.  Moreover, the improvement is higher in New York City than in Washington DC which can be attributed to the higher demand there. 
%The performance improvement of the proposed model is higher d

\subsubsection{Window movement}

The proposed model constrains the search space within which the route is recommended by creating a window around the source node and slides this window until the destination node is reached.
The number of times window is slid is an important factor that determines the complexity of the proposed approach. Theoretically, the movements are of the order of $n$, where $n$ denotes the number of grid cells. 
In this section, we display through experiments the movement of windows over different time frames and with different distances. % depends upon two factors: the demand at that point in time, and the distance between the source and destination points of the passenger. %Figure \ref{} shows the window movement %at different hours of the day.
%at different distances. When the distance is low, the window movement is very less 
Figure \ref{img:window_distance} displays the movement of the window over different distances (distance between source and destination of passengers) in New York and Washington DC datasets. The default value of $k$ is kept as $5$. When the distance between source and destination points is less than $8\,km$, i.e., the passenger has to travel within $8\, km$,  these points lie within the same window, and in this case window is not slid forward or backward. When the distance between source and destination points increases the window movement increases. %This is intuitive that the increase is distance will lead to more window movement %, as the destination of passengers is far away from their source, and the window . %is between $8$ and $15\,km$ the window is slid about times in New York and Washington DC respectively. This is intuitive as the increase in distance  between source and destination points leads to the higher movement of the window. 
With the same distance, the window movement is more in Washington DC than in New York. This is because of the difference in demand arrival patterns in these two cities. In New York City, the requests are higher and the routes with the highest expected requests are found within the window frequently which results in fewer movements of the window whereas in Washington DC the requests are scattered around different places which shows up in the higher movement of the window. Apart from distance, the window movement also depends upon the time of the day. Figure \ref{img:window_time} shows the movement of the window over different times of the day. During the night the demand is less, which results in the less movement of the window. The demand increases during the morning hours which results in a higher movement of window in order to cover the demand at different places. Its movement further increases during the day hours. The reason behind this is attributed to the demand scattered around different places which results in the frequent movement of the window in order to cover the demand over different places. %The window movement is highest during the evening hours, which is mainly due to fact that
Although the demand is near about the same in the morning and evening rush hours, the window movement is much higher in the evening rush hours than that of morning hours. The analysis behind this functioning of model is the difference in travelling behavior of people. During morning rush hours people prioritize speed over convenience and try to reach their destination quickly. Whereas during evening hours, people usually prefer to follow the desired routes which results in the higher movement of the window. %patterns during these hours %. In morning hours people are moving from homes to offices. The office of different passengers are similar so the routes with multiple passengers get recommended quickly, which results in less movement of window, whereas during evening hours demand arises at offices and people are travelling to their respective homes which are scattered around different places so the window movement is more during those hours.
Moreover, the figure displays that window movement is higher in New York than in Washington DC. This is mainly because the area covered by the New York dataset is higher than the Washington DC. There are $576$ grid cells in New York, whereas Washington DC contains only $99$ grid cells. The higher area leads to higher movement of window in New York than in Washington DC. With similar distances, the window movement was higher in Washington DC due to demand spread over the grid cells as was displayed through Figure \ref{img:window_distance}. However, due to the higher area, the movement of window is more frequent in New York than in Washington DC. %Through the analysis of window movement we can say the proposed model %It also depends upon the demand patterns in New York. 

\looseness=-1 

\subsubsection{Approximation ratio}

\begin{figure}[t!]
\vspace*{-29mm}
    \centering
        \includegraphics[width=0.48\textwidth]{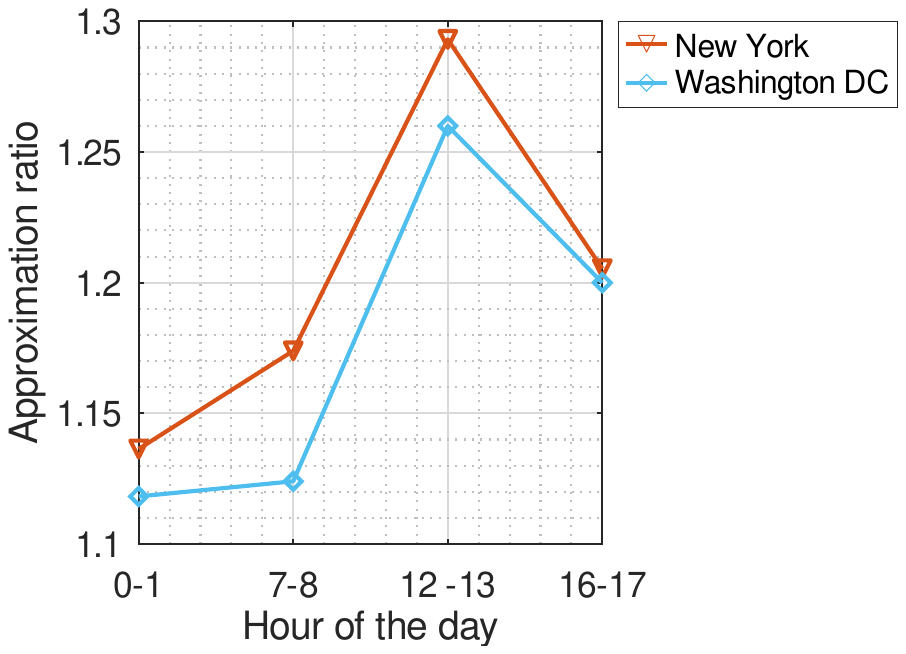}
    \vspace*{-28mm}
    \caption{Approximation ratio at different times of the day}
    \label{img:approxratio}
    \vspace*{-4.5mm}

\end{figure}

The proposed model approximates the highest expected request path by using the sliding window based greedy algorithm which is submodular. To determine the approximation quality, we plot the approximation ratio over different times of the day. The approximation ratio is quantified as $\frac{vu_{prop}}{vu^*}$, where $vu_{prop}$ %e_{prop}$ denotes the
denotes the vehicle utilization of the proposed model, and $vu^*$ denotes the optimal vehicle utilization. Since computing $vu^*$ is NP-Hard, we perform the computation on a smaller area of $6*6\,km$ on both datasets. Figure \ref{img:approxratio} shows the approximation ratio over different times of the day.
It can be seen that the proposed model works effectively during different times of the day and its approximation value lies within $1.3$ times the optimal value. The approximation value is high during the day hours ($12-13)$ which can be attributed to the demand scattered over different grid cells. Over the other times, the proposed model performs well which is displayed through the approximation ratio being close to $1$ over those times. %The approximation ratio is particy
%The approximation ratio is close to the optimal value during the night hours of $0-1$ since the demand is low during those times, and the vehicle is not effectively utilized during those hours by the proposed model and the optimal model. The approximation ratio comes close to $1$ during the morning and evening rush hours when the demand is very high and the proposed model effectively utilizes the vehicle during those hours which shows in its value being close to optimal value. The value is relatively low during the day hours when requests are scattered around different places. However, even in this case, it lies within $1.5$ times the optimal value. Through these cases, we can say the proposed model works effectively during different times of the day with varying demands.  %, and it comes out high during the morning and evening rush hours when demand is high. This value remains within $1.5$ times the optimal value.

\subsubsection{Fleet Size estimation}

%\vspace*{-1mm}

\begin{figure*}[t!]
  %\vspace*{-18mm}
    \begin{minipage}[b]{1\textwidth}
    \vspace*{-30mm}
    \centering
    \begin{subfigure}[b]{0.45\textwidth}
% \vspace*{-5mm}      
\includegraphics[width=1\textwidth]{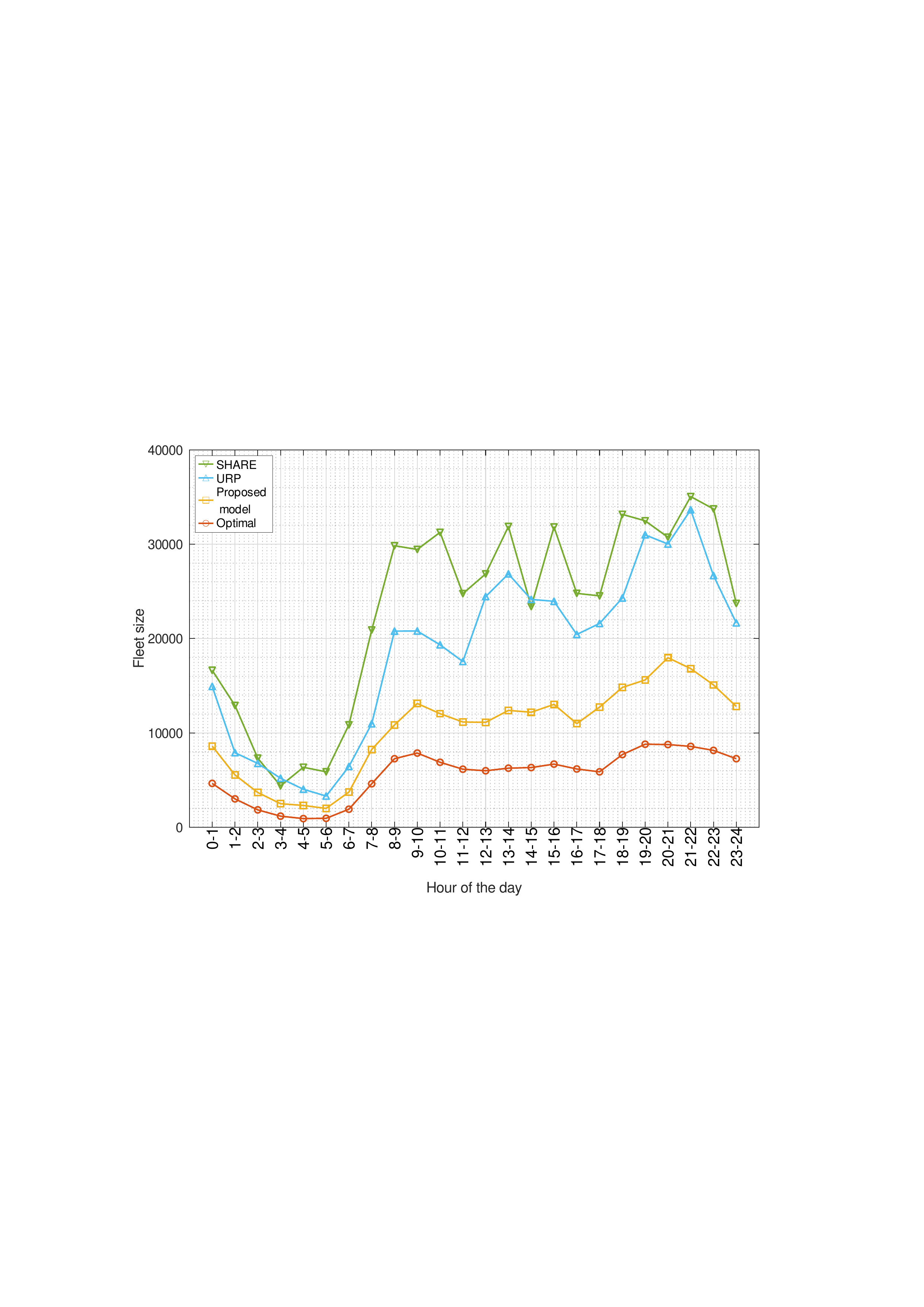}
      \vspace*{-32mm}
      \caption{New York dataset}
      \label{img:fleet_ny}
    \end{subfigure}
    \hfill
    \begin{subfigure}[b]{0.45\textwidth}
      \includegraphics[width=1\textwidth]{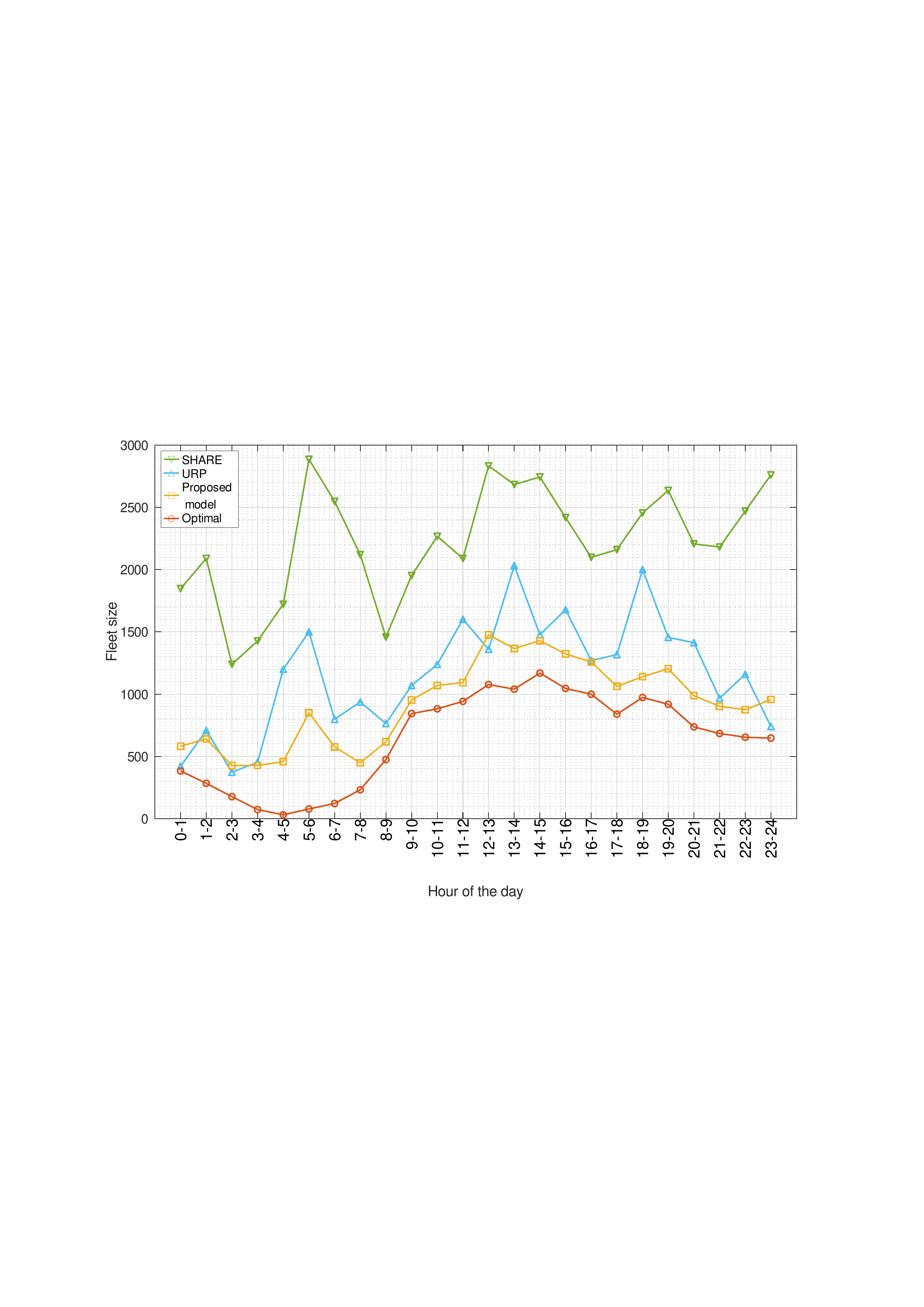}
    %  \vspace*{-2mm}
      \vspace*{-32mm}
      \caption{Washington DC dataset}
      \label{img:fleet_dc}
    \end{subfigure} %\vspace*{-3.5mm} 
    \caption{Fleet size  at different times of the day}
  \label{img:fleetsize}
  \end{minipage}
   %\hfill
\end{figure*}

\begin{figure*}
\begin{minipage}[b]{1\textwidth}
    \vspace*{-29mm}
    \centering
    \begin{subfigure}[b]{0.45\textwidth}
      \includegraphics[width=1.1\textwidth]{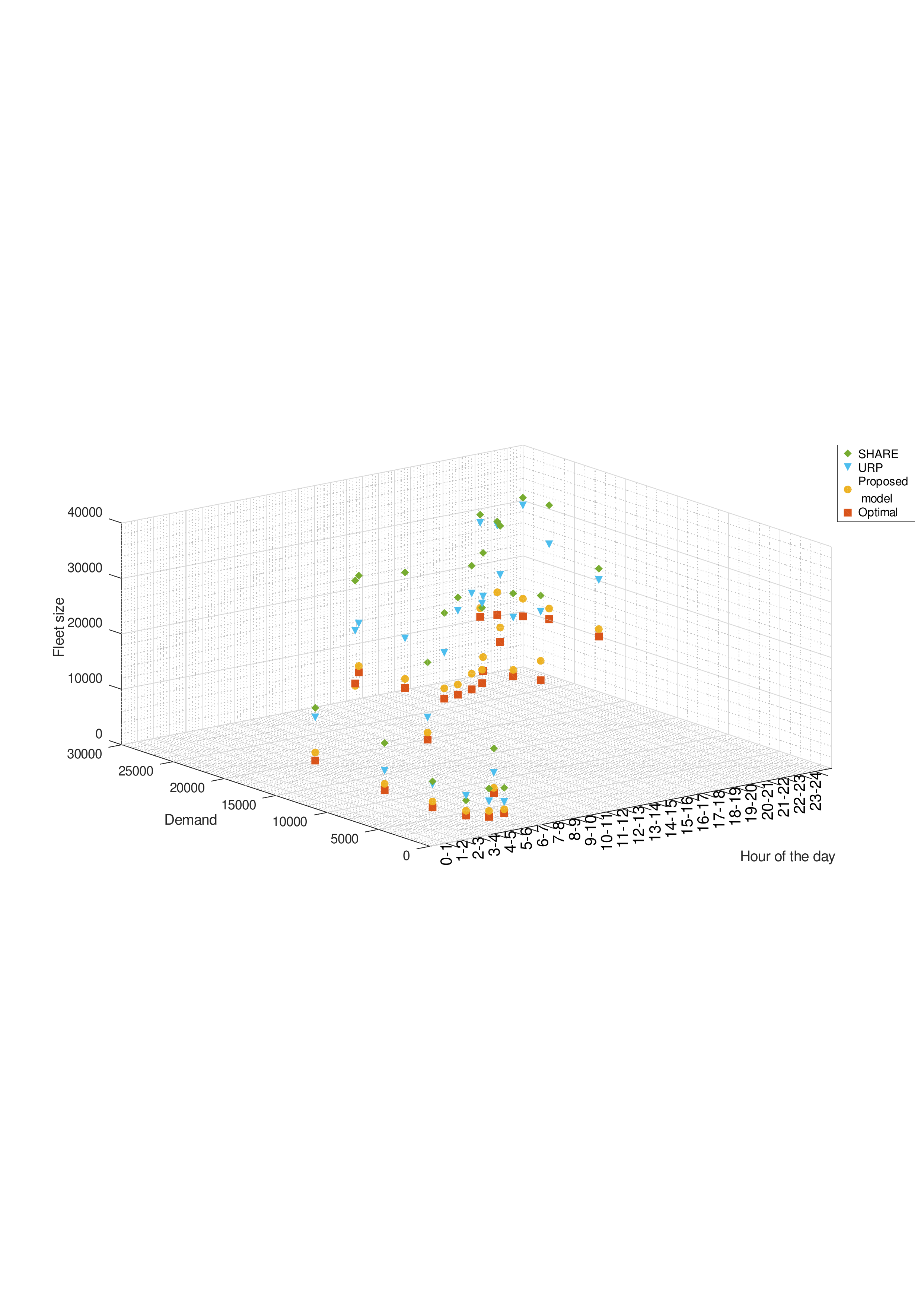}
      \vspace*{-32mm}
      \caption{New York dataset}
      \label{img:fleet_demand_ny}
    \end{subfigure}
    \hfill
    \begin{subfigure}[b]{0.45\textwidth}
      \includegraphics[width=1.1\textwidth]{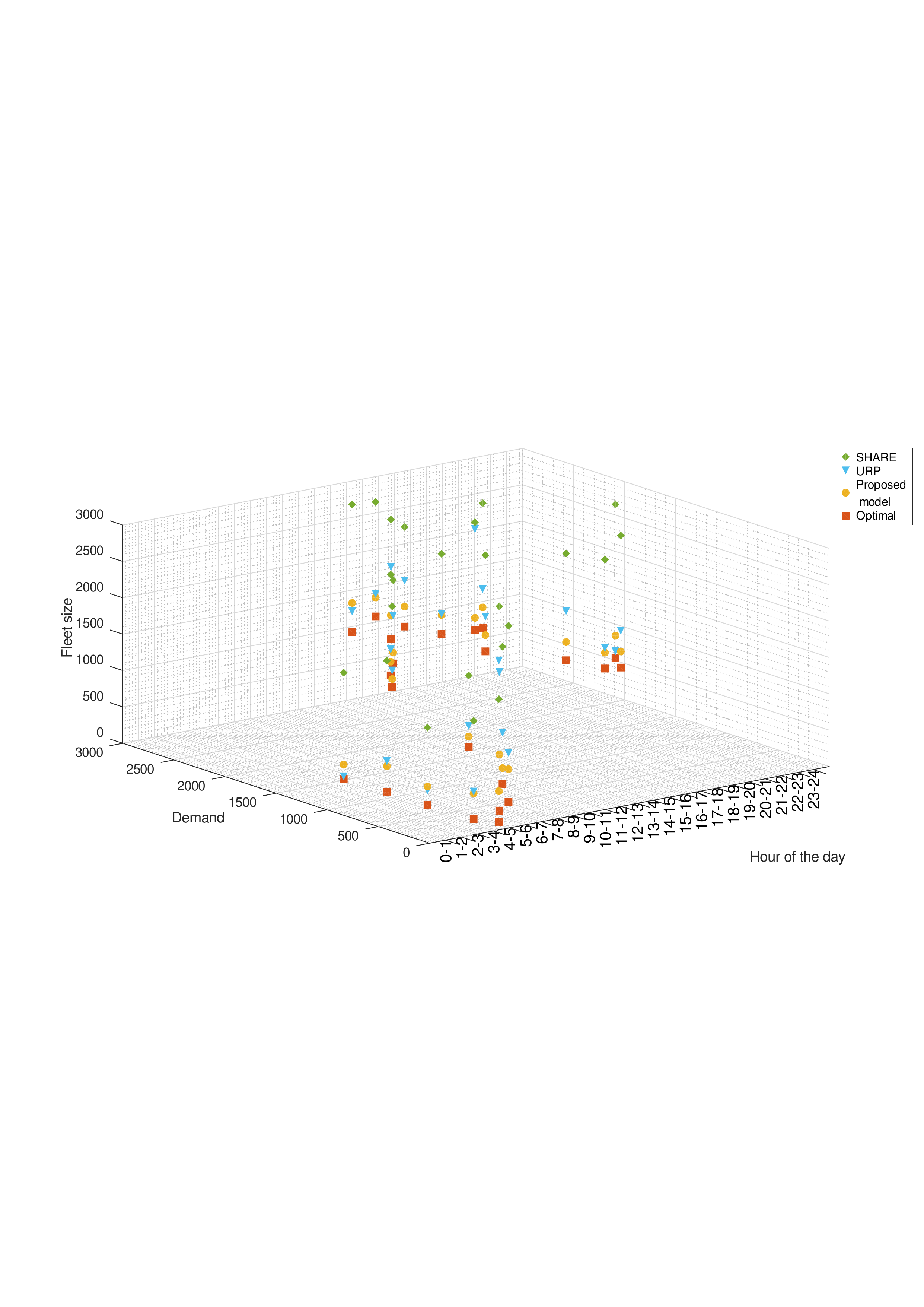}
      \vspace*{-32mm}
   %   \vspace*{-10mm}
      \caption{Washington DC dataset}
      \label{img:fleet_demand_dc}
    \end{subfigure} \vspace*{-1mm} \caption{Evaluation of fleet size over different times of the day with different demands}
  \label{img:fleetsize}
  \end{minipage}

  \vspace{-2mm}
\end{figure*}

The fleet size determines the number of vehicles required by
ridesharing companies to service passenger requests over different
time periods of the day.
%To determine the fleet size for our proposed model, we analyze the number of vehicles required to service passenger requests during different hours of the day. 
Figures \ref{img:fleet_ny} and  \ref{img:fleet_dc} show the fleet size of
the proposed model considering the backward greedy approach,
existing baselines, and the optimal fleet size over different hours
of the day in New York City and Washington DC. %show the fleet size of the proposed model, existing baselines and the optimal fleet size over different time periods of the day in New York City and Washington DC. %The vehicle capacity is assumed to be $2$.
These figures illustrate the dynamic nature of fleet size throughout the day. Notably, during late-night hours (from $2$  to $7$), the fleet size remains relatively low, while it experiences an increase during the morning rush hours (from $7$  to $11$) and the evening rush hours (from $16$ to $20$).
This variation in fleet size is directly influenced by the fluctuating demand throughout the day. During late-night hours, the demand decreases as most people have already reached their respective homes. Conversely, the demand surges during the morning and evening rush hours as people commute from their homes to offices in the morning and vice versa in the evening.
This can be verified through Figures \ref{img:fleet_demand_ny} and \ref{img:fleet_demand_dc}, which plot the number of vehicles required over different times of the day with the demand over different time frames.  For instance, the data point $(0-1, 5000, 3000)$ signifies that between $12\,am$  and $1\,am$, there were $5000$ passenger requests, and our proposed model utilized $3000$ vehicles to service those requests. 
This graph displays that the demand is different over different time periods of the day and the vehicle count changes accordingly.

 Through these figures, it can be seen that the proposed model uses fewer vehicles in comparison to the existing baselines. This is primarily due to the reason that the proposed model effectively utilizes the available vehicle capacity which results in a decrease in the count of vehicles on the road.  
% This results in the superior performance of the proposed model in comparison to the existing baselines. 

 \begin{figure}[h!]
  \vspace*{-29mm}
  \centering
        \includegraphics[width=0.6\textwidth]{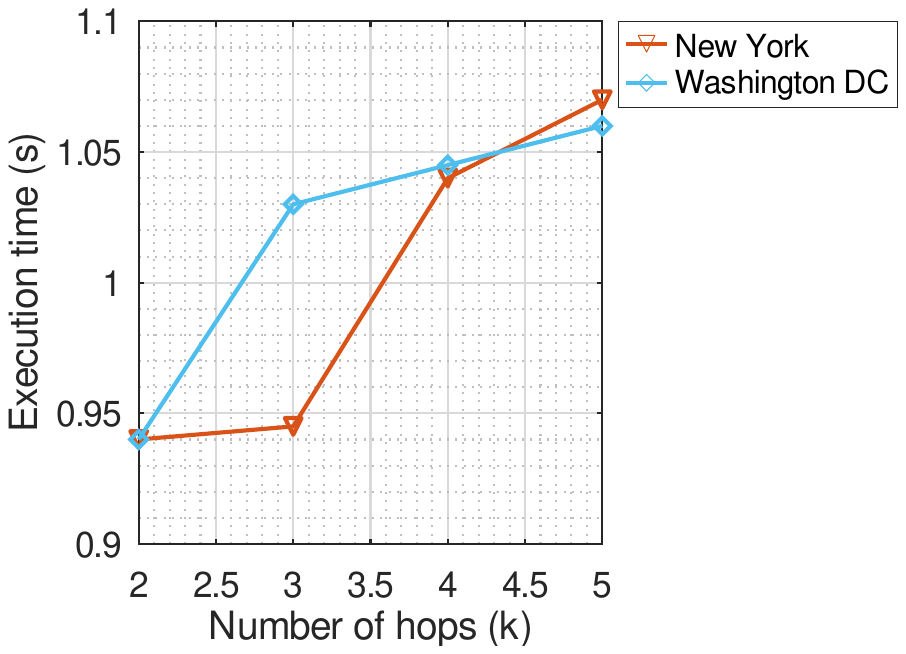}
         \vspace*{-32mm}
    \caption{Scalability of the proposed model}
         \label{img:exec_time}
         \vspace*{-4mm}
    \end{figure}

\subsubsection{Discussion}

After examining the performance of the proposed model over different detours, time frames, and examining the other aspects of the system like the window movements we can conclude that our proposed model is: \looseness=-1

\textbf{Eco-friendly and Cost-efficient:} As the proposed model is found to perform better on all the evaluation metrics in comparison to the existing baselines, it displays that the drivers are provided with a route that has higher passenger density which makes them cruise for passengers less and utilize the vehicle in an efficient manner. This results in eco-friendly rides and it also increases the profit of drivers as they are provided with the passengers earlier on the road. \looseness=-1

\textbf{Reduces fleet size: } Since the proposed model recommends the routes that have higher count of passengers, it results in the efficient utilization of vehicle and decreases the number of vehicles on the road.
    
\textbf{Scalable:}
%\textcolor{cyan}{Scalability point reframed}
As can be seen through Figure \ref{img:exec_time},  our proposed model responds to any query in nearly about $1.1$ seconds which makes it useful for real-time recommendation.

\looseness=-1 

%\vspace*{-4mm}
\section{Conclusion}
In this paper, we establish that it is possible to use the origin-destination request patterns for route recommendation systems. The proposed approach does this by applying a greedy-based algorithm to maximize the underlying objective function. Further, it overcomes the scalability challenges posed by the NP-Hard nature of these problems, by reducing the search space to a $ k$-hop-based sliding window. Through this novel approach, the proposed model achieves better performance than the state-of-the-art techniques and reduces the fleet size on the real network taxi datasets generated from New York City and Washington DC. Moreover, it overcomes the computational complexity and runs the simulations in ~$1.1$ seconds which implies the model can be run in real time.

\bibliographystyle{ACM-Reference-Format}
\bibliography{sample-base}


\begin{thebibliography}{39}

%%% the Web address.

\ifx \showCODEN    \undefined \def \showCODEN     #1{\unskip}     \fi
\ifx \showDOI      \undefined \def \showDOI       #1{#1}\fi
\ifx \showISBNx    \undefined \def \showISBNx     #1{\unskip}     \fi
\ifx \showISBNxiii \undefined \def \showISBNxiii  #1{\unskip}     \fi
\ifx \showISSN     \undefined \def \showISSN      #1{\unskip}     \fi
\ifx \showLCCN     \undefined \def \showLCCN      #1{\unskip}     \fi
\ifx \shownote     \undefined \def \shownote      #1{#1}          \fi
\ifx \showarticletitle \undefined \def \showarticletitle #1{#1}   \fi
\ifx \showURL      \undefined \def \showURL       {\relax}        \fi
% invisible to TeX
\providecommand\bibfield[2]{#2}
\providecommand\bibinfo[2]{#2}
\providecommand\natexlab[1]{#1}
\providecommand\showeprint[2][]{arXiv:#2}

\bibitem[Gur(2018)]%
        {Gurumurthy:Elsevier_2018}
 \bibinfo{year}{2018}\natexlab{}.
\newblock \showarticletitle{Analyzing the dynamic ride-sharing potential for
  shared autonomous vehicle fleets using cellphone data from Orlando, Florida}.
\newblock \bibinfo{journal}{\emph{Computers, Environment and Urban Systems}}
  \bibinfo{volume}{71} (\bibinfo{year}{2018}), \bibinfo{pages}{177--185}.
\newblock
\showISSN{0198-9715}


\bibitem[Vos(2019)]%
        {Vosooghi:Elsevier_2019}
 \bibinfo{year}{2019}\natexlab{}.
\newblock \showarticletitle{Shared autonomous vehicle simulation and service
  design}.
\newblock \bibinfo{journal}{\emph{Transportation Research Part C: Emerging
  Technologies}}  \bibinfo{volume}{107} (\bibinfo{year}{2019}),
  \bibinfo{pages}{15--33}.
\newblock
\showISSN{0968-090X}


\bibitem[Alonso-Mora et~al\mbox{.}(2017)]%
        {Alonso:NAS_2017}
\bibfield{author}{\bibinfo{person}{Javier Alonso-Mora},
  \bibinfo{person}{Samitha Samaranayake}, \bibinfo{person}{Alex Wallar},
  \bibinfo{person}{Emilio Frazzoli}, {and} \bibinfo{person}{Daniela Rus}.}
  \bibinfo{year}{2017}\natexlab{}.
\newblock \showarticletitle{On-demand high-capacity ride-sharing via dynamic
  trip-vehicle assignment}.
\newblock \bibinfo{journal}{\emph{Proceedings of the National Academy of
  Sciences}} \bibinfo{volume}{114}, \bibinfo{number}{3} (\bibinfo{year}{2017}),
  \bibinfo{pages}{462--467}.
\newblock


\bibitem[Cai et~al\mbox{.}(2019)]%
        {Cai:ElsevierEnergy_2019}
\bibfield{author}{\bibinfo{person}{Hua Cai}, \bibinfo{person}{Xi Wang},
  \bibinfo{person}{Peter Adriaens}, {and} \bibinfo{person}{Ming Xu}.}
  \bibinfo{year}{2019}\natexlab{}.
\newblock \showarticletitle{Environmental benefits of taxi ride sharing in
  Beijing}.
\newblock \bibinfo{journal}{\emph{Energy}}  \bibinfo{volume}{174}
  (\bibinfo{year}{2019}), \bibinfo{pages}{503--508}.
\newblock
\showISSN{0360-5442}


\bibitem[C{\'a}p and Alonso~Mora(2018)]%
        {Cap:RSS_2018}
\bibfield{author}{\bibinfo{person}{Michal C{\'a}p} {and}
  \bibinfo{person}{Javier Alonso~Mora}.} \bibinfo{year}{2018}\natexlab{}.
\newblock \showarticletitle{Multi-objective analysis of ridesharing in
  automated mobility-on-demand}.
\newblock \bibinfo{journal}{\emph{Proceedings of RSS 2018: Robotics-Science and
  Systems XIV}} (\bibinfo{year}{2018}).
\newblock


\bibitem[Cheng et~al\mbox{.}(2017)]%
        {Cheng:ACMSIGMOD_2017}
\bibfield{author}{\bibinfo{person}{Peng Cheng}, \bibinfo{person}{Hao Xin},
  {and} \bibinfo{person}{Lei Chen}.} \bibinfo{year}{2017}\natexlab{}.
\newblock \showarticletitle{Utility-Aware Ridesharing on Road Networks}. In
  \bibinfo{booktitle}{\emph{Proceedings of the 2017 ACM International
  Conference on Management of Data}} (Chicago, Illinois, USA)
  \emph{(\bibinfo{series}{SIGMOD '17})}. \bibinfo{publisher}{Association for
  Computing Machinery}, \bibinfo{address}{New York, NY, USA},
  \bibinfo{pages}{1197–1210}.
\newblock
\showISBNx{9781450341974}


\bibitem[Conforti and Cornuéjols(1984)]%
        {Conforti:ElsevAppliedmaths_1984}
\bibfield{author}{\bibinfo{person}{Michele Conforti} {and}
  \bibinfo{person}{Gérard Cornuéjols}.} \bibinfo{year}{1984}\natexlab{}.
\newblock \showarticletitle{Submodular set functions, matroids and the greedy
  algorithm: Tight worst-case bounds and some generalizations of the
  Rado-Edmonds theorem}.
\newblock \bibinfo{journal}{\emph{Discrete Applied Mathematics}}
  \bibinfo{volume}{7}, \bibinfo{number}{3} (\bibinfo{year}{1984}),
  \bibinfo{pages}{251--274}.
\newblock
\showISSN{0166-218X}


\bibitem[Cormen et~al\mbox{.}(2022)]%
        {Cormen:book_2022}
\bibfield{author}{\bibinfo{person}{Thomas~H Cormen}, \bibinfo{person}{Charles~E
  Leiserson}, \bibinfo{person}{Ronald~L Rivest}, {and}
  \bibinfo{person}{Clifford Stein}.} \bibinfo{year}{2022}\natexlab{}.
\newblock \bibinfo{booktitle}{\emph{Introduction to algorithms}}.
\newblock \bibinfo{publisher}{MIT press}.
\newblock


\bibitem[Ding et~al\mbox{.}(2013)]%
        {Ding:IEEConf_2013}
\bibfield{author}{\bibinfo{person}{Ye Ding}, \bibinfo{person}{Siyuan Liu},
  \bibinfo{person}{Jiansu Pu}, {and} \bibinfo{person}{Lionel~M. Ni}.}
  \bibinfo{year}{2013}\natexlab{}.
\newblock \showarticletitle{HUNTS: A Trajectory Recommendation System for
  Effective and Efficient Hunting of Taxi Passengers}. In
  \bibinfo{booktitle}{\emph{2013 IEEE 14th International Conference on Mobile
  Data Management}}, Vol.~\bibinfo{volume}{1}. \bibinfo{pages}{107--116}.
\newblock
\showISSN{2375-0324}


\bibitem[Fagnant and Kockelman(2018)]%
        {Fagnant:SpringerTranportation_2018}
\bibfield{author}{\bibinfo{person}{Daniel~J Fagnant} {and}
  \bibinfo{person}{Kara~M Kockelman}.} \bibinfo{year}{2018}\natexlab{}.
\newblock \showarticletitle{Dynamic ride-sharing and fleet sizing for a system
  of shared autonomous vehicles in Austin, Texas}.
\newblock \bibinfo{journal}{\emph{Transportation}}  \bibinfo{volume}{45}
  (\bibinfo{year}{2018}), \bibinfo{pages}{143--158}.
\newblock


\bibitem[Garg and Ranu(2018)]%
        {Garg:ACMKDD_2018}
\bibfield{author}{\bibinfo{person}{Nandani Garg} {and} \bibinfo{person}{Sayan
  Ranu}.} \bibinfo{year}{2018}\natexlab{}.
\newblock \showarticletitle{Route Recommendations for Idle Taxi Drivers: Find
  Me the Shortest Route to a Customer!}. In
  \bibinfo{booktitle}{\emph{Proceedings of the 24th ACM SIGKDD International
  Conference on Knowledge Discovery \& Data Mining}}
  \emph{(\bibinfo{series}{KDD '18})}. \bibinfo{publisher}{Association for
  Computing Machinery}, \bibinfo{address}{New York, NY, USA},
  \bibinfo{pages}{1425–1434}.
\newblock
\showISBNx{9781450355520}


\bibitem[Guo et~al\mbox{.}(2021)]%
        {Guo:ACM_Trans_2021}
\bibfield{author}{\bibinfo{person}{Pengzhan Guo}, \bibinfo{person}{Keli Xiao},
  \bibinfo{person}{Zeyang Ye}, {and} \bibinfo{person}{Wei Zhu}.}
  \bibinfo{year}{2021}\natexlab{}.
\newblock \showarticletitle{Route Optimization via Environment-Aware Deep
  Network and Reinforcement Learning}.
\newblock \bibinfo{journal}{\emph{ACM Trans. Intell. Syst. Technol.}}
  \bibinfo{volume}{12}, \bibinfo{number}{6}, Article \bibinfo{articleno}{74}
  (\bibinfo{date}{dec} \bibinfo{year}{2021}), \bibinfo{numpages}{21}~pages.
\newblock
\showISSN{2157-6904}


\bibitem[Ji et~al\mbox{.}(2020)]%
        {Ji:ElsevierKnowledge_2020}
\bibfield{author}{\bibinfo{person}{Shenggong Ji}, \bibinfo{person}{Zhaoyuan
  Wang}, \bibinfo{person}{Tianrui Li}, {and} \bibinfo{person}{Yu Zheng}.}
  \bibinfo{year}{2020}\natexlab{}.
\newblock \showarticletitle{Spatio-temporal feature fusion for dynamic taxi
  route recommendation via deep reinforcement learning}.
\newblock \bibinfo{journal}{\emph{Knowledge-Based Systems}}
  \bibinfo{volume}{205} (\bibinfo{year}{2020}), \bibinfo{pages}{106302}.
\newblock
\showISSN{0950-7051}


\bibitem[Jiang et~al\mbox{.}(2018)]%
        {Jiang:Springer_2018}
\bibfield{author}{\bibinfo{person}{Xiaoting Jiang}, \bibinfo{person}{Yanyan
  Shen}, {and} \bibinfo{person}{Yanmin Zhu}.} \bibinfo{year}{2018}\natexlab{}.
\newblock \showarticletitle{Cruising or Waiting: A Shared Recommender System
  for Taxi Drivers}. In \bibinfo{booktitle}{\emph{Advances in Knowledge
  Discovery and Data Mining}}, \bibfield{editor}{\bibinfo{person}{Dinh Phung},
  \bibinfo{person}{Vincent~S. Tseng}, \bibinfo{person}{Geoffrey~I. Webb},
  \bibinfo{person}{Bao Ho}, \bibinfo{person}{Mohadeseh Ganji}, {and}
  \bibinfo{person}{Lida Rashidi}} (Eds.). \bibinfo{publisher}{Springer
  International Publishing}, \bibinfo{address}{Cham},
  \bibinfo{pages}{418--430}.
\newblock
\showISBNx{978-3-319-93037-4}


\bibitem[Li et~al\mbox{.}(2020)]%
        {Li:ElsevierTransport_2020}
\bibfield{author}{\bibinfo{person}{Yuanyuan Li}, \bibinfo{person}{Yang Liu},
  {and} \bibinfo{person}{Jun Xie}.} \bibinfo{year}{2020}\natexlab{}.
\newblock \showarticletitle{A path-based equilibrium model for ridesharing
  matching}.
\newblock \bibinfo{journal}{\emph{Transportation Research Part B:
  Methodological}}  \bibinfo{volume}{138} (\bibinfo{year}{2020}),
  \bibinfo{pages}{373--405}.
\newblock
\showISSN{0191-2615}


\bibitem[Liang and Wang(2018)]%
        {Liang:SIGIR_2018}
\bibfield{author}{\bibinfo{person}{Hongwei Liang} {and} \bibinfo{person}{Ke
  Wang}.} \bibinfo{year}{2018}\natexlab{}.
\newblock \showarticletitle{Top-k Route Search through Submodularity Modeling
  of Recurrent POI Features}. In \bibinfo{booktitle}{\emph{The 41st
  International ACM SIGIR Conference on Research \&amp; Development in
  Information Retrieval}} (Ann Arbor, MI, USA) \emph{(\bibinfo{series}{SIGIR
  '18})}. \bibinfo{publisher}{Association for Computing Machinery},
  \bibinfo{address}{New York, NY, USA}, \bibinfo{pages}{545–554}.
\newblock
\showISBNx{9781450356572}


\bibitem[Liu et~al\mbox{.}(2023)]%
        {Liu:IEEETransMC_2023}
\bibfield{author}{\bibinfo{person}{Linfeng Liu}, \bibinfo{person}{Yaoze Zhou},
  {and} \bibinfo{person}{Jia Xu}.} \bibinfo{year}{2023}\natexlab{}.
\newblock \showarticletitle{A Cloud-edge-end Collaboration Framework for
  Cruising Route Recommendation of Vacant Taxis}.
\newblock \bibinfo{journal}{\emph{IEEE Transactions on Mobile Computing}}
  (\bibinfo{year}{2023}), \bibinfo{pages}{1--16}.
\newblock
\showISSN{1558-0660}


\bibitem[Makhdomi and Gillani(2023)]%
        {Ashraf_archive:2022}
\bibfield{author}{\bibinfo{person}{Aqsa~Ashraf Makhdomi} {and}
  \bibinfo{person}{Iqra~Altaf Gillani}.} \bibinfo{year}{2023}\natexlab{}.
\newblock \showarticletitle{GNN-based passenger request prediction}.
\newblock \bibinfo{journal}{\emph{Transportation Letters}} \bibinfo{volume}{0},
  \bibinfo{number}{0} (\bibinfo{year}{2023}), \bibinfo{pages}{1--15}.
\newblock
\urldef\tempurl%
\url{https://doi.org/10.1080/19427867.2023.2283949}
\showDOI{\tempurl}


\bibitem[Petit(2020)]%
        {Article:UberPollution_2020}
\bibfield{author}{\bibinfo{person}{Yoann~Le Petit}.}
  \bibinfo{year}{2020}\natexlab{}.
\newblock \bibinfo{title}{Uber pollutes more than the cars it replaces–US
  scientists}.
\newblock
  \bibinfo{howpublished}{\url{https://www.transportenvironment.org/discover/uber-pollutes-more-cars-it-replaces-us-scientists/}}.
\newblock
\newblock
\shownote{Accessed: 2022-02-28}.


\bibitem[Qu et~al\mbox{.}(2022)]%
        {Qu:IEEETrans_2022}
\bibfield{author}{\bibinfo{person}{Boting Qu}, \bibinfo{person}{Linran Mao},
  \bibinfo{person}{Zhenzhou Xu}, \bibinfo{person}{Jun Feng}, {and}
  \bibinfo{person}{Xin Wang}.} \bibinfo{year}{2022}\natexlab{}.
\newblock \showarticletitle{How Many Vehicles Do We Need? Fleet Sizing for
  Shared Autonomous Vehicles With Ridesharing}.
\newblock \bibinfo{journal}{\emph{IEEE Transactions on Intelligent
  Transportation Systems}} \bibinfo{volume}{23}, \bibinfo{number}{9}
  (\bibinfo{date}{Sep.} \bibinfo{year}{2022}), \bibinfo{pages}{14594--14607}.
\newblock
\showISSN{1558-0016}


\bibitem[Qu et~al\mbox{.}(2020)]%
        {Qu:IEEETrans_2020}
\bibfield{author}{\bibinfo{person}{Boting Qu}, \bibinfo{person}{Wenxin Yang},
  \bibinfo{person}{Ge Cui}, {and} \bibinfo{person}{Xin Wang}.}
  \bibinfo{year}{2020}\natexlab{}.
\newblock \showarticletitle{Profitable Taxi Travel Route Recommendation Based
  on Big Taxi Trajectory Data}.
\newblock \bibinfo{journal}{\emph{IEEE Transactions on Intelligent
  Transportation Systems}} \bibinfo{volume}{21}, \bibinfo{number}{2}
  (\bibinfo{date}{Feb} \bibinfo{year}{2020}), \bibinfo{pages}{653--668}.
\newblock
\showISSN{1558-0016}


\bibitem[Qu et~al\mbox{.}(2014)]%
        {Qu:ACMKDD_2014}
\bibfield{author}{\bibinfo{person}{Meng Qu}, \bibinfo{person}{Hengshu Zhu},
  \bibinfo{person}{Junming Liu}, \bibinfo{person}{Guannan Liu}, {and}
  \bibinfo{person}{Hui Xiong}.} \bibinfo{year}{2014}\natexlab{}.
\newblock \showarticletitle{A Cost-Effective Recommender System for Taxi
  Drivers}. In \bibinfo{booktitle}{\emph{Proceedings of the 20th ACM SIGKDD
  International Conference on Knowledge Discovery and Data Mining}} (New York,
  New York, USA) \emph{(\bibinfo{series}{KDD '14})}.
  \bibinfo{publisher}{Association for Computing Machinery},
  \bibinfo{address}{New York, NY, USA}, \bibinfo{pages}{45–54}.
\newblock
\showISBNx{9781450329569}


\bibitem[Schreieck et~al\mbox{.}(2016)]%
        {SCHREIECK:ElsevierTransport_2016}
\bibfield{author}{\bibinfo{person}{Maximilian Schreieck},
  \bibinfo{person}{Hazem Safetli}, \bibinfo{person}{Sajjad~Ali Siddiqui},
  \bibinfo{person}{Christoph Pflügler}, \bibinfo{person}{Manuel Wiesche},
  {and} \bibinfo{person}{Helmut Krcmar}.} \bibinfo{year}{2016}\natexlab{}.
\newblock \showarticletitle{A Matching Algorithm for Dynamic Ridesharing}.
\newblock \bibinfo{journal}{\emph{Transportation Research Procedia}}
  \bibinfo{volume}{19} (\bibinfo{year}{2016}), \bibinfo{pages}{272--285}.
\newblock
\showISSN{2352-1465}


\bibitem[Shi et~al\mbox{.}(2021)]%
        {Shi:KDD_21}
\bibfield{author}{\bibinfo{person}{Dingyuan Shi}, \bibinfo{person}{Yongxin
  Tong}, \bibinfo{person}{Zimu Zhou}, \bibinfo{person}{Bingchen Song},
  \bibinfo{person}{Weifeng Lv}, {and} \bibinfo{person}{Qiang Yang}.}
  \bibinfo{year}{2021}\natexlab{}.
\newblock \showarticletitle{Learning to Assign: Towards Fair Task Assignment in
  Large-Scale Ride Hailing}. In \bibinfo{booktitle}{\emph{Proceedings of the
  27th ACM SIGKDD Conference on Knowledge Discovery \& Data Mining}}
  \emph{(\bibinfo{series}{KDD '21})}. \bibinfo{publisher}{Association for
  Computing Machinery}, \bibinfo{address}{New York, NY, USA},
  \bibinfo{pages}{3549–3557}.
\newblock
\showISBNx{9781450383325}


\bibitem[Sun et~al\mbox{.}(2022)]%
        {Sun:KDD_2022}
\bibfield{author}{\bibinfo{person}{Jiahui Sun}, \bibinfo{person}{Haiming Jin},
  \bibinfo{person}{Zhaoxing Yang}, \bibinfo{person}{Lu Su}, {and}
  \bibinfo{person}{Xinbing Wang}.} \bibinfo{year}{2022}\natexlab{}.
\newblock \showarticletitle{Optimizing Long-Term Efficiency and Fairness in
  Ride-Hailing via Joint Order Dispatching and Driver Repositioning}. In
  \bibinfo{booktitle}{\emph{Proceedings of the 28th ACM SIGKDD Conference on
  Knowledge Discovery and Data Mining}}. \bibinfo{pages}{3950--3960}.
\newblock


\bibitem[Ta et~al\mbox{.}(2018)]%
        {Ta:IEEETrans_2018}
\bibfield{author}{\bibinfo{person}{Na Ta}, \bibinfo{person}{Guoliang Li},
  \bibinfo{person}{Tianyu Zhao}, \bibinfo{person}{Jianhua Feng},
  \bibinfo{person}{Hanchao Ma}, {and} \bibinfo{person}{Zhiguo Gong}.}
  \bibinfo{year}{2018}\natexlab{}.
\newblock \showarticletitle{An Efficient Ride-Sharing Framework for Maximizing
  Shared Route}.
\newblock \bibinfo{journal}{\emph{IEEE Transactions on Knowledge and Data
  Engineering}} \bibinfo{volume}{30}, \bibinfo{number}{2}
  (\bibinfo{year}{2018}), \bibinfo{pages}{219--233}.
\newblock


\bibitem[Thangaraj et~al\mbox{.}(2017)]%
        {Thangaraj:IEEEConf_2017}
\bibfield{author}{\bibinfo{person}{Raja~Subramaniam Thangaraj},
  \bibinfo{person}{Koyel Mukherjee}, \bibinfo{person}{Gurulingesh Raravi},
  \bibinfo{person}{Asmita Metrewar}, \bibinfo{person}{Narendra Annamaneni},
  {and} \bibinfo{person}{Koushik Chattopadhyay}.}
  \bibinfo{year}{2017}\natexlab{}.
\newblock \showarticletitle{Xhare-a-Ride: A Search Optimized Dynamic Ride
  Sharing System with Approximation Guarantee}. In
  \bibinfo{booktitle}{\emph{2017 IEEE 33rd International Conference on Data
  Engineering (ICDE)}}. \bibinfo{pages}{1117--1128}.
\newblock
\showISSN{2375-026X}


\bibitem[Tong et~al\mbox{.}(2017)]%
        {Tong:VLDB_2017}
\bibfield{author}{\bibinfo{person}{Yongxin Tong}, \bibinfo{person}{Libin Wang},
  \bibinfo{person}{Zhou Zimu}, \bibinfo{person}{Bolin Ding},
  \bibinfo{person}{Lei Chen}, \bibinfo{person}{Jieping Ye}, {and}
  \bibinfo{person}{Ke Xu}.} \bibinfo{year}{2017}\natexlab{}.
\newblock \showarticletitle{Flexible online task assignment in real-time
  spatial data}.
\newblock \bibinfo{journal}{\emph{Proceedings of the VLDB Endowment}}
  \bibinfo{volume}{10}, \bibinfo{number}{11} (\bibinfo{year}{2017}),
  \bibinfo{pages}{1334--1345}.
\newblock


\bibitem[Tong et~al\mbox{.}(2022)]%
        {Tong:ACMTrans_2022}
\bibfield{author}{\bibinfo{person}{Yongxin Tong}, \bibinfo{person}{Yuxiang
  Zeng}, \bibinfo{person}{Zimu Zhou}, \bibinfo{person}{Lei Chen}, {and}
  \bibinfo{person}{Ke Xu}.} \bibinfo{year}{2022}\natexlab{}.
\newblock \showarticletitle{Unified Route Planning for Shared Mobility: An
  Insertion-Based Framework}.
\newblock \bibinfo{journal}{\emph{ACM Trans. Database Syst.}}
  \bibinfo{volume}{47}, \bibinfo{number}{1}, Article \bibinfo{articleno}{2}
  (\bibinfo{date}{may} \bibinfo{year}{2022}), \bibinfo{numpages}{48}~pages.
\newblock
\showISSN{0362-5915}


\bibitem[Tong et~al\mbox{.}(2018)]%
        {Tong:vldb_2018}
\bibfield{author}{\bibinfo{person}{Yongxin Tong}, \bibinfo{person}{Yuxiang
  Zeng}, \bibinfo{person}{Zimu Zhou}, \bibinfo{person}{Lei Chen},
  \bibinfo{person}{Jieping Ye}, {and} \bibinfo{person}{Ke Xu}.}
  \bibinfo{year}{2018}\natexlab{}.
\newblock \showarticletitle{A unified approach to route planning for shared
  mobility}.
\newblock \bibinfo{journal}{\emph{Proceedings of the VLDB Endowment}}
  \bibinfo{volume}{11}, \bibinfo{number}{11} (\bibinfo{year}{2018}),
  \bibinfo{pages}{1633}.
\newblock


\bibitem[Vazifeh et~al\mbox{.}(2018)]%
        {Vazifeh:Nature_2018}
\bibfield{author}{\bibinfo{person}{Mohammad~M Vazifeh}, \bibinfo{person}{Paolo
  Santi}, \bibinfo{person}{Giovanni Resta}, \bibinfo{person}{Steven~H
  Strogatz}, {and} \bibinfo{person}{Carlo Ratti}.}
  \bibinfo{year}{2018}\natexlab{}.
\newblock \showarticletitle{Addressing the minimum fleet problem in on-demand
  urban mobility}.
\newblock \bibinfo{journal}{\emph{Nature}} \bibinfo{volume}{557},
  \bibinfo{number}{7706} (\bibinfo{year}{2018}), \bibinfo{pages}{534--538}.
\newblock


\bibitem[Verma et~al\mbox{.}(2017)]%
        {Verma:aai_2017}
\bibfield{author}{\bibinfo{person}{Tanvi Verma}, \bibinfo{person}{Pradeep
  Varakantham}, \bibinfo{person}{Sarit Kraus}, {and}
  \bibinfo{person}{Hoong~Chuin Lau}.} \bibinfo{year}{2017}\natexlab{}.
\newblock \showarticletitle{Augmenting decisions of taxi drivers through
  reinforcement learning for improving revenues}. In
  \bibinfo{booktitle}{\emph{Proceedings of the International Conference on
  Automated Planning and Scheduling}}, Vol.~\bibinfo{volume}{27}.
  \bibinfo{pages}{409--417}.
\newblock


\bibitem[Wang et~al\mbox{.}(2022)]%
        {Wang:VLDB_2022}
\bibfield{author}{\bibinfo{person}{Jiachuan Wang}, \bibinfo{person}{Peng
  Cheng}, \bibinfo{person}{Libin Zheng}, \bibinfo{person}{Lei Chen}, {and}
  \bibinfo{person}{Wenjie Zhang}.} \bibinfo{year}{2022}\natexlab{}.
\newblock \showarticletitle{Online Ridesharing with Meeting Points}.
\newblock \bibinfo{journal}{\emph{Proceedings of the VLDB Endowment}}
  \bibinfo{volume}{15}, \bibinfo{number}{13} (\bibinfo{year}{2022}),
  \bibinfo{pages}{3963--3975}.
\newblock


\bibitem[Wang et~al\mbox{.}(2020)]%
        {Wang:VLDB_2020}
\bibfield{author}{\bibinfo{person}{Jiachuan Wang}, \bibinfo{person}{Peng
  Cheng}, \bibinfo{person}{Libin Zheng}, \bibinfo{person}{Chao Feng},
  \bibinfo{person}{Lei Chen}, \bibinfo{person}{Xuemin Lin}, {and}
  \bibinfo{person}{Zheng Wang}.} \bibinfo{year}{2020}\natexlab{}.
\newblock \showarticletitle{Demand-Aware Route Planning for Shared Mobility
  Services}.
\newblock \bibinfo{journal}{\emph{Proc. VLDB Endow.}} \bibinfo{volume}{13},
  \bibinfo{number}{7} (\bibinfo{date}{mar} \bibinfo{year}{2020}),
  \bibinfo{pages}{979–991}.
\newblock
\showISSN{2150-8097}


\bibitem[Wang et~al\mbox{.}(2019)]%
        {Wang:ACMKDD_2019}
\bibfield{author}{\bibinfo{person}{Yuandong Wang}, \bibinfo{person}{Hongzhi
  Yin}, \bibinfo{person}{Hongxu Chen}, \bibinfo{person}{Tianyu Wo},
  \bibinfo{person}{Jie Xu}, {and} \bibinfo{person}{Kai Zheng}.}
  \bibinfo{year}{2019}\natexlab{}.
\newblock \showarticletitle{Origin-Destination Matrix Prediction via Graph
  Convolution: A New Perspective of Passenger Demand Modeling}. In
  \bibinfo{booktitle}{\emph{Proceedings of the 25th ACM SIGKDD International
  Conference on Knowledge Discovery \& Data Mining}} (Anchorage, AK, USA)
  \emph{(\bibinfo{series}{KDD '19})}. \bibinfo{publisher}{Association for
  Computing Machinery}, \bibinfo{address}{New York, NY, USA},
  \bibinfo{pages}{1227–1235}.
\newblock
\showISBNx{9781450362016}


\bibitem[Wang et~al\mbox{.}(2021)]%
        {Wang:ACMTrans_2022}
\bibfield{author}{\bibinfo{person}{Yuandong Wang}, \bibinfo{person}{Hongzhi
  Yin}, \bibinfo{person}{Tong Chen}, \bibinfo{person}{Chunyang Liu},
  \bibinfo{person}{Ben Wang}, \bibinfo{person}{Tianyu Wo}, {and}
  \bibinfo{person}{Jie Xu}.} \bibinfo{year}{2021}\natexlab{}.
\newblock \showarticletitle{Passenger Mobility Prediction via Representation
  Learning for Dynamic Directed and Weighted Graphs}.
\newblock \bibinfo{journal}{\emph{ACM Trans. Intell. Syst. Technol.}}
  \bibinfo{volume}{13}, \bibinfo{number}{1}, Article \bibinfo{articleno}{2}
  (\bibinfo{date}{nov} \bibinfo{year}{2021}), \bibinfo{numpages}{25}~pages.
\newblock
\showISSN{2157-6904}


\bibitem[Xu et~al\mbox{.}(2020)]%
        {Xu:ACM_SSDM_2020}
\bibfield{author}{\bibinfo{person}{Yixin Xu}, \bibinfo{person}{Jianzhong Qi},
  \bibinfo{person}{Renata Borovica-Gajic}, {and} \bibinfo{person}{Lars Kulik}.}
  \bibinfo{year}{2020}\natexlab{}.
\newblock \showarticletitle{Geoprune: Efficiently matching trips in
  ride-sharing through geometric properties}. In \bibinfo{booktitle}{\emph{32nd
  International Conference on Scientific and Statistical Database Management}}.
  \bibinfo{pages}{1--12}.
\newblock


\bibitem[Yengejeh and Smith(2021)]%
        {Yengejeh:IEEETrans_2021}
\bibfield{author}{\bibinfo{person}{Armin~Sadeghi Yengejeh} {and}
  \bibinfo{person}{Stephen~L. Smith}.} \bibinfo{year}{2021}\natexlab{}.
\newblock \showarticletitle{Rebalancing Self-Interested Drivers in Ride-Sharing
  Networks to Improve Customer Wait-Time}.
\newblock \bibinfo{journal}{\emph{IEEE Transactions on Control of Network
  Systems}} \bibinfo{volume}{8}, \bibinfo{number}{4} (\bibinfo{date}{Dec}
  \bibinfo{year}{2021}), \bibinfo{pages}{1637--1648}.
\newblock
\showISSN{2325-5870}


\bibitem[Yuen et~al\mbox{.}(2019)]%
        {Yuen:ACMWWW_2019}
\bibfield{author}{\bibinfo{person}{Chak~Fai Yuen},
  \bibinfo{person}{Abhishek~Pratap Singh}, \bibinfo{person}{Sagar Goyal},
  \bibinfo{person}{Sayan Ranu}, {and} \bibinfo{person}{Amitabha Bagchi}.}
  \bibinfo{year}{2019}\natexlab{}.
\newblock \showarticletitle{Beyond Shortest Paths: Route Recommendations for
  Ride-Sharing}. In \bibinfo{booktitle}{\emph{The World Wide Web Conference}}
  \emph{(\bibinfo{series}{WWW '19})}. \bibinfo{publisher}{Association for
  Computing Machinery}, \bibinfo{address}{New York, NY, USA},
  \bibinfo{pages}{2258–2269}.
\newblock
\showISBNx{9781450366748}


\end{thebibliography}

\end{document}